\documentclass[twocolumn,showpacs,preprintnumbers,amsmath,amssymb]{revtex4}
\usepackage{graphicx}
\usepackage{epstopdf}
\usepackage{dcolumn}
\usepackage{bm}
\usepackage[french]{babel}
\usepackage[applemac]{inputenc}
\usepackage{undertilde}
\newcommand{\csEho}{{{}^1 \! E_{1/2}}}
\newcommand{\csEht}{{{}^2 \! E_{1/2}}}
\newcommand{\cC}{{\cal C}}
\newcommand{\cB}[1]{{\cal B}_{#1}}
\newcommand{\cS}[1]{{\cal S}_{#1}}
\newcommand{\wGam}{\widetilde{\Gamma}}

\begin{document}
\preprint{APS/123-QED}
\title{Maximal symetrization and reduction of fields: \\ application to wavefunctions in solid state nanostructures}
\author{S. Dalessi}
\altaffiliation[Present address: ]{Computational Biology Group, Department of Medical Genetics, University of Lausanne, Rue du Bugnon 27, CH-1005 Lausanne, Switzerland }
\author{M.-A. Dupertuis}
\email[E-mail me at: ]{Marc-Andre.Dupertuis@epfl.ch}
\homepage[]{http://people.epfl.ch/marc-andre.dupertuis}
\affiliation{Laboratory of Physics of Nanostructures, Ecole Polytechnique F\'{e}d\'{e}rale de Lausanne (EPFL), CH-1015 Lausanne, Switzerland }
\date{\today}
\begin{abstract}
A novel general formalism for the maximal symetrization and reduction of fields (MSRF) is proposed and applied to wavefunctions in solid state nanostructures. Its primary target is to provide an essential tool for the study and analysis of the electronic and optical properties of semiconductor quantum heterostructures with relatively high point-group symmetry, and studied with the $k\cdot p$ formalism. Nevertheless the approach is valid in a much larger framework than $k\cdot p$ theory, it is applicable to arbitrary systems of coupled partial differential equations (e.g. strain equations or Maxwell equations). This general MSRF formalism makes extensive use of group theory at {\em all} levels of analysis. For spinless problems (scalar equations), one can use a systematic Spatial Domain Reduction (SDR) technique which allows, for every irreducible representation, to reduce the set of equations on a minimal domain with automatic incorporation of the boundary conditions at the border, which are shown to be non-trivial in general. For a vectorial or spinorial set of functions, the SDR technique must be completed by the use of an optimal basis in vectorial or spinorial space (in a crystal we call it the optimal Bloch function basis (OBB)). The full MSR formalism thus consists of three steps : 1) explicitly separate spatial (or Fourier space) and vectorial (spinorial) part of the operators and eigenstates, 2) choose, according to the symmetry and well defined prescriptions (e.g. specific transformation properties), optimal fully symmetrized basis for both spatial and vector (or spin) space, and 3) finally apply the SDR to every individual scalar ultimate component function. We show that with such a formalism the coupling between different vectorial (spinorial) components by symmetry operations becomes minimized and every ultimately reduced envelope function (UREF) acquires a well-defined specific symmetry. The advantages are numerous: sharper insights on the symmetry properties of every eigenstate, minimal coupling schemes (analytically and computationally exploitable at the component function level), minimal computing domains. The formalism can be applied also as a postprocessing operation, offering all subsequent analytical and computationnal advantages of symmetrization. The specific case of a quantum wire (QWRs) with $C_{3v}$ point group symmetry is used as a concrete illustration of the application of MSRF. %
\end{abstract}
\maketitle
\section{introduction}
\label{sec_intro}
The study of low dimensional solid state nanostructures is a very interesting and promising domain. Indeed a good knowledge of electronic and optical properties of nanostructures like metallic~\cite{GIR} or semiconductor~\cite{YOF} nanostructures, or photonic crystals~\cite{SAK}, is essential for many applications in advanced lasers, photonics and telecommunications. New truly quantum applications like quantum cryptography also make extensive use of semiconductor quantum heterostructures like quantum wells, quantum wires and quantum dots~\cite{Stevenson2006,YOUNG}. The quality of semiconductor quantum wires and dots have been extensively improved during the last fifteen years, and one is now able to produce high quality structures with higher and higher point-group symmetries (e.g. $C_{6v}$ quantum dots~\cite{Kako2004,OReilly2000}). In such a case a group-theoretical approach is usually the most powerful tool for describing the effects resulting from symmetry on the electronic states, their optical properties in particular. However the problem is rather complicated since in heterostructures one must take into account both the underlying microscopical crystalline structure and the mesoscopic heterostructure confinement potential.

The theoretical study of low symmetry effects in semiconductor heterostructures (like quantum wires with $C_s$ symmetry, e.g. T- and V-shaped quantum wires~\cite{Akiyama98,Runge02}) is already well developed~\cite{MAD00} and has lead to fundamental conclusions regarding their electronic and optical properties. First electronic and excitonic states can be labelled with respect to their characteristic transformation properties under symmetry operations, second rigorous and important selection rules were readily obtained on the basis of such a classification, useful even in a low symmetry case~\cite{MAD00,MAD00_2}. The effects of lateral confinement to the polarization anisotropy were largely studied~\cite{MAD98, SERC,SERC1, BOC91,CIT1}. However, it should be pointed out that up to now only very little work has been devoted to higher symmetries (called here HSH: High Symmetry Heterostructure), for example $C_{3v}$ structures~\cite{HAR99,FAB04}, or even $C_{6v}$~\cite{TRO}. A particularity of HSH is to allow the existence of symmetry-induced degenerate eigenstates, related to irreducible representations (irreps) with dimension greater than one~\cite{COR, ALT}, and which display a much more complex behavior under symmetry operations. Symmetry-induced degenerate eigenstates play an important role in the generation of entangled photon pairs from quantum dots~\cite{YOUNG}.

The electronic structure of semiconductor heterostructures are very often studied in the frame of the $k \cdot p$ envelope function approach for heterostructures~\cite{BAST}, with at least four bands when describing the valence band. In such a frame the different envelope functions (components of the spinorial eigenstates) become entangled under symmetry operations: their shapes are therefore {\em mutually coupled}, and their behavior is complex. Up to now there has been very few attempts to try to use an Optimal Bloch function Basis (OBB). In~\cite{MAD00} we tried to rely on the concept of an 'Optimal Quantization Axis (OQA) direction' (A Bloch function basis which diagonalizes the component of angular momentum in the chosen optimal direction). In fact we shall show in the following that such a method is of limited interest: it is optimally adapted only in very low symmetry cases like $C_s$ structures! For Quantum Wires (QWR) with a higher symmetry group, the previously defined OQA direction may only be an improved choice, not the optimal choice. 

In this paper, we propose the optimal and systematic solution to this problem: a general \textit{Maximal Symmetrization and Reduction (MSRF) formalism}, perfectly adapted to the study of scalar or spinorial HSH problems. We will show how to find true OBBs which minimize the coupling between envelope functions, and which truly maximize their individual symmetry. Moreover we will show how to systematically compute the whole solution on a reduced {\em minimal domain}. With the improvement of growth techniques increasingly high symmetries can indeed be produced {\em enhancing the need for novel tools allowing to fully take into account the symmetry properties} and to significantly simplify, theoretically and numerically, the understanding and the computation of electronic and optical properties. We originally developed the MSRF formalism to study a $C_{3v}$ quantum wire (QWR), and therefore we shall often use it as an example, but one should stress that the method is general, applicable to other groups and to widely different cases since in fact its possible scope of application is much wider: it could be applied in its full generality to arbitrary tensorial fields obeying a set partial differential equations characterized by any given point group symmetry. In particular the method is independent of the number of coupled bands kept in the problem, and independent of the specific terms kept, provided the global symmetry is conserved. For example we could easily take into account interface terms~\cite{FOR93,LAS04} or strain terms in an eight-band approach~\cite{STIER,SAUV}, but in case of strain we would also need to treat in the same way the elasticity equations for the strain tensor (this could be done most conveniently by post-symmetrization of the elasticity calculation, see end of Section~\ref{sec_SDR_spin}). The MSRF method is also independent of dimension, it applies equally well for two or three spatial dimensions, i.e. to QWR or to quantum dot (QD) heterostructures with a given point-group symmetry.

Let us now shortly present the heart of the MSRF formalism, which is threefold. First for every quantum states one performs an \textit{explicit} separation of the spatial character (3D orbital motion, eventually treated in Fourier space) and of the field character (e.g. spinorial). Second one selects \textit{optimal fully symmetrized bases}, both for spinorial space (the Optimal Bloch function Basis (OBB)) and orbital space, which minimize the coupling between different spinorial components. These two ingredients allow to obtain every spinorial component of the field as a sum of \textit{symmetrized scalar functions of spatial coordinates}. Third \textit{for every irrep} one identifies minimum sets of independent parameters (orbital reduced domains) which form the \textit{systematic Spatial (or Fourier) Domain Reduction technique (SDR)} and which allow to obtain reduced Hamiltonians with respect to reduced domains, and systematically minimally reduce the computing requirements. 

The advantages of the new MSRF formalism are manifold. Indeed besides the possibility of performing SDR we shall show that there are also many advantages from the analytical point of view: first the Hamiltonian operator usually takes a simpler form in the adapted fully symmetrized basis (OBB), second the spinorial components of eigenstates (as well as the components of any operator in the spinorial basis) can be treated in a similar way and easily decomposed into fundamental parts to which single group irreps can be associated (and for which "sub-selection rules" can be applied at an intermediate calculational level). In this way particularly simple analytical expressions can be obtained for certain operator matrix elements, which allow to find, for example, new analytical ratios in the polarization anisotropy that were previously unnoticed in the numerics~\cite{ZHU06}. Further insight can also be gained from the fact that this symmetry-based technique simplifies the expression of coupling matrix elements. Most notably weak symmetry breaking mechanisms can be understood more deeply at the analytical level~\cite{DAL1}. From the numerical point of view, the systematic SDR technique will enable one to solve independently for every irrep on a minimally reduced solution domain. The SDR technique not only allows to find the boundary conditions at the boundary of such a domain, shown below to be non-trivial, but it also allows to eliminate the need to explicitely care for them! It should be pointed out that the MSRF method can also be used as a {\em post symmetrization technique} on numerical results obtained without taking into account any symmetry. In such a case it not only allows to classify all eigenfunctions and symmetrize them within the OBB, and benefit of an in-depth symmetry analysis, but in all subsequent computations symmetrized wavefunctions on reduced domain can then be used, which may still represent a further significant potential gain.

Let us now detail the necessary procedures of the proposed MSRF technique. 

For scalar problems, like the single band $k\cdot p$ spinless conduction band Hamiltonian, it reduces to our systematic SDR procedure. The SDR procedure involves two fundamental steps: first the spatial domain must be decomposed into a minimal number of disjoint sub-domains which map into each other through symmetry operations (including borders as separate domains), second the set of domains and wavefunctions must be projected on the relevant irreps. This last step allows to find the critical geometrical features of all states by identifying for every irrep the minimal independent parts of any function of a given symmetry, to which a corresponding reduced subdomain can therefore be associated. At this stage non-trivial boundary conditions can be derived if necessary. The same procedure can then be carried out for all relevant functional operators like the Hamiltonian. The reduced Hamiltonian reflects directly the coupling between different sub-domains and does incorporate {\em automatically} the restrictions implied by non-trivial geometrical boundary conditions.\\
For spin dependent problems, like the $ 4\times 4$ $k\cdot p$ Luttinger Hamiltonian describing the valence band in diamond semiconductors, or the much used eight-band $k\cdot p$ Hamiltonian~\cite{SAUV}, the OBB basis functions must be first found. They transform like an irrep of the (double) group, and allow to block-diagonalize the corresponding matrix representation of the double-group. By choosing the OBB, one enforces a minimal coupling between different spinorial function components under symmetry operations. Since every component can then be decomposed in a simple way into scalar envelope functions labelled with single groups irreps, the SDR technique is applicable to every component, and reduced Hamiltonians can be found for every double-group irrep.

In section~\ref{sec_form} we shall recall in more details essential results obtained in low symmetry heterostructures which are needed to understand the HSH challenge and establishing the basis for the development of the MSRF formalism. In Section~\ref{sec_theo0}, transformation laws in both ordinary space and spin space are studied, together with fundamental group-theoretical results needed for the definition of the fully symmetrized OBB basis and for the separation of spinorial and spatial parts. These goals are attained in Section~\ref{sec_OBB}). In the next two Sections, we first develop the SDR formalism for single group spinless (scalar) functions (Sec.~\ref{sec_SDR}), and in the second we apply the SDR to the symmetrized envelope functions created by the OBB (Sec.~\ref{sec_SDR_spin}). In addition we show how the technique leads to reduced Hamiltonians, and how it can be used as a post-symmetrization technique. Finally, in Section~\ref{sec_SR} we demonstrate how selection rules can be applied at an intermediate level - a specific feature of the MSRF formalism - to compute the matrix elements of operators. As a result we find novel strong analytical results (polarization anisotropy in $C_{3v}$ structures) which can be interpreted with the help of the generalized Wigner-Eckart theorem for point groups. Finally, in Section~\ref{sec_autres_ex}, an outlook is given on other symmetry groups (the hexagonal $C_{6v}$ group, an approximate zone-center symmetry group $D_{3h}$, the (commutative) $C_n\, ,\, n\in \mathbb{N}$ subgroup of the rotation group and the $C_s$ group). The potential of the MSRF formalism for different problems is shortly described in Section~\ref{orig_MSRF} as well as its relationship with the most close works found in the litterature on heterostructures which exploit symmetry.

\section{Envelope-function theory of low symmetry heterostructures and the HSH challenge}
\label{sec_form}
The MSRF formalism developed in this paper is built upon specific techniques previously developed for low symmetry heterostructures. The main goal of this section is not only to introduce the basic envelope function $k\cdot p$ Hamiltonians for the conduction and valence band that will be used throughout the paper, but also to recall fundamental results on low symmetry heterostructures~\cite{MAD00} which are at the origin of MSRF. We will also show explicitly the limitations of these techniques which do not allow to reach {\em maximal} symmetrization for higher symmetry heterostructures, which thus represent a main challenge. These results will form an essential basis for the systematic study of transformation laws developed in Section~\ref{sec_theo0} and the development of the cornerstone of the MSRF in Section~\ref{sec_OBB}.

\subsection{Introduction to envelope function models}
\label{sub_form_env}
Multiband $k \cdot p$ Hamiltonians are extensively used for the study of the electronic structure in semiconductor heterostructures~\cite{BAST}. Such models allow to introduce all the relevant physics close to a high symmetry point of the band-structure, whilst keeping maximum simplicity. For many applications in III-V zincblende semiconductors like $AlGaAs/GaAs$ it is possible to treat separately the conduction and the valence band problems. Nevertheless the method presented in the following is generalizable to more complex multiband schemes which treat simultaneously the coupling between these bands, like eight-band or fourteen-band $k \cdot p$ Hamiltonians.

Let us also assume a heterostructure translation with translation invariance in some spatial directions and which are confining in the remaining directions, like quantum wells or quantum wires (see Fig.~\ref{fig_vQWRpersp}), and split the coordinate system into $\textbf r_\parallel$ and $\textbf r_\bot$ respectively. 

For the isolated conduction band the $k \cdot p$ approach gives rise to the simple effective-mass model when one ignores spin-splitting terms. In the case of QWRs or QWs the conduction band Hamiltonian operator $H$ is defined by its action $H[\psi]$ on any electron state $\psi$. In the $\textbf{r}$ representation, this amounts to apply the following differential operator
\begin{equation}\label{eqt_bc}
H(\textbf{r}_\bot,\textbf k_{{}^\parallel}) = 
- \frac{\hbar^2}{2}\nabla_\bot\frac{1}{m\left(\mathbf{r}_\bot\right)}\nabla_\bot 
+ \frac{\hbar^2 \, \textbf k_{{}^\parallel}^2}{2 \, 
m\left(\mathbf{r}_\bot\right)} + V_c\left(\mathbf{r}_\bot\right)
\end{equation}
on the electron wavefunction $\psi(\textbf r_\bot)$, which is a simple scalar function of position. Here the perpendicular gradient $\nabla_\bot$ is of course related to the confined directions, and translation invariance of $H$ implies a translation invariant confining potential $V_c\left(\mathbf{r}_\bot\right)$ and effective mass $m\left(\mathbf{r}_\bot\right)$, both independent of $\textbf r_\parallel$, and the appearance of the corresponding good quantum number $\textbf k_{{}^\parallel}$, which can be interpreted as the electron momentum in the free directions. Note that we have used for convenience in Eq.~(\ref{eqt_bc}) the somewhat clumsy notation $H(\textbf{r}_\bot,\textbf k_{{}^\parallel})$, which comprises differential operators like $\nabla_\bot$ and variables like $\textbf{r}_\bot$ and $k_{{}^\parallel}$, keeping in mind that, in the following, eventual transformations on the argument $\textbf{r}_\bot$ of $H(\textbf{r}_\bot,\textbf k_{{}^\parallel})$ must also be applied consistently to $\nabla_\bot$. Let us also denote $\psi_{\textbf k_{{}^\parallel}}(\textbf r_\bot)$, the eigenstates of $H(\textbf{r}_\bot,\textbf k_{{}^\parallel})$ associated with the eigenvalue $E_{\textbf k_{{}^\parallel}}$. 

For the valence-band let us use the minimal four band Luttinger Hamiltonian~\cite{LUT55, LUT56,BAST,FISH} which is required when one wants to have a good estimate of the optical polarization anisotropy. Such model also provides a fairly good description of the QWR valence subband energy dispersion close to the $\Gamma$ zone-center. 

In its original form the bulk Luttinger Hamiltonian can be written as~\cite{LUT55, LUT56} 
\begin{eqnarray}
H_L &= &- \frac{\hbar^2}{m_0} \left\{ \frac{1}{2}(\gamma_1+\frac{5}{2}\gamma_2) k^2 - \gamma_2 \sum_i k_i^2J_i^2 \right. \nonumber \\
& &\left. - \gamma_e \sum_\circlearrowleft k_ik_j(J_iJ_j+J_jJ_i) \right\}
\label{eqt_inv_ok}
\end{eqnarray}
where $\circlearrowleft$ denotes all cyclic permutations of the indices ($i\neq j\in\{1,2,3\}$) of the $4\times 4$ matrix representations of the components of quasi-angular momentum $J_i, i=1,2,3$. The corresponding envelope function Hamiltonian for the QWR valence band reads:
\begin{equation}\label{eqt_bv} 
H\left(\mathbf{r}_\bot,k \right) = H_L\left(\mathbf{r}_\bot,k \right) +I_4 \, V_v(\mathbf{r}_\bot) 
\end{equation}
where $I_4 \, V_v(\mathbf{r}_\bot)$ is the diagonal part incorporating the effect of the heterostructure confinement potential, here $k\equiv{\textbf k}_\parallel$ is the wave vector along the wire, and $H_L\left(\mathbf{r}_\bot,k \right)$ is the kinetic part as given by the bulk Luttinger Hamiltonian where $\mathbf{r}_\bot\rightarrow\ -i\mathbf{\nabla}_\bot$. As a result the kinetic part can always be put in the following standard form~\cite{FISH}:
\begin{equation}\label{eqt_lutt}
H_L = -\frac{\hbar^2}{m_0}
\left(
\begin{array}{cccc}
 p+q  &  -s  &  r   &  0   \\
 -s^+ &  p-q &  0   &  r   \\
 r^+  &  0   &  p-q &  s   \\
 0    &  r^+ &  s^+ &  p+q 
\end{array}
\right)
\end{equation}
In this equation the $p,q,r,s$ matrix elements are $k$-dependent partial differential operators obtained from the Luttinger quadratic polynomials~(\ref{eqt_inv_ok}). Eq.~(\ref{eqt_lutt}) may also involve spatially-dependent Luttinger parameters $\gamma_i(\mathbf{r}_\bot), i=1,2,3$ corresponding to the spatial composition dependance of the heterostructure. The eigenstates of $H\left(\mathbf{r}_\bot,k \right)$ can be considered as 4D spinorial fields (related to the $j=3/2$ Bloch function basis at the top of the valence-band), and each component of this field is a $k$-dependent scalar field of $\mathbf{r}_\bot$, the so-called envelope functions.

At this stage a few fundamental comments are in order. First the actual form of the $p,q,r,s$ coefficients is a function of the Bloch function basis chosen. The standard form~\cite{FISH}, corresponding to the main crystal directions $[100],[010],[001]$, is obtained using standard matrix representations~\cite{MES} of $J_i, i=1,2,3$. However for heterostructures whose main symmetry elements differ, one should usually define a rotated cartesian frame with a new $z$-axis oriented along adapted directions, e.g. see~\cite{FISH} for $[hhk]$ directions, such that the corresponding implicit Bloch function basis diagonalizes the new $J_z$ component of the quasi-angular momentum. 

It is however important to point out that {\em the shape of each envelope function is basis-dependent}, even though {\em the expectation value of any physical quantity remains basis-independent}! This point, which will be illustrated in the next subsection, was not explicitely recognized for a long time, and complicates a lot the question of the symmetry of the envelope functions in the presence of valence band mixing. 

\subsection{The effects of symmetry on the envelope functions}\label{sym_effects}
It is well known~\cite{BAS} that the existence of a symmetry group for an Hamiltonian allows to classify each kind of eigenstates, to deduce their degeneracies, and to specify their possible transformation properties, however in the case of a spinorial set of envelope functions, the analysis of the symmetry of every individual envelope and its transformation properties has not been addressed and requires much more work (Section~\ref{sec_theo0}). Let us give here an intuitive justification for this extensive effort. We shall first discuss the effects of a single symmetry plane, already quite well-known (see~\cite{MAD00}, and subsequent works~\cite{MAD00_2,MAD02,MAD05,MAD05_2}). Then we shall enlighten in detail the difficulties involved in applying these results, based on rather elementary concepts, to higher symmetries. The discussion will both develop intuition and a new understanding of the difficulties involved, demonstrating the need for MSRF.
\subsubsection{A low symmetry case: $C_s$ symmetry}
\label{lowCSsymm}
\begin{figure}[hbt]
\includegraphics[scale = 0.65]{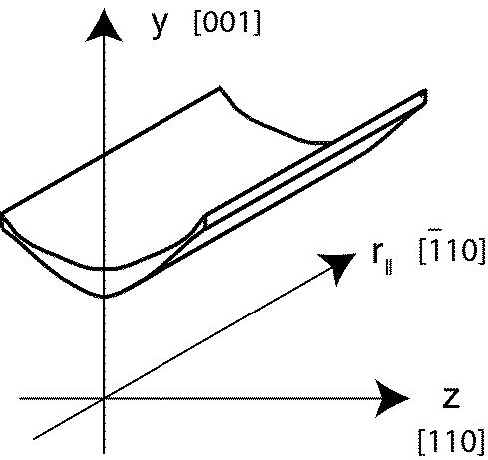}
\caption{View in perspective of a typical V-shaped $Al_xGa_{1-x}As$ QWR, with coordinate system. The central part is typically pure $GaAs$, whilst the surrounding bulk part is typically $Al_x Ga_{1-x} As$ with $x \approx 30\%$.}
\label{fig_vQWRpersp}
\end{figure}

\begin{figure}[hbt]
\begin{tabular}{cc}
\includegraphics[scale = 0.9]{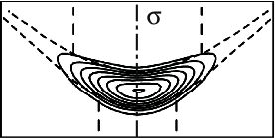} &
\includegraphics[scale = 0.9]{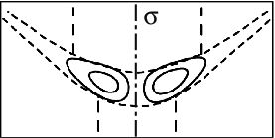} \\
\textbf{(a)} & \textbf{(b)} \\
\end{tabular}
\caption{Contour plots of conduction band envelope functions of the first two electronic states in a typical V-shaped QWR  \\
\textbf{(a)} Even function (ground state); \textbf{(b)} Odd function (first excited state). \\
The V-shaped QWR $GaAs$ potential well is shown with a doted line, the 
vertical quantum well with $x\approx 20\%$ is also visible.}
\label{fig_etat_V_BC}
\end{figure}

Let us now assume a heterostructure with the simplest symmetry, like the typical V-shaped QWR shown in Fig.~\ref{fig_vQWRpersp}, with a single symmetry plane. The corresponding symmetry point group is $C_s$. Let us take the coordinate system such that $\textbf r_\bot=(y,z)$. The symmetry of the heterostructure with respect to the plane implies that $V_c\left(\mathbf{r}_\bot\right)$ and $m\left(\mathbf{r}_\bot\right)$ are both invariant with respect to the symmetry plane operation $\sigma: \mathbf{r}_\bot \mapsto \sigma(\mathbf{r}_\bot)=(y,-z)$. 
The wavefunction profiles of the first two eigenstates of the stationary Schr\"{o}dinger equation are shown on Fig.~\ref{fig_etat_V_BC}.
It is easy to show that the symmetry of the structure implies that the eigenstate symmetry can be labelled as being either ``even'' or ``odd''. Indeed we see on Fig.~\ref{fig_etat_V_BC} that the ground state wavefunction $\psi^{(e)}_{\textbf k_{{}^\parallel}}(\textbf r_\bot)$ is strictly {\em even} with respect to the symmetry plane $\sigma$, whilst the first excited state wavefunction $\psi^{(o)}_{\textbf k_{{}^\parallel}}(\textbf r_\bot)$ is strictly {\em odd}. The higher states will all display one character or the other unless there would be an accidental degeneracy, which would then allow an accidental mixing within the degenerate subspace. To summarize: the eigenfunctions all obey one of the following transformation rules under $\vartheta_\sigma$, the operation of reversing a wavefunction with respect to the $\sigma$-symmetry plane
\begin{eqnarray}\label{even_odd_bc}
\vartheta_\sigma \, \psi^{(e)}_{\textbf k_{{}^\parallel}}(\textbf r_\bot) 
= \psi^{(e)}_{\textbf k_{{}^\parallel}}(\sigma^{-1}\textbf r_\bot) 
= + \psi^{(e)}_{\textbf k_{{}^\parallel}}(\textbf r_\bot) \nonumber\\
\vartheta_\sigma \, \psi^{(o)}_{\textbf k_{{}^\parallel}}(\textbf r_\bot) 
= \psi^{(o)}_{\textbf k_{{}^\parallel}}(\sigma^{-1}\textbf r_\bot) 
= - \psi^{(o)}_{\textbf k_{{}^\parallel}}(\textbf r_\bot) 
\end{eqnarray}
It is also obvious that relations~(\ref{even_odd_bc}) do translate into stringent conditions for the properties of the wavefunctions on the symmetry axis:
\begin{eqnarray}\label{bc_bc}
\partial_z \psi^{(e)}_{\textbf k_{{}^\parallel}}(y,z=0) = 0 \nonumber\\
 \psi^{(o)}_{\textbf k_{{}^\parallel}}(y,z=0) = 0
\end{eqnarray}
Such relations are very useful because they can be used as boundary conditions to reduce the domain of solution on the left or right halfplane, which is therefore the natural reduced domain of solution of the stationary Schr\"{o}dinger equation in this case. In most solution schemes, e.g. real space methods like finite element (FE) or finite differences (FD) approaches, it is easy to obtain odd and even solutions {\em separately} by solving two times the eigenproblem with different Dirichlet/Neumann boundary conditions on the symmetry axis boundary. Let us shortly have a group-theoretical approach: the even and odd wavefunctions shown in Fig.~\ref{fig_etat_V_BC} correspond respectively to the $A'$ and $A''$ irreducible representations (irreps) of the $C_s$ group. $A'$ and $A''$ are simply new group-theoretical labels meaning ``even'' or ``odd'', and no further insight arises. Nevertheless our considerations related to the symmetry of conduction band wavefunctions in $C_s$ symmetry - despite their trivial aspect - will prove important to better appreciate the differences occurring for valence-band envelope functions.

The case of the valence band eigenstates in $C_s$ symmetry is much more complex because of their (four-dimensional) spinorial character, this is why point group theory will immediately become an unvaluable asset. It tells us~\cite{ALT} that the spinorial eigenstates bare two labels again, but this time corresponding to the {\em double group} irreps  ${}^i\!E_{1/2}, i=1,2$ instead of $A'$ and $A''$. The spinors bearing these labels display a much more subtle and disturbing behaviour: {\em there is not any more any simple intrinsic symmetry for each of the envelope function components $\psi_{k,m}(y,z)$} [where $k=|{\textbf k_{{}^\parallel}}|$]. Their individual symmetry may indeed depend on the basis chosen, and, although there are two kinds of eigenstates, there is no such simple geometrical interpretation of the symmetry of such states by words of every-day life like ``even'' or ``odd''. The role of their more complex label ${}^i\!E_{1/2}$ is precisely to convey the nature of these more complex transformation laws under mirror symmetry. 

\begin{figure}[hbt]
\includegraphics[scale = 0.75]{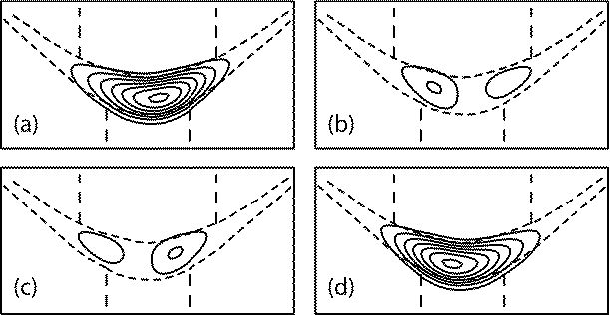}
\caption{Contour plots of valence band envelope functions $\psi_{k,m}^{\csEho}(y,z)$ of the ground state spinor (symmetry ${}^1\!E_{1/2}$) in a typical V-shaped QWR, when the Bloch function basis diagonalizes $J_y$ corresponding to the $[001]$ crystal direction (same QWR as in Fig.~\ref{fig_etat_V_BC}). \textbf{(a-d)} $m = 3/2,1/2,-1/2,-3/2$ function components respectively ($J_y$).}
\label{fig_etat_V_BV_001}
\end{figure}

Let us now illustrate this behaviour with the valence band eigenstates of the V-shaped QWR shown in Fig.~\ref{fig_vQWRpersp}. With the standard Bloch function basis used in the early works on the subject~\cite{BOC91,MAD98}, the ground state envelope functions are those of Fig.~\ref{fig_etat_V_BV_001}, where one clearly sees that none of the envelope functions is either perfectly symmetric or antisymmetric with respect to the symmetry plane! This ``standard'' choice of Bloch function basis was at the time guided by the fact that the shape of a V-shaped QWR is close to a deformed quantum well, and it was indeed reasonable since many qualitative features of the optical absorption spectrum could be understood~\cite{BOC91,MAD98} on the basis of ``quantum well light and heavy holes''. This is why this basis is the one diagonalizing the component of pseudo-angular momentum $\mathbf{J}$ aligned with the $[001]$ crystalline direction (note: with the choice of labels as in Fig.~\ref{fig_vQWRpersp} this actually correspond the vertical direction $y$, i.e. $J_y$). 

Even if the envelope functions of Fig.~\ref{fig_etat_V_BV_001} are not symmetric, intuitively one still expects some symmetries induced by the QWR symmetry. Indeed a closer analytical look reveals that there are still some symmetry relations linked with ${}^1\!E_{1/2}$ or ${}^2\!E_{1/2}$ eigenstate. They can be formulated as follows:
\begin{eqnarray}
\psi_{k,m}^{\csEho }(y,z) & =  & + \, \psi_{k,-m}^{\csEho }(y,-z)  \nonumber\\
\psi_{k,m}^{\csEht }(y,z) & =  & - \, \psi_{k,-m}^{\csEht}(y,-z) 
\label{Eq:wavefctCs001}
\end{eqnarray}
Clearly such symmetry relations, which only couple $\pm m$ envelope functions, cannot enforce the individual symmetry of every envelope functions in the spinor, and are nevertheless {\em awkward} from the numerical point of view since they do not allow to reduce the domain of solution on the half-plane as in the spinless case! 

\begin{figure}[hbt]
\includegraphics[scale = 0.75]{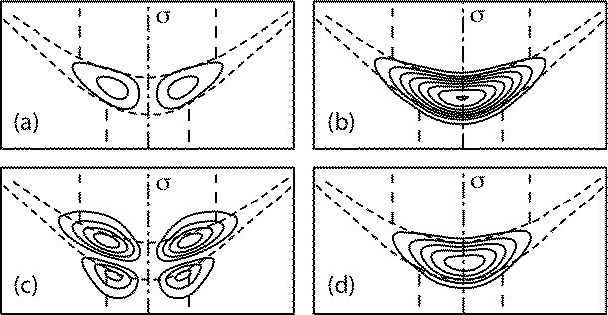}
\caption{Contour plots of valence band envelope functions $\Psi_{k,m}^{\csEho}(y,z)$ of the ground state of symmetry ${}^1\!E_{1/2}$ in a typical V-shaped QWR, when the Bloch function basis diagonalizes the $J_z$ component corresponding to the $[110]$ crystal direction (same QWR as in Fig.~\ref{fig_etat_V_BC}). \textbf{(a-d)} $m = 3/2,1/2,-1/2,-3/2$ function components respectively ($J_z$).}
\label{fig_etat_V_BV_110}
\end{figure}

The clue to this problem was found in~\cite{MAD00} by choosing a different Bloch function basis which diagonalizes the component $J_z$ oriented along the $[110]$ crystalline direction defined in Fig.~\ref{fig_vQWRpersp}. In such a case one could find novel envelope functions $\Psi_{k,m}(y,z)$ associated with every quantum state, with the following symmetry for ${}^1\!E_{1/2}$ or ${}^2\!E_{1/2}$ states respectively: 
\begin{eqnarray}
\Psi_{k,m}^{\csEho }(y,z) & =  &(-1)^{j+m} \; \Psi_{k,m}^{\csEho }(y,-z) \nonumber\\
\Psi_{k,m}^{\csEht }(y,z) & =  &(-1)^{j+m+1} \,  \Psi_{k,m}^{\csEht}(y,-z) 
\label{Eq:wavefctCs110}
\end{eqnarray}
To illustrate this we display in Fig.~\ref{fig_etat_V_BV_110} the contour plots for {\em the same} ground state as in Fig.~\ref{fig_etat_V_BV_001}. Although seemingly different, this new envelope function representation for the eigenstate carries exactly the same physics, i.e. gives the same expectation values for all physical observables.
It should be mentioned that the basic reason for the $m$-behaviour of the envelope functions in Eqs.~(\ref{Eq:wavefctCs110}) compared to Eqs.~(\ref{Eq:wavefctCs001}), which might seem surprising at first, can in fact be explained in a very intuitive way by looking at the behaviour of angular momentum components through a planar reflection $\sigma$: indeed the sign of the in-plane components of the angular momentum are reversed, while the perpendicular component is conserved, i.e.
\begin{equation}\label{eqt_amsigma}
\sigma \mathbf{J} \sigma^{-1} = -\mathbf{J} + 2 J_z \widehat e_z
\end{equation}
Therefore it is obvious that if one uses a Bloch basis diagonalizing the component of the pseudo-angular momentum $\mathbf{J}$ perpendicular to the symmetry plane (i.e. $J_z$), every envelope function will be mapped onto itself (either in a symmetric way or antisymmetric way), whilst the $+m$ and $-m$ components will be mapped onto each other if one diagonalizes $J_y$. To show that the envelope function spinors linked with the two double group irreps ${}^1\!E_{1/2}$ and ${}^2\!E_{1/2}$ have opposite alternating parity in Eqs.~(\ref{Eq:wavefctCs110}) requires a more detailed analysis. However we prefer to relegate this discussion after the presentation of the general theory, since most properties will become obvious.

\subsubsection{From low to high symmetries: $C_{2v}$,  $C_{3v}$ and higher}
We have thus shown that a careful choice of basis allows, in the case of $C_s$, to symmetrize individual envelope functions. Would this be possible in the case of higher symmetry? We intend now to clearly demonstrate that the approach suggested in Ref.~\cite{MAD00}, where we introduced the concept of Optimal Quantization Axis (OQA), has limits, motivating the more elaborate MSRF approach. We shall now take quantum dots as simple examples.

\begin{figure}[hbt]
\includegraphics[scale = 0.65]{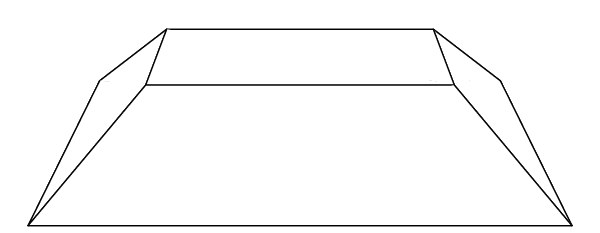}
\caption{Perspective view of a $C_{2v}$ QD}\label{fig_C2v}
\label{fig_C2vQD}
\end{figure}

Let us start with $C_{2v}$, the next higher symmetry depicted on Fig.~\ref{fig_C2vQD}. A practical example is pyramidal $InAs$ QDs which are have the shape of a rhombus-based pyramid. We shall denote the two perpendicular symmetry planes $\sigma_y$ and $\sigma_z$ for compatibility with the axes used in this paper. Since $C_s$ is a sub-group of $C_{2v}$, we could use the same basis diagonalizing $\sigma_z$, and reduce the problem to the half-domain. However it would be highly desirable to use the additional symmetry with respect to $\sigma_y$ to further split the domain of solution. 
No problems for electrons, but problems would arise for holes, since a 4 band $k \cdot p$ model would be required, and from our previous discussion of $C_s$ symmetry we immediately see that if we would take the spin quantization axis along $z$, the spinors envelope functions would become alternatingly even/odd with respect to $z$, but with respect to $y$ would necessarily obey a symmetry relation coupling $+m$ and $-m$ (c.f. Eq.~(\ref{Eq:wavefctCs001}). Oppositely if one would have chosen the $y$ basis, the symmetry relations with respect to $z$ would have become badly behaved. Therefore it is apparently never possible to obtain symmetric envelope functions in the two directions simultaneously, and solve on the half domain in the two directions! 

There is a rather simple explanation at a more fundamental level for this different behaviour. For the electron symmetry, described by the single group, $\sigma_y$ and $\sigma_z$ commute, and therefore one can in principle diagonalize the two operations simultaneously, and get simultaneous good quantum numbers linked with them. For the hole spinorial symmetry, described by the double group, $\sigma_y$ and $\sigma_z$ {\em do not commute}, indeed the general commutator can be written as:
\begin{equation}
\left[\sigma_y,\sigma_z\right]=C_2 \left(1+\left(-1\right)^{2j}\right)
\end{equation}
showing that when $j$ is half-integer (here $j=3/2$), it is {\em never} possible to diagonalize simultaneously both symmetry operations (such a fact can also be related to the appearance of a 2D irrep for the double group and the properties of its corresponding $2\times 2$ unitary matrix representation $D^{E_{1/2}}(g), g\in C_{2v}$). As a result of this analysis we suggested in~\cite{MAD00} that the optimal basis was naturally the one diagonalizing the projection of angular momentum along the third perpendicular axis, allowing to treat $\sigma_y$ and $\sigma_z$ on an equal footing, and diagonalizing the rotation $C_2 = \sigma_y \cdot \sigma_z$. However a closer inspection reveals that no solution on the quarter of the domain is yet allowed in $C_{2v}$ by this idea, this will only become possible with MSRF.

\begin{figure}[hbt]
\includegraphics[scale = 0.65]{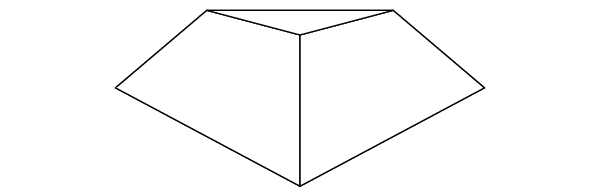}
\caption{Perspective view of a $C_{3v}$ QD}\label{fig_C3v}
\label{fig_C3vQD}
\end{figure}

\begin{figure}[hbt]
\includegraphics[scale = 0.75]{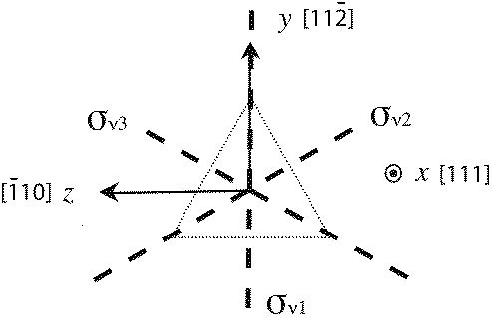}
\caption{Axes and cross section  of the $Al_xGa_{1-x}As$ $C_{3v}$ QD}
\label{fig_section}
\end{figure}

Even more challenging is the next higher symmetry: $C_{3v}$ symmetry, with three symmetry planes like in Fig.~\ref{fig_C3v}. Such a symmetry is also of practical interest~\cite{HAR99,FAB04}.
The reference axes of the crystal, and our $x,y,z$ labels, are shown in more detail in Fig.~\ref{fig_section}, together with the three vertical symmetry planes $\sigma_{vi}, i=1,2,3$. 
In addition to the three improper rotations (vertical symmetry planes), called $\sigma_{vi},\,\ i=1,2,3$, the $C_{3v}$ group includes as additional symmetry operations two rotation of $\pm 120°$ 
($C_3^\pm$) and the identity ($E$).

The $C_{3v}$ group displays two 1D irreps $A_1$ and $A_2$, and one 2D irrep ($E$), even for the single group, leading to two basis function (partner functions) for the subspace related to degenerate eigenvalues. There is a supplementary difficulty linked with this 2D irrep, in particular the corresponding 2D matrix representation explicitly depend on the basis functions. The simpler 1D single group irreps $A_1$ and $A_2$ are respectively even and odd with respect to all the symmetry planes. Therefore, for electrons it is straightforward to see that one can compute easily $A_1$ and $A_2$ eigenstates via the solutions on 1/6 of the domain by imposing respectively Neumann or Dirichlet boundary conditions on the two symmetry planes. For the degenerate $E$ irrep one can choose the two basis function such that they are either even or odd with respect to one of the symmetry plane (mirror), for instance $\sigma_{v1}$, but it will not be possible to diagonalize simultaneously any two of the mirrors at the same time, for instance $\sigma_{v1}$ and $\sigma_{v2}$, since they do not commute, as can be seen in the multiplication table (see annexe~\ref{appendix}). Therefore it is not obvious how to use the mirror symmetries to solve on a reduced domain smaller than one half. One may conclude that in a $C_{3v}$ heterostructure the problem that appeared only for valence-band holes in the case of $C_{2v}$ symmetry already appears for spinless electrons: it is not possible to solve with this technique on the most reduced domain, which, in $C_{3v}$, should be smaller than one half of the full domain! For holes in $C_{3v}$ there is also a 2D faithful self-conjugated double group irrep ($E_{1/2}$), but also two 1D mutually conjugated irreps (${}^i\!E_{3/2}\ , \ i=1,2$)~\cite{ALT}.

\begin{figure}[hbt]
\includegraphics[scale = 0.65]{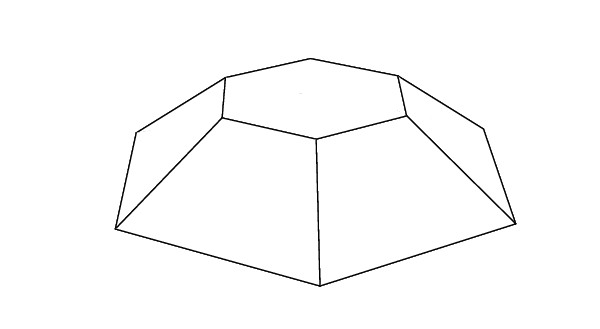}
\caption{Perspective view of a $C_{6v}$ QD}\label{fig_C6v}
\label{fig_C6vQD}
\end{figure}

Finally all these questions become more severe in higher symmetry, like the hexagonal $C_{6v}$ symmetry (c.f. perspective view of a $C_{6v}$ QD in Fig.~\ref{fig_C6v}) which is also of practical interest~\cite{SIM,TRO}. A characteristic of this group is that the double group displays only 2D degenerate irreps. Although such structures have been discussed to some extent in the literature with the help of group theory~\cite{TRO}, the symmetry properties of the envelope functions have never been studied and discussed. With MSR it is possible to approach systematically all higher symmetries, and for $C_{6v}$ the results will be given in subsection~\ref{subsec_C6v}.

To summarize, we can identify a limit between low and high symmetry groups in our sense: the appearance of a 2D irrep, which is a manifestation of the non-abelian character of groups like the dihedral groups ($C_{nv}$) that we have considered. Such irreps with dimension greater than 1 complicate a lot the question of the symmetry of the basis functions, making it non-trivial. 

Finally, the concept of spin quantization axis direction, linked with optimal pure rotations of the original spinorial basis, is not a suitable concept to tackle such higher symmetries (with spin, already $C_{2v}$!). Clearly maximal symmetrization of the envelope functions could not be achieved, and computation on a reduced domain was not enabled. In the following, a radically new approach for HSH is presented which will fulfill these goals. The optimal spinorial basis is obtained by a more general unitary transformation corresponding to the reduction (block-diagonalization) of the spinorial representation and related to the choice of double group labelled basis functions.

In the next chapter we shall start developing the MSRF formalism from beginning, by looking at transformation laws in ``orbital'' and ``spin'' space. Whenever needed the general theory will be  illustrated by the typical case of a $C_{3v}$ QWR (2D problem), either for the spinless conduction band, or for the valence band, again with the four-band Luttinger Hamiltonian. Numerical examples were worked out in real space with a FE approach incorporating linear elements only.

\section{General transformation laws, in ordinary and spinorial space}
\label{sec_theo0}
The study of transformation laws under symmetry operations is a prerequisite for the efficient use of symmetry and group theory explicitely on a given problem. One needs to know exactly how conduction and valence band envelope functions do transform under symmetry operations. This is of course related to the corresponding $k\cdot p$ Hamiltonians presented in Sec.~\ref{sec_form}. In this section we introduce in details transformation laws, which are also a prerequisite for the development of our new theory, independently of any symmetry consideration. For clarity we treat seperately scalar functions and spinors. In the last part of the section only we introduce symmetry, and look at the resulting constraints on envelope functions encountered in $k\cdot p$ theory.
 \subsection{Transformation laws}
Let us first introduce transformation laws for simple spinless scalar functions (typically the quantum wavefunction of an electron in a conduction band), and then in a subsequent step expose the transformation laws in the spinorial case (typically a hole in the valence band). To eliminate any ambiguity in the following we shall always use a passive point of view for the symmetry operations, i.e. the operations are always considered as coordinate transformations linked with a change in reference frame, i.e. they are not rotations of the physical system.
\subsubsection{Coordinate transformations}
Let us first define a basic transformation of coordinates $c$ linked with a change in orthonormal cartesian reference frame. It is defined by an orthogonal $3\times 3$ matrix $\Re(c)$ belonging to $O(3)$ and defining the basis vectors of the new frame with respect to the old one:
\begin{equation}\label{eqt_changBase}
\hat{e}_i'=\sum_{j}\Re_{ji}^{-1}(c) \hat{e}_j
\end{equation}
where $i,j\in \{ x,y,z\}$. It is either a rotation or an improper rotation, any rotation can be parametrized by its Euler angles $\alpha$, $\beta$ and $\gamma$, and the corresponding set of rotation matrices $\Re\left( g \right)\equiv \Re\left( \alpha\,\beta\,\gamma \right)$  define a representation of the rotation group ($SO(3)$). These matrices can be systematically constructed using the generators of the rotations~\cite{MES}: 
\begin{equation}\label{eqt_mtx_R}
\Re\left( \alpha\,\beta\,\gamma \right) = e^{+i\gamma J_z}e^{+i\beta J_y}e^{+i\alpha J_z}
\end{equation}
where $J_x,J_y,J_z$ are the components of the angular momentum pseudo-vector $\mathbf{J}$. 
Improper rotations can always be decomposed as the product of the spatial inversion $i$ with a proper rotation, therefore their matrix representation is written as the product of $\Re\left( i \right) = -I_3$ with the corresponding $\Re\left( \alpha\,\beta\,\gamma \right)$ as defined in Eq.~(\ref{eqt_mtx_R}. A typical example is $\sigma_{\hat{s}}$, a mirror symmetry with normal $\hat{s}$, which is the product of the inversion $i$ with $C_2(\hat{s})$, a $\pi$-rotation around the axis $\hat{s}$ . Therefore $\Re\left(\sigma_{\hat{s}}\right) = - \Re\left(\pi,\hat{s}\right)$, where temporarily the Euler angle notation is left out. 
Similarly, in the following we shall use $c$ to denote the change in coordinates linked with {\em any arbitrary change of orthonormal reference frame}, and often $\Re\equiv\Re(c)$ (without argument) will denote implicitely the {\em corresponding} orthogonal matrix representation (the argument of $\Re(c)$ will be restored only when absolutely needed for clarity).

Let us now recall how the set of components $\textbf{r} = (x,y,z)$ of a vector $\vec r$ with respect to a given basis $\left\{\hat{e}_i\right\}, i \in \{ x,y,z\}$ transform under a change in coordinates (new basis $\left\{\hat{e}'_i\right\}$):
\begin{equation}
\label{eqt_TransfVect}
x'_i=\sum_j \Re_{ij}x_j \iff \textbf{r}' = \Re\textbf{r}
\end{equation}
which is a contragredient law with respect to eq.(\ref{eqt_changBase}), as it should for a passive point of view.\\

\subsubsection{Transformation of scalar functions}
In the Hilbert space $\mathcal{H} = \mathcal{L}^2(\mathbb{R}^d)$ corresponding to the set of possible electronic wavefunctions in $d$ confined dimensions (d=1,2,3), one can associate linear operators $\vartheta_{c}$ to every possible coordinate transformation $c$:
\begin{eqnarray}\label{eqt_Op3D}
\vartheta_{c}:  && \mathcal{H} \longrightarrow \mathcal{H}  \nonumber \\
&& \psi \longmapsto \psi' = \vartheta_{c}\left[\psi\right]
\end{eqnarray}
The new mathematical function $\psi'$ is a function of the new coordinates, but is defined as representing the {\em same} quantum state, i.e.
\begin{equation}\label{transfFct}
\psi'(\textbf{r}')=\psi(\textbf{r}) \Rightarrow \psi'(\textbf{r}') = \vartheta_{c}\left[\psi\right](\textbf{r}') = \psi(\Re^{-1}\textbf{r}') 
\end{equation}
It is easy to check that this definition leads to the expected multiplication rule
$\vartheta_{c_2c_1}\left[\psi\right](\textbf{r}) = \vartheta_{c_2}\circ\vartheta_{c_1}\left[\psi\right](\textbf{r}) = \vartheta_{c_2}\left[\psi\circ\Re^{-1}(c_1)\right](\textbf{r}) = \psi((\Re^{-1}(c_1)\Re^{-1}(c_2))\textbf{r}) = \psi((\Re^{-1}(c_2c_1))\textbf{r})$ for two successive coordinate transformations $c_1$ and $c_2$.

Let us now consider the generic form $H(\textbf{r},\textbf{k})$ of the scalar $k\cdot p$ Hamiltonian in mixed position and momentum representation (of complementary dimensions $d$ and $d_f=3-d$ respectively) which appeared in Section~\ref{sub_form_env} (Eq.~(\ref{eqt_bc})). This generic form is applicable for heterostructures of any dimensionality $d_f$, from 3D to 0D ($d_f$ is the dimensionality in $k$-space related to the ``free-like'' motion of charge carriers in the directions with full translationnal invariance at the heterostructure level). The arguments $(\textbf{r},\textbf{k})$ should thus be understood as follows:
\begin{eqnarray}
\text{bulk (3D, d=0):} (\textbf{r},\textbf{k}) &\longrightarrow &(\textbf{k}=(k_x,k_y,k_z))  \nonumber\\
\text{quantum well (2D, d=1):} (\textbf{r},\textbf{k}) &\longrightarrow & (z,\textbf{k}_{{}^\parallel}=(k_x,k_y))  \nonumber\\
\text{quantum wire (1D, d=2):} (\textbf{r},\textbf{k}) &\longrightarrow &(\textbf{r}_\bot =(x,y),k_{{}^\parallel})  \nonumber\\
\text{quantum dot (0D, d=3):} (\textbf{r},\textbf{k}) &\longrightarrow &(\textbf{r}=(x,y,z))   \nonumber\\
& & \label{eqt_genericH}
\end{eqnarray}
The new transformed Hamiltonian operator is obtained by enforcing (in Dirac notation) $\left< \psi \right| H \left| \psi \right>=\left< \psi' \right| H' \left| \psi' \right>$, which leads to
\begin{equation}\label{transfH}
H'(\textbf{r}',\textbf{k}') = \vartheta_{c}\, H(\textbf{r}',\textbf{k}') \, \left[\vartheta_{c}\right]^{-1}= H(\Re^{-1}\textbf{r}',\Re^{-1}\textbf{k}') 
\end{equation}
Note that $\textbf{k}$ is implicitely understood as the vectorial components of the wavevector (covector). In Eq.~(\ref{transfH}) the operations $c$ are always considered in 3D (i.e. belonging to $O(3)$), therefore the eigenfunctions of the generic Hamiltonian must have an auxiliary index in order to transform consistently, i.e.
\begin{equation}\label{transfFct2}
\psi'_{\textbf{k}'}(\textbf{r}') = \vartheta_{c}\left[\psi_{\textbf{k}'}\right](\textbf{r}') = \psi_{\Re^{-1}\textbf{k}'}(\Re^{-1}\textbf{r}') 
\end{equation}
Such a generic scalar Hamiltonian is typically a quadratic form of the momentum, i.e. of the components of $\textbf{k}_{{}^\parallel}$ along the non-confined directions (translational invariance), and of differential operators which are kept in the confined directions (according to the correspondance $\textbf{k}_\bot=-i\nabla_\bot$). The parameters of this Hamiltonian, i.e. the effective masses and the confinement potential, are functions of positions like in Eq.~(\ref{eqt_bc}). It may also include anisotropic masses as follows: $H(\textbf{r},\textbf{k}) = H_0(\textbf{r}) + \textbf{k}^t\, \textbf{C}(\textbf{r})\, \textbf{k}$, where $\mathbf{C}(\textbf{r})$ is the ``matrix'' (tensor) of coefficients and $H_0(\textbf{r})$ is a scalar operator. In this rather general case the new Hamiltonian is obtained by 
\begin{equation}\label{eqt_H_cond}
H'(\textbf{r},\textbf{k}) = H(\Re^{-1}\textbf{r},\Re^{-1}\textbf{k}) = H_0(\Re^{-1}\textbf{r}) + \textbf{k}^t \, \textbf{C}'(\textbf{r})\, \textbf{k}
\end{equation}
where
\begin{equation}\label{eqt_C}
\textbf{C}'(\textbf{r}) = \Re \,\textbf{C}(\Re^{-1}\textbf{r})\, \Re^{-1}
\end{equation}
Is easy to understand equation~(\ref{eqt_C}) when $\textbf{C}$ is independent of $\textbf{r}$: $\textbf{k}^t\, \textbf{C}\, \textbf{k}$ is a scalar, invariant under a passive transformation and $\textbf{k}$ are the vectorial components of wave vector, then $\textbf{C}$ is a tensor two times covariant and, for this kind of tensors, the tranformation laws give $C'^{ij} = \sum_{rs} \Re_{ir} \Re_{js} C^{rs} = \sum_{rs} \Re_{ir} C^{rs} \,\Re^{-1}_{sj}$.
\subsubsection{Transformation of spinors and spinorial fields}
In a model including the spin of the charge carriers, the quantum state may generally be described by a spinorial field $\underline{\psi}(\textbf{r})$. In this section, for clarity, we will use the underscore to explicitely denote the spinorial character, which is assumed to be of dimension $2j+1$ where $j$ is a strictly positive {\em half-integer}. In the $k\cdot p$ multiband envelope function formalism, the spinorial components are related to the Bloch function basis at a high symmetry point of the Brillouin zone, diagonalizing the Hamiltonian of the charge carrier in the bulk crystal structure~\cite{BAS}. \\
The corresponding Hilbert space is spanned by the tensor products of spinors belonging to $\mathbb{C}^{(2j+1)}$ with the envelope functions belonging to $\mathcal{L}^2(\mathbb{R}^d)$. Hence when considering truly arbitrary coordinate transformations one should consider to perform the transformations independently in both spaces. \\
Let us first look solely to $SU(2)$ spinor transformations. By contrast to vectors in normal cartesian 3D space (c.f. Eq.~(\ref{eqt_changBase})), spinors obey different transformation rules under rotations, called spinorial transformation rules~\cite{APPEL,CARMELI}. For pure rotations the transformation rules can always be related by a similarity transformation to a set  of matrices $\underline{\underline{W}}^{(j)}\left(\alpha\,\beta\,\gamma\right)$, indexed by the Euler angles, and called the Wigner representation~\cite{MES,WIG}, which is a $(2j+1)$-dimensional projective representation of $SO(3)$:
\begin{equation}\label{eqt_mtx_W}
\underline{\underline{W}}^{(j)}\left( \alpha\,\beta\,\gamma \right) = e^{+i\gamma \underline{\underline{J}}_z^{(j)}} e^{+i\beta \underline{\underline{J}}_y^{(j)}} e^{+i\alpha \underline{\underline{J}}_z^{(j)}}
\end{equation}
When $j$ is half-integer it is a spinor representation of $SU(2)$ (instead of $SO(3)$). The double underscore notation introduced here always denotes a square matrix character in spinorial space. A typical feature of spinor representations is that a $2 \pi$ rotation around {\em any axis} $\hat{s}$ will always be associated to a sign change, i.e. $\underline{\underline{W}}^{(j)}\left(2 \pi,\hat{s}\right) = -\underline{\underline{1}}$. For the representation of improper rotations one can again use the factorization of the inversion, however a small complication is the Wigner representation of the inversion, which can be either $\pm 1$. We shall not discuss here the proper choice of the latter sign since it will not matter within this paper. In some cases it might be important, see~\cite{DAL1}.

A crucial feature of our approach is now to consider {\em more general unitary change in coordinates} $c$, not linked with 3D rotations, characterized by the general set of square matrices $\underline{\underline{V}}(c)$ of dimension $2j+1$ belonging to the group $U(2j+1)$. \\
Let us start with the $(2j+1)$-dimensional Bloch functions basis denoted $\{ \left| j,m \right> \}$, as is customary in the field, and a new basis $\{ \left| j,m \right>' \}$ differing by a rotation $c = (\alpha\,\beta\,\gamma)$, so that:
\begin{equation}\label{eqt_changBaseSpinRot}
\left| j,m \right>' = \sum_n W^{(j)}_{nm}(c^{-1}) \left| j,n \right>\; ,
\end{equation}
which is similar to the standard transformation law of partner functions linked with the irrep of dimension $2j+1$ of the rotation group under $c^{-1}$. It is important to make here two remarks for clarity. First we stress that for such Bloch functions $j$ simply correspond to a label related to the transformation law~(\ref{eqt_changBaseSpinRot}) and to the dimensionality of the Bloch functions basis, but in this context it is {\em not a true angular momentum quantum number} (it just refers to the transformation properties, and not truly to rotations). A second important remark is that the $\{ \left| j,m \right> \}$ basis, and its corresponding transformation law under rotations~(\ref{eqt_changBaseSpinRot}), already incorporate, in addition to pure spin, an orbital part, since the ``bulk'' spin-orbit interaction is already diagonalized by the Bloch function basis. Later below a purely orbital transformation, operating solely on the envelope function, will have to be simultaneously added. When one considers more general transformations of coordinates in spinorial space ($c\in U(2j+1)$):
\begin{eqnarray}
\vartheta^{(j)}_{c}:  && \mathbb{C}^{(2j+1)} \longrightarrow \mathbb{C}^{(2j+1)}  \nonumber\\
&& \underline{\psi} \longmapsto \underline{\psi}'=\vartheta^{(j)}_{c}\left[\,\underline{\psi}\,\right]
\label{eqt_OpSpin}
\end{eqnarray}
the spinor components $\underline{\psi}'$ can be transformed using the matrix equation
\begin{equation}\label{eqt_transf_spin}
\underline{\psi}' = \underline{\underline{V}}(c) \underline{\psi}
\end{equation}
which is anologous to Eq.~(\ref{eqt_TransfVect}). The matrix $\underline{\underline{V}}(c)$ defines a more general basis change than the one specifed by Eq.~(\ref{eqt_changBaseSpinRot}). \\
Let us now consider all the possible transformations of spinorial fields, which are defined by two separate coordinates transformations in both spaces $c_1\in O(3)$ in real space and $c_2 \in U(2j+1)$ and combine them in $\vartheta_{c}$ associated with $c = (c_1,c_2)$ acting in the tensor product Hilbert space:
\begin{equation}\label{eqt_generalTransf}
\vartheta_{c} = \vartheta^{(3D)}_{c_1} \otimes \vartheta^{(j)}_{c_2} 
\end{equation}
where $\vartheta^{(3D)}_{c_1}$ is the operator defined by Eq.~(\ref{eqt_Op3D}) and $\vartheta^{(j)}_{c_2}$ by Eq.~(\ref{eqt_OpSpin}).
Here we have kept the possibility of arbitrary and different changes of coordinates in both spaces, which will prove crucial later. Under a coordinate transformation $\vartheta_{c}$ an arbitrary quantum state of our system, described by {\em a spinorial field}, transforms therefore like
\begin{eqnarray}
\underline{\psi}_{\textbf{k}}'(\textbf{r}) &= & \vartheta_{c} \,\underline{\psi}_{\textbf{k}}(\textbf{r}) = \vartheta^{(j)}_{c_2} \left[\underline{\psi}_{\Re^{-1}(c_1)\textbf{k}}\right]\left(\Re^{-1}(c_1)\,\textbf{r}\right) \nonumber\\
&= &\underline{\underline{V}}(c_2) \,\underline{\psi}_{\Re^{-1}(c_1)\textbf{k}}\left(\Re^{-1}(c_1)\textbf{r}\right)
\label{eqt_TranfVal}
\end{eqnarray}
We see clearly in Eq.~(\ref{eqt_TranfVal}) that the spinorial character of the transformation does couple different envelope function components through the $\underline{\underline{V}}$ matrix, and is more complicated than the simple 3D transformation of the envelope functions appearing in Eq.~(\ref{transfFct2}). This complication is one of the roots of the difficulty in interpreting the individual symmetry of envelope functions in a heterostruture with a given symmetry.
\\
In the spinorial case one can transform the Hamiltonian operator matrix too, but one must take into account the presence of the $\underline{\underline{V}}$ matrix in a similar way:
\begin{eqnarray}
\underline{\underline{H}}'(\textbf{r},\textbf{k}) &= &\vartheta_{c} \, \underline{\underline{H}}(\textbf{r},\textbf{k}) \,\left[\vartheta_{c}\right]^{-1} \nonumber\\
&= & \vartheta^{(j)}_{c_2} \,\underline{\underline{H}}(\Re^{-1}\textbf{r},\Re^{-1}\textbf{k}) \,\left[\vartheta^{(j)}_{c_2}\right]^{-1} \nonumber \\
&= &\underline{\underline{V}} \,\underline{\underline{H}}(\Re^{-1}\textbf{r},\Re^{-1}\textbf{k}) \,\underline{\underline{V}}^{-1}
\label{eqt_transfoHgen}
\end{eqnarray}
where, in the last two lines, we have used again the shorthand notation $\Re \equiv \Re(c_1)$ and $\underline{\underline{V}} \equiv \underline{\underline{V}}(c_2)$. 

\subsection{Symmetry and resulting constraints on eigenstate envelope functions}
Let us now assume a heterostructure with a given symmetry group $\mathcal{G} = \left\{ g \right\}$ of spatial transformations. More precisely, if there is translationnal invariance (i.e. $d_f>0$), let us take $\mathcal{G}$ as the {\em small point group of $\textbf{k}\equiv\textbf{k}_\parallel$} of the structure, with cardinality $\left| \mathcal{G} \right|$, and defined by the restriction:
\begin{equation}\label{eqt_invK}
\Re(g) \,\textbf{k} = \textbf{k}
\end{equation}
The operations $g\in \mathcal{G}$ are either pure rotations or roto-inversions (mirrors). The heart of the passive point of view is just to express than any full coordinate transformation corresponding to a symmetry element $g$ of the structure will leave {\em invariant} the form of the $\textbf{k}$-restriction of the conduction and valence band Hamiltonians (hence from now on we shall leave out the implicit $\textbf{k}$-subscript on eigenstates and envelope functions appearing in Eqs.~(\ref{transfFct2}) and~(\ref{eqt_TranfVal}).\\
For the conduction band the product $g''$ of two symmetry operations $g$ and $g'$ will simply follow the multiplication table of the single point group $\mathcal{G}$, whilst for the valence band one must necessarily use a double group notation of symmetry operations $\tilde{g} \in \widetilde{\mathcal{G}}$ due to its spinorial nature (please note that here we use the notation $\tilde{g}$ to denote {\em any element} of $\widetilde{\mathcal{G}}$, instead of the single group element $g$ multiplied by a $2\pi$-rotation, as is quite standard (see e.g.~\cite{ALT})). The composite index $c$ for the corresponding full coordinate transformation can be identified with $\tilde{g}$, provided one understands it as $(g,\tilde{g})$ where $g$ is the single point group image of $\tilde{g}$. Clearly this composite index obeys also the double point group multiplication table since the product is defined as $(g,\tilde{g})\cdot(g',\tilde{g}') = (g.g',\tilde{g}.\tilde{g}') = (g'',\tilde{g}'')$. 
\\
Using Eq.~(\ref{eqt_transfoHgen}) let us now express the invariance of the $\textbf{k}$-restricted Hamiltonian with respect to a given symmetry operation $\tilde{g}$ of the coordinate system: 
\begin{equation}\label{eqt_invHL}
\underline{\underline{H}}'(\textbf{r},\textbf{k}) = \underline{\underline{V}} \,\underline{\underline{H}}(\Re^{-1} \textbf{r},\textbf{k}) \,\underline{\underline{V}}^{-1} \equiv \underline{\underline{H}}(\textbf{r},\textbf{k})
\end{equation}
where shorthand notations have been used. Eq.~(\ref{eqt_invHL}) is equivalent to state that every symmetry operation $\tilde{g}$ commutes with the $\textbf{k}$-restricted Hamiltonian, which allows the use of a well known theorem~\cite{BAS} that states that every eigenspace of the Hamiltonian can then be labeled by an irreducible representation (irrep) $\widetilde{\Gamma}$ of $\widetilde{\mathcal{G}}$ (meaning that except for accidental degeneracies its dimension $d_{\widetilde{\Gamma}}$ is necessarily the dimension of the irrep), and that a basis of partner eigenstates $\underline{\psi}^{\widetilde{\Gamma}}_\mu(\textbf{r})$ ($\mu = 1,\ldots, d_{\widetilde{\Gamma}}$ is called the partner function index) can be found such that under a basis change $\vartheta_{\tilde{g}}$ it transforms according to
\begin{equation}\label{eqt_transfFctPart}
\vartheta_{\tilde{g}}\underline{\psi}^{\widetilde{\Gamma}}_\mu(\textbf{r}) = \sum_{\nu =1}^{d_{\widetilde{\Gamma}}} \left[ D^{\widetilde{\Gamma}}(\tilde{g}) \right]_{\nu \mu} \,\underline{\psi}^{\widetilde{\Gamma}}_\nu(\textbf{r})
\end{equation}
where the set $\left\{D^{\widetilde{\Gamma}}(\tilde{g})\right\}$ form a unitary irreducible matrix representation of the irrep $\widetilde{\Gamma}$.\\

From the physical point of view, Eq.~(\ref{eqt_transfFctPart}) means that a transformed symmetrized eigenstate $\underline{\psi}^{\widetilde{\Gamma}}_\mu(\textbf{r})$ under $\vartheta_{\tilde{g}}$ does not only have the same energy but can also be developed on its partners, so that there are constraints on individually transformed envelope functions since they must ``reconnect'' on all the other ones in a very intricate fashion. Using Eq.~(\ref{eqt_TranfVal}) we find 
\begin{equation}\label{eqt_reconnect}
\sum_{\nu =1}^{d_{\widetilde{\Gamma}}} \left[ D^{\widetilde{\Gamma}}(\tilde{g}) \right]_{\nu \mu} \,\underline{\psi}^{\widetilde{\Gamma}}_\nu(\textbf{r}) = \underline{\underline{V}}(\tilde{g}) \,\underline{\psi}^{\widetilde{\Gamma}}_\mu(\Re^{-1}(\tilde{g})\textbf{r})
\end{equation}
\\
This equation forms the starting point of our theory. It will be exploited in the next section where we use the explicit separation of the spatial and the spinorial part of the operators in Eq.~(\ref{eqt_generalTransf}), and then seeks the basis that will minimally reconnect envelope functions according to~(\ref{eqt_reconnect}).\\

\section{The fully symmetrized OBB basis and the separation of spinorial and spatial parts}
\label{sec_OBB}
By definition, a 3D representation of any point group can be obtained simply from the analytical expression of the $\Re(\alpha,\beta,\gamma)$ matrices~(\ref{eqt_mtx_R}) and by factoring out the inversion for improper operations. In spinorial space things are slightly more complicated since, depending on the basis used to express the Hamiltonian, the $\underline{\underline{V}}(\widetilde{g})$ matrices do form a spinorial $(2j+1)$-dimensional representation of the double group $\widetilde{\mathcal{G}}$, but not necessarily given by Wigner matrices~(\ref{eqt_mtx_W}).

In this section we shall start from a spinorial Bloch function basis which, at high symmetry points of the Brillouin zone, is often denoted $\left\{ \left| j,m \right> \right\}$, due to its symmmetry transformation properties. For the valence-band of semiconductors like $GaAs$, at the so-called $\Gamma$-point, one restricts to $j=3/2$~\cite{BAST} giving rise to the Luttinger Hamiltonian (c.f. Eqs.~(\ref{eqt_bv}) and~(\ref{eqt_lutt})). However the formalism holds for an arbitrary large number of bands. The heterostructure symmetry is kept general, but illustrated with $C_{3v}$ symmetry. A few other cases, in particular the $C_{6v}\, ,\, D_{3h}\, ,\, C_n\, ,\, C_s$ symmetry groups, will be shortly discussed in Sec.~\ref{sec_autres_ex}.\\
To introduce the concept of the Optimal Bloch function Basis (OBB) with respect to a heterostructure with a given symmetry, we must heavily rely on the explicit separation of orbital and spinorial part carried out in the last section, and on the possibility of separate coordinate transformations in both spaces. \\
\subsection{The optimal Bloch function basis}
Our main goal is to simplify Eq.~(\ref{eqt_reconnect}) and minimize the coupling between different envelope functions. The main novel idea is to perform once for all a unitary coordinate transformation $c=(1,c_2), c_2\in U(2j+1)$, i.e. purely in spinorial space, corresponding to a best choice of the Bloch function basis, such that the set of reducible matrices $\underline{\underline{V}}(\tilde{g})$ appearing in~(\ref{eqt_reconnect}) would become block-diagonal!

Let us first note that some aspects of this idea are not completely new. Up to now one as used 3D rotations of the quantization axis (direction of $J_z$), parametrized by the Euler angles $c_2=(\alpha,\beta,\gamma)$ such that $J_z$  would be transformed towards $J'_z$, diagonalized by $\left| j,m \right>'$~\cite{FISH,MAD00}. The corresponding image of $\hat{e}_z$ is the so-called Optimal Quantization Axis (OQA)~\cite{MAD00}. The more trivial case of quantum wells grown in $[hhk]$ direction, where the OQA is always $[hhk]$, is treated in~\cite{FISH}. The OQA was supposedly considered the best choice~\cite{MAD00} to simplify the Luttinger Hamiltonian. Indeed a change in basis $\left| j,m \right>'=\sum_n U_{nm}(c_2^{-1})\left| j,n \right>$ has lead to a new form of the Hamiltonian expressed in the new basis as 
\begin{equation}\label{eq_newH}
H'(\textbf{r},\textbf{k}) = \underline{\underline{U}}(c_2) \,H(\textbf{r},\textbf{k}) \,\left[\underline{\underline{U}}(c_2)\right]^{-1}
\end{equation} 
which lead to simpler wavefunctions in the $C_s$ case. It also lead to a corresponding new representation of the symmetry operations
\begin{equation}\label{eq_newW}
\underline{\underline{V}}'(\tilde{g}) = \underline{\underline{U}}(c_2) \,\underline{\underline{V}}(\tilde{g}) \,\left[\underline{\underline{U}}(c_2)\right]^{-1}
\end{equation}
The optimal choice of angles $c_2=(\alpha\,\beta\,\gamma)$ was carefully made~\cite{FISH,MAD00}, and discussed partly in Sec.~\ref{sec_form}, but we have seen that it cannot be efficient for higher symmetries.
\\

In the present approach, the novelty is to consider {\em more general unitary transformations $c$ that cannot be represented as rotations of the original reference frame}. For this purpose we shall seek the coordinate transformation $\bar{c}=(1,\bar{c}_2), \bar{c}_2\in U(2j+1)$, {\em but possibly} $\bar{c}_2\notin SU(2)$, towards new Bloch states $\left| \widetilde{\Gamma},\mu \right>$, {\em where $\widetilde{\Gamma}$ is the label of a double group irrep, and $\mu$ a corresponding partner function label}. This set of states, that we shall call Optimal Bloch function Basis (OBB), is a fully symmetrized Bloch basis, {\em i.e. symmetrized according to the symmetry of the quantum heterostructure}. From its definition  
\begin{equation}\label{eq_newBasis}
\left| \widetilde{\Gamma}_b,\beta \right> =\sum_m U_{m;\widetilde{\Gamma}_b,\beta}(\bar{c}_2^{-1}) \left| j,m \right>
\end{equation}
The notation makes it clear that every new basis state $\left| \widetilde{\Gamma}_b,\beta \right>$ will map under passive symmetry operations like the standard set of partner functions of an irrep of the double group:
\begin{equation}\label{eqt_transfOBB}
\vartheta^{(j)}_{\tilde{g}} \left| \widetilde{\Gamma}_b,\beta \right> =
\sum_{\beta' =1}^{d_{\widetilde{\Gamma}_b}} \left[ D^{\widetilde{\Gamma}_b}(\tilde{g}^{-1}) \right]_{\beta' \beta}  \left| \widetilde{\Gamma}_b,\beta'\right>  \; ,
\end{equation}
but with $\tilde{g}^{-1}$ on the right hand side. Let us now look at the transformation properties of the new spinorial components of the field $\underline{\psi}^B(\textbf{r}) = \vartheta_{\bar{c}}\underline{\psi}(\textbf{r}) = \underline{\underline{U}}(\bar{c}_2) \,\underline{\psi}(\textbf{r})$, transforming as
\begin{equation}\label{eqt_TranfNewSpin}
\vartheta_{\tilde{g}}\,\underline{\psi}^B(\textbf{r}) = \underline{\underline{V}}^B(\tilde{g}) \,\underline{\psi}^B(\Re^{-1}(\tilde{g})\textbf{r})
\end{equation}
From Eq.~(\ref{eqt_transfOBB}) it is clear that the change in basis specified by the $\underline{\underline{U}}(\bar{c}_2)$ matrix induces automatically a new block-diagonal representation of the symmetry operations $\underline{\underline{V}}^B(\tilde{g})$ where Eq.~(\ref{eq_newW}) then becomes
\begin{equation}\label{eqt_transfoVmat}
\underline{\underline{V}}^B(\tilde{g}) = \underline{\underline{U}}(\bar{c}_2) \,\underline{\underline{V}}(\tilde{g}) \,\left[\underline{\underline{U}}(\bar{c}_2)\right]^{-1}
\end{equation}
The $n_B = \sum_b n_{\widetilde{\Gamma}_b}$ blocks of dimension $d_{\widetilde{\Gamma}_b}$ of $\underline{\underline{V}}^B(\tilde{g})$ are labelled by $\widetilde{\Gamma}_b$ and a possible multiplicity index runnig from $1$ to $n_{\widetilde{\Gamma}_b}$. Indeed a given representation $\widetilde{\Gamma}_b$ may appear more than once (a pratical example is the $C_s$ group rediscussed later), however for simplicity we shall forget it since it can be restored without difficulty.\\
Let us now separate every irreducible blocks $\widetilde{\Gamma}_b$ of $\underline{\underline{V}}^B(\tilde{g})$, and of every partner eigenstate $\left.\underline{\psi}^B\right.^{\widetilde{\Gamma}}_{\mu}(\textbf{r})$ of the irrep $\widetilde{\Gamma}$:
\begin{eqnarray}
\label{eqt_blocksVB}
\underline{\underline{V}}^B(\tilde{g}) &= &\sum_{\widetilde{\Gamma}_b} \underline{\underline{V}}^{\widetilde{\Gamma}_b}(\tilde{g}) \\
\label{eqt_blocksPsiV}
\left.\underline{\psi}^B\right.^{\widetilde{\Gamma}}_{\mu}(\textbf{r}) &= &\sum_{\widetilde{\Gamma}_b} \underline{\psi}^{\widetilde{\Gamma},\widetilde{\Gamma}_b}_{\mu}(\textbf{r})
\end{eqnarray}
It is understood here that every $\underline{\underline{V}}^{\widetilde{\Gamma}_b}(\tilde{g})$ has only a single non-zero square block on its diagonal, which we shall denote  $\utilde{\utilde{V}}^{\widetilde{\Gamma}_b}(\tilde{g}) \equiv D^{\widetilde{\Gamma}_b}(\tilde{g})$ to pinpoint that its dimension is  $d_{\widetilde{\Gamma}_b}$ instead of $(2j+1)$. Similarly $\underline{\psi}^{\widetilde{\Gamma},\widetilde{\Gamma}_b}_{\mu}(\textbf{r})$ is defined as having only a single subset of $d_{\widetilde{\Gamma}_b}$ relevant components which we shall denote $\utilde{\psi}^{\widetilde{\Gamma},\widetilde{\Gamma}_b}_{\mu}(\textbf{r})$ (all other components are zero). 
If one now applies the transformation $\bar{c}_2$ to Eq.~(\ref{eqt_reconnect}), and uses Eq.~(\ref{eqt_transfoVmat}), and identifies every subblock labelled by $\widetilde{\Gamma}_b$, one finds:
\begin{equation}\label{eqt_newReconnect}
\sum_{\nu =1}^{d_{\widetilde{\Gamma}}} \left[ D^{\widetilde{\Gamma}}(\tilde{g}) \right]_{\nu \mu} \,\utilde{\psi}^{\widetilde{\Gamma},\widetilde{\Gamma}_b}_\nu(\textbf{r}) = \utilde{\utilde{V}}^{\widetilde{\Gamma}_b}(\tilde{g}) \utilde{\psi}^{\widetilde{\Gamma},\widetilde{\Gamma}_b}_\mu(\Re^{-1}(\tilde{g})\,\textbf{r})
\end{equation}
Clearly this equation shows that our goal to minimize absolutely the coupling between different spinorial components under symmetry operations is achieved: the coupled components are reduced and grouped according to the irreps $\widetilde{\Gamma}_b$. However this is not sufficient, and more work needs to be done, and will be carried out in subsection~\ref{subsec_symEnvOBB}, to ensure that the resulting envelope functions have maximum symmetry. Let us first illustrate the development of this section with a concrete example.

\subsection{The OBB for the Luttinger Hamiltonian in $C_{3v}$ symmetry}

The matrix representation of point group symmetry operations for the standard $4\times 4$ Luttinger problem at hand are in fact naturally the original Wigner representation~\cite{MES}, i.e. $\underline{\underline{V}}(\tilde{g})\equiv\underline{\underline{W}}^{(j)}\left(\alpha\,\beta\,\gamma\right)$, since they are consistent with the $J$ matrices appearing in~(\ref{eqt_inv_ok}). 
One just needs the reduction in block diagonal form for $C_{3v}$ symmetry which can be, in this case, efficiently carried out with the help of the matrix traces. One then finds that this matrix representation of $C_{3v}$ is reducible to ${}^1\!E_{3/2} \oplus E_{1/2} \oplus {}^2\!E_{3/2}$.
The ad-hoc $\underline{\underline{U}}(\bar{c}_2)$ transformation can be found with standard techniques~\cite{ALT} and leads to the following fully symmetrized Bloch function basis 
\begin{eqnarray}
\left|{}^1\!E_{3/2}\right> &= & \frac{1}{2} \,e^{i\,\frac{\pi}{4}} \,\left( \left|\frac{3}{2},\frac{3}{2}\right> + \sqrt{3} \,\left|\frac{3}{2},-\frac{1}{2}\right> \right) \nonumber \\
\left|E_{1/2},1\right> &= & \frac{1}{2} \,e^{i\,\frac{\pi}{4}}  \,\left( \left|\frac{3}{2},\frac{1}{2}\right> - \sqrt{3} \,\left|\frac{3}{2},-\frac{3}{2}\right> \right) \nonumber \\
\left|E_{1/2},2\right> &= & \frac{1}{2\,i} \,e^{i\frac{\pi}{4}}  \,\left( \left|\frac{3}{2},-\frac{1}{2}\right> - \sqrt{3} \,\left|\frac{3}{2},\frac{3}{2}\right>  \right) \nonumber \\
\left|{}^2\!E_{3/2}\right> &= & \frac{1}{2\,i} \,e^{i\,\frac{\pi}{4}}   \,\left( \left|\frac{3}{2},-\frac{3}{2}\right> + \sqrt{3} \,\left|\frac{3}{2},\frac{1}{2}\right>  \right) \; ,
\label{eq_OBB}
\end{eqnarray}
where the $\left|j,m\right>$ basis pertains to the $x,y,z$ axes in Fig.~\ref{fig_section}. The reduced block-diagonal form of the $\underline{\underline{V}}^B(\tilde{g})$ matrices is then:
\begin{equation}\label{eq_W32}
\underline{\underline{V}}^B(\tilde{g}) = 
\left(
\begin{array}{cccc}
\chi^{{}^1\!E_{3/2}}(\tilde{g}) & {} & \mathbf{0} \\
{} & D^{E_{1/2}}(\tilde{g}) & {} \\
\mathbf{0} & {} & \chi^{{}^2\!E_{3/2}}(\tilde{g})
\end{array}
\right)
\end{equation}
Here $\chi^{{}^i\!E_{3/2}}(\tilde{g}), i=1,2$ are the characters of the 1D irrep ${}^i\!E_{3/2}, i=1,2$ respectively, and $D^{E_{1/2}}(\tilde{g})$ is a 2D matrix representation for $E_{1/2}$ that one can freely fix by choosing the two partner functions. We have specifically chosen also  
\begin{equation}\label{eq_DE}
D^{E_{1/2}}(\tilde{g}) = D^{E}(\tilde{g}) \,\chi^{{}^2\!E_{3/2}}(\tilde{g}) \; ,
\end{equation}
where $D^{E}(\tilde{g})$ is the 2D representation for the single group irrep $E$, according to the product of representations $E\otimes {}^2\!E_{3/2} \approx E_{1/2}$ (see appendix~\ref{appendix}). It is important to realize at this point that there is a degree of arbitrariness that cannot be avoided: for example the order of appearance of the blocks in~(\ref{eq_W32}), or the fact that both matrix representations $D^{E}(\tilde{g})$ and $D^{E_{1/2}}(\tilde{g})$ can in principle be chosen independently. To justify our specific choice two remarks are in order: first it is desirable to keep the $p,q,r,s$-form of the Luttinger Hamiltonian (linked with the form of the time-reversal operator), and this can be achieved only with a few possible orderings of basis states $\left| \widetilde{\Gamma}_b,\beta \right>$, second Eq.~(\ref{eq_DE}) is {\em a particular restriction} whose motivation will appear more clearly later.\\
An alternative group-theoretical view on the decomposition of the Wigner representation into block-diagonal form~(\ref{eq_W32}), is to consider the diamond group $O_h$, which is the symmetry group of the Luttinger Hamiltonian linked with the underlying bulk semiconductor. With respect to the symmetry $O_h$, the $C_{3v}$ heterostructure can be viewed as a {\em symmetry breaking perturbation} (due to the mesoscopic heterostructure potential), therefore a good basis within the 4D subspace linked with the irrep $F_{3/2,g}$ of $O_h$ will be found using subduction tables~\cite{ALT} which give $F_{3/2,g}\to {}^1\!E_{3/2} \oplus E_{1/2} \oplus {}^2\!E_{3/2}$.\\

It is interesting to note that the preliminary coordinate transformation $\bar{c}=(1,\bar{c}_2)$ is here purely in spinorial space, but in fact this is not a general feature. One can explain it in the following way: for $C_{3v}$ the 3D matrix representation $\left\{ \Re(g) \right\}$ is already in a reduced block-diagonal form ($A_1\oplus E$) with the basis presented in Fig.~\ref{fig_section} ($\hat e_x$ is invariant respect to every symmetry operations of the group and $\hat e_y$,$\hat e_z$ are mutually coupled according to $E$ irrep). In fact here one has directly implicitly chosen the vectorial basis according to the symmetry of the heterostructure analogously to the Bloch function basis in Eq.~(\ref{eq_W32})! A counterexample would be in the $C_n$ group, which will be shortly discussed in section~\ref{subsec_Cn}: one indeed needs to introduce a slightly more complex preliminary coordinate transformation of the form $\bar{c}=(\bar{c}_1,\bar{c}_2)$ such that the matrix representations $\left\{ \Re(g) \right\}$ and $\left\{ \underline{\underline{V}}(\tilde{g}) \right\}$ would become simultaneously block-diagonalized according to the irreps of $C_n$. \\

\subsection{Ultimately reduced envelope functions in the OBB basis}
\label{subsec_symEnvOBB}

We now come to the last positive by-product of the introduction of the unitary coordinate transformation $\bar{c}$ towards the OBB: the ability to define ``ultimately'' reduced envelope functions (UREF). 

Let us revert Eq.~(\ref{eqt_newReconnect}) by multiplying it by $\underline{\underline{V}}^{\widetilde{\Gamma}_b}(\tilde{g}^{-1})$. We now look at each individual envelope function components $\psi^{\widetilde{\Gamma},\widetilde{\Gamma}_b}_{\mu,\beta}(\textbf{r})$ of $\utilde{\psi}^{\widetilde{\Gamma},\widetilde{\Gamma}_b}_{\mu}(\textbf{r})$:
\begin{equation}\label{eqt_LoiTFctEnv}
\psi^{\widetilde{\Gamma},\widetilde{\Gamma}_b}_{\mu,\beta}(\Re^{-1}(\tilde{g})\,\textbf{r}) = \sum_{\beta'=1}^{d_{\widetilde{\Gamma}_b}} \sum_{\nu =1}^{d_{\widetilde{\Gamma}}} \left[ D^{\widetilde{\Gamma}}(\tilde{g}) \right]_{\nu \mu} \,V^{\widetilde{\Gamma}_b}_{\beta,\beta'}(\tilde{g}^{-1}) \,\psi^{\widetilde{\Gamma},\widetilde{\Gamma}_b}_{\nu,\beta'}(\textbf{r})
\end{equation}
where the envelope function components are now clearly related to the OBB basis~(\ref{eq_newBasis}) through the indices $\widetilde{\Gamma}_b$ and $\beta$. One should now make a fundamental remark, namely that Eq.~(\ref{eqt_LoiTFctEnv}) can be interpreted as a {\em reducible} transformation law for individual envelope function components under the symmetry group of the heterostructure! Indeed the transformation matrix element can be written as $D^{\widetilde{\Gamma}}_{\nu \mu}(\tilde{g}) \,\left[D^{\widetilde{\Gamma}_b}_{\beta',\beta}(\tilde{g})\right]^*$ which contains all elements of the product of representations 
\begin{equation}\label{eqt_lastDisentangl}
\widetilde{\Gamma} \otimes \widetilde{\Gamma}_b^* = \bigoplus_{a} n_{\widetilde{\Gamma},\widetilde{\Gamma}_b^*;\Gamma_a} \,\Gamma_a 
\end{equation}
where $n_{\widetilde{\Gamma},\widetilde{\Gamma}_b^*;\Gamma_a}$ is the multiplicity of the $\Gamma_a$ irrep in the product. Moreover, since $\Gamma_a$ appears in a tensor product of the {\em double} group representations, $\Gamma_a$ is necessarily a {\em single} group representation! (note also that for point groups $n_{\widetilde{\Gamma},\widetilde{\Gamma}_b^*;\Gamma_a}\leq 2$, however we shall deliberately ignore in the following the corresponding additionnal multiplicity index: first it is trivial to restore it if needed, second one does not usually need it (simple-reducible point groups). 

Let us therefore introduce unitary generalized Clebsch-Gordan coefficients $C^{\widetilde{\Gamma},\widetilde{\Gamma}_b^*;\Gamma_a}_{\mu,\beta;\alpha}$ which perform in practice the block decomposition of the reducible matrix appearing in~(\ref{eqt_LoiTFctEnv}), i.e.
\begin{eqnarray}\label{eqt_blockRedD}
& & D^{\Gamma_a}_{\alpha' \alpha}(\tilde{g}) = \\
& & \quad\sum_{\beta',\beta =1}^{d_{\widetilde{\Gamma}_b}} \sum_{\mu', \mu =1}^{d_{\widetilde{\Gamma}}} 
\left[ C^{\widetilde{\Gamma},\widetilde{\Gamma}_b^*;\Gamma_a}_{\mu',\beta';\alpha'} \right]^* D^{\widetilde{\Gamma}}_{\mu' \mu}(\tilde{g}) \,\left[D^{\widetilde{\Gamma}_b}_{\beta',\beta}(\tilde{g})\right]^* \,C^{\widetilde{\Gamma},\widetilde{\Gamma}_b^*;\Gamma_a}_{\mu,\beta;\alpha} \nonumber 
\end{eqnarray}
This equation naturally leads us to introduce, by the use of Eq.(\ref{eqt_lastDisentangl}), {\em a set of Ultimately Reduced Envelope Function components} (UREFs), denoted  $\phi^{\widetilde{\Gamma},\Gamma_a}_{\widetilde{\Gamma}_b,\alpha}(\textbf{r})$, and associated with every subspace of the nanostructure Hamiltonian of symmetry $\widetilde{\Gamma}$ and every block $\widetilde{\Gamma}_b$:
\begin{equation}\label{eqt_ultEnvFctComp}
\phi^{\widetilde{\Gamma},\Gamma_a}_{\widetilde{\Gamma}_b,\alpha}(\textbf{r}) = \sum_{\mu=1}^{d_\Gamma} \sum_{\beta =1}^{d_{\widetilde{\Gamma}_b}}  C^{\widetilde{\Gamma},\widetilde{\Gamma}_b^*;\Gamma_a}_{\mu,\beta;\alpha} \,\psi^{\widetilde{\Gamma},\widetilde{\Gamma}_b}_{\mu,\beta}(\textbf{r})
\end{equation}
This equation can be reversed easily by using the unitarity of the Clebsch-Gordan matrix, which gives a development of the original envelope functions in terms of UREF's:
\begin{equation}\label{eqt_fixedVarDev}
\psi^{\widetilde{\Gamma},\widetilde{\Gamma}_b}_{\mu,\beta}(\textbf{r}) = \sum_{\Gamma_a,\alpha} \left[ C^{\widetilde{\Gamma},\widetilde{\Gamma}_b^*;\Gamma_a}_{\mu,\beta;\alpha} \right]^*  \,\,\phi^{\widetilde{\Gamma},\Gamma_a}_{\widetilde{\Gamma}_b,\alpha}(\textbf{r})
\end{equation}

The definition of UREF's, Eqs.~(\ref{eqt_ultEnvFctComp}) and~(\ref{eqt_fixedVarDev}), deserve a number of important comments. First, the UREF decomposition~(\ref{eqt_ultEnvFctComp}) is {\em totally general}. Second, for every subspace of the Hamiltonian related to the symmetry $\widetilde{\Gamma}$ there are $n_{\widetilde{\Gamma}} = \sum_{\Gamma_a,\widetilde{\Gamma}_b} n_{\widetilde{\Gamma}_b} n_{\widetilde{\Gamma},\widetilde{\Gamma}_b^*;\Gamma_a}$ sets of UREFs, in particular $n_{\widetilde{\Gamma}}$ can be higher than $(2j+1)$, note also that the UREFs are characterizing strictly $\widetilde{\Gamma}$ and $\widetilde{\Gamma}_b$, i.e. they are {\em independent} of the partner function indices $\mu$ and $\beta$. Third, from Eq.~(\ref{eqt_lastDisentangl}) one can show that every set of UREFs indeed transform {\em minimally under symmetry operations}, simply like the partner functions of the single group irreps $\Gamma_a$:
\begin{eqnarray}\label{eqt_TrsfUltEnvFct}
\vartheta^{(3D)}_{\tilde{g}} \,\left[ \phi^{\widetilde{\Gamma},\Gamma_a}_{\widetilde{\Gamma}_b,\alpha} \right](\textbf{r}) &= &\phi^{\widetilde{\Gamma},\Gamma_a}_{\widetilde{\Gamma}_b,\alpha}(\Re^{-1}(\tilde{g})\,\textbf{r}) \\
&= &\sum_{\alpha'} \left[ D^{\Gamma_a}(\tilde{g}) \right]_{\alpha' \alpha} \,\phi^{\widetilde{\Gamma},\Gamma_a}_{\widetilde{\Gamma}_b,\alpha'}(\textbf{r})
\nonumber 
\end{eqnarray}
Fourth, one interest of Eq.~(\ref{eqt_fixedVarDev}) is to suggest an alternative way to arrive at UREF's by imposing for every eigenstate a {\em fixed variance development} on the fully symmetrized OBB (a generalization of the invariant development approach for Hamiltonians~\cite{BIRPIKUS}). Fifth  and last, the definition of the Clebsch-Gordan matrix entails additionnal phase factors due to the conjugation of $\widetilde{\Gamma}_b$ (c.f. Eq.~(\ref{eqt_blockRedD})).

\subsection{The UREF's in $C_{3v}$ symmetry}

We now illustrate the general results of the preceding section by studying again the specific case of a $C_{3v}$ heterostructure.
Let us consider first the non-degenerate irrep $\widetilde{\Gamma}={}^2\!E_{3/2}$. The first component of $\underline{\psi}^{{}^2\!E_{3/2}}(\textbf{r})$, related to the $\left|{}^1\!E_{3/2}\right>$ basis function, reduces directly, via Eq.~(\ref{eqt_LoiTFctEnv}), to the simple uncoupled expression:
\begin{eqnarray}
\psi^{{}^2\!E_{3/2}}_1(\Re^{-1}(\tilde{g})\,\textbf{r}) &= &\chi^{{}^2\!E_{3/2}}(\tilde{g}) \,\left[\chi^{{}^1\!E_{3/2}}(\tilde{g})\right]^* \,\,\psi^{{}^2\!E_{3/2}}_1(\textbf{r}) \nonumber \\
&= &\chi^{A_2}(\tilde{g}) \,\psi^{{}^2\!E_{3/2}}_1(\textbf{r})
\label{eqt_2E32_fct1}
\end{eqnarray}
where we have used the fact that ${}^i\!E_{3/2},i=1,2$ are mutually conjugated irreps, and that the direct product representation ${}^2\!E_{3/2} \otimes {}^1\!E_{3/2}^* \equiv A_2$ (c.f. table in appendix~\ref{appendix}). Therefore we finally obtain, in agreement with Eq.~(\ref{eqt_TrsfUltEnvFct}) that $\psi^{{}^2\!E_{3/2}}_1(\textbf{r})$ transforms like the single group irrep $A_2$. In a completely similar way, it is possible to show that the last component of $\underline{\psi}^{{}^2\!E_{3/2}}(\textbf{r})$ transforms like $A_1$ and that the two central functions like the 2D mutual partner functions of the irrep $E$. Therefore one can write the eigenstate $\underline{\psi}^{{}^2\!E_{3/2}}(\textbf{r})$ in most simple symmetrized form as:
\begin{equation}\label{eqt_2E32}
\underline{\psi}^{{}^2\!E_{3/2}}(\textbf{r}) =
\left(
\begin{array}{c}
\phi^{A_2}(\textbf{r}) \\
\phi^{E}_1(\textbf{r}) \\
\phi^{E}_2(\textbf{r}) \\
\phi^{A_1}(\textbf{r}) 
\end{array}
\right)
\end{equation}
where we have left out for clarity the $\widetilde{\Gamma}_b$ label on the $\phi$ functions (it can be easily restored by using the consecutive labels of the Bloch function basis).
For the ${}^1\!E_{3/2}$ irrep we obtain a similar decomposition:
\begin{equation}\label{eqt_1E32}
\underline{\psi}^{{}^1\!E_{3/2}}(\textbf{r}) =
\left(
\begin{array}{c}
\phi^{A_1}(\textbf{r}) \\
-\phi^{E}_2(\textbf{r}) \\
\phi^{E}_1(\textbf{r}) \\
-\phi^{A_2}(\textbf{r}) 
\end{array}
\right)
\end{equation}
It should be pointed out, that the $\phi$-functions appearing in Eq.~(\ref{eqt_1E32}) are not identical to those appearing in Eq.~(\ref{eqt_2E32}), this becomes obvious if one restores the implicit $\widetilde{\Gamma}$ label. However time-reversal will induce a mapping between these functions (see e.g.~\cite{DAL1}), but simultaneously with a change of the $\textbf{k}$-index to $-\textbf{k}$ (we have not used here the complicated notations of the preceding section for the $\phi$-functions because they are unnecessary in this fairly trivial case). For this reason we also included a minus sign in $\phi^{A_2}(\textbf{r})$ in Eq.~(\ref{eqt_1E32}).

For the irrep $E_{1/2}$ nothing is so trivial since {\em it is not possible to introduce a single group label for every component} because of the bidimensionality of the irrep $E_{1/2}$, however the general approach developed in the previous section will show its relevance in opening the way.\\
One must take into account the fact that the $\underline{\underline{V}}^B(\tilde{g})$ matrices are block-diagonal with dimensions $1,2,1$. First from Eq.~(\ref{eqt_LoiTFctEnv}) we easily deduce directly that the first component of the two mutual partner eigenstates of the $E_{1/2}$ irrep ($\psi^{E_{1/2}}_{\mu,1}(\textbf{r}), \,\mu=1,2$), are mutual partner functions of the single group 2D irrep $E$! The same holds also for the last components $\psi^{E_{1/2}}_{\mu, 4}, \,\mu=1,2$, but with a reversed association of the partners with the eigenstates. These results can be relatively simply understood by considering the representation product ${}^i\!E_{3/2}\otimes E_{1/2} \equiv E$, for $i=1,2$.\\
Much more intricate is the behaviour of the two central components of the two partner functions, i.e. $\psi^{E_{1/2}}_{\mu,2}(\textbf{r}), \,\mu=1,2$ and $\psi^{E_{1/2}}_{\mu,3}(\textbf{r}), \,\mu=1,2$. Indeed, there one must resort to the full power of Eq.~(\ref{eqt_ultEnvFctComp}) to find the set of symmetrized envelope functions that form a basis for the 4D reducible representation $E_{1/2}\otimes E_{1/2}$. Using the products $E_{1/2}\otimes E_{1/2} = A_1 \oplus A_2 \oplus E$ and $E\otimes E = A_1 \oplus A_2 \oplus E$ it is possible to disentangle the central components into corresponding single group irreps. The full result reads as:
\begin{eqnarray}
\underline{\psi}^{E_{1/2}}_1(\textbf{r}) &= &
\left(
\begin{array}{c}
-\phi_2^E(\textbf{r}) \\
\frac{1}{\sqrt{2}}\left(\phi^{A_1}(\textbf{r}) + \Phi^E_1(\textbf{r})\right) \\
-\frac{1}{\sqrt{2}}\left(\phi^{A_2}(\textbf{r}) + \Phi^E_2(\textbf{r})\right) \\
\varphi_1^E(\textbf{r})
\end{array}
\right) \nonumber \\
\underline{\psi}^{E_{1/2}}_2(\textbf{r}) &= &
\left(
\begin{array}{c}
\phi_1^E(\textbf{r}) \\
\frac{1}{\sqrt{2}}\left(\phi^{A_2}(\textbf{r}) - \Phi^E_2(\textbf{r})\right) \\
\frac{1}{\sqrt{2}}\left(\phi^{A_1}(\textbf{r}) - \Phi^E_1(\textbf{r})\right) \\
\varphi_2^E(\textbf{r})
\end{array}
\right) \label{eqt_E12}
\end{eqnarray}
where again a minus sign was added on $\phi^{A_2}(\textbf{r})$.
There are three different $E$ representations which appear in the expressions above. Let us identify clearly the different sets of $\phi^E$-functions with the help of the general theory: 
\begin{eqnarray}
\phi_\alpha^E(\textbf{r}) &\equiv & \phi^{E_{1/2},E}_{{}^2\!E_{3/2},\alpha}(\textbf{r}) \nonumber \\
\varphi_\alpha^E(\textbf{r}) &\equiv & \phi^{E_{1/2},E}_{{}^1\!E_{3/2},\alpha}(\textbf{r}) \nonumber \\
\Phi_\alpha^E(\textbf{r}) &\equiv & \phi^{E_{1/2},E}_{E_{1/2},\alpha}(\textbf{r}) \label{eqt_identE12}
\end{eqnarray}
which helps to keep track of the different possible origins of the $E$ representation. In total we see that with $\widetilde{\Gamma}=E_{1/2}$ we can associate the five independent UREF $\phi^{A_1}(\textbf{r}),\phi^{A_2}(\textbf{r})\phi_1^E(\textbf{r}),\varphi_1^E(\textbf{r})$ and $\Phi_1^E(\textbf{r})$ (the partner functions $\phi_2^E(\textbf{r}),\varphi_2^E(\textbf{r}),\Phi_2^E(\textbf{r})$ are not independent, they are related to $\phi_1^E(\textbf{r}),\varphi_1^E(\textbf{r}),\Phi_1^E(\textbf{r})$ by a symmetry operation). Moreover multiplicities $n_{\widetilde{\Gamma},\widetilde{\Gamma}_b^*;\Gamma_a}$ larger than $1$ cannot be illustrated here because they would only occur in the cubic and icosahedral point groups which are the only point groups which are not simple-reducible~\cite{ALT}.

Despite of the fact that five independent UREF functions are now required for $\widetilde{\Gamma}=E_{1/2}$ (instead of the four independent envelope functions related to the original spinor $\underline{\psi}^{E_{1/2}}_1(\textbf{r})$) we shall see in Section~\ref{sec_SDR_spin} that, thanks to the enabled SDR technique for spinors, the use of five independent UREF's indeed leads to a maximum reduction of computing time and computing memory requirements.

\subsection{The symmetry of the Luttinger matrix elements in the OBB}
\label{subsec_Lutt_OBB}

In this last subsection we shall shortly comment on the fact that simpler symmetry properties of the envelope functions goes hand-in-hand with simpler and more elegant expressions for the matrix elements of the $k.p$ Hamiltonian. It is possible to obtain scalar transformation laws for every matrix elements by considering again the expression of invariance of the $\bar{c}_2$-transformed Luttinger matrix under symmetry operations, and taking advantage of the explicit separation of symmetry operations into spatial and spinorial parts (c.f. Eq.~(\ref{eqt_generalTransf})). \\
It can be shown that, in the OBB basis, the Luttinger Hamiltonian takes a simpler form, and also that every $p,q,r,s$ operator can carry a single group irrep label: $p$ and $q$ are scalar operators transforming under symmetry operations like $A_1$, whilst the two operators $\left( r,s \right)\equiv \left( \theta^E_1,\theta^E_2 \right)$ form a set of irreducible tensorial operator (ITO) transforming like the partner functions of the irrep $E$, i.e. 
\begin{equation}\label{eqt_transfRS}
\vartheta^{(3D)}_{g}\, \theta^E_\mu\, \vartheta^{(3D)}_{g^{-1}} = \sum_{\nu=1}^2 D^E_{\nu\mu}(g)\,\theta^E_\nu
\end{equation}
We refer the reader to~\cite{DAL1} for more details.\newline

To shortly summarize the whole section~\ref{sec_OBB}, let us emphasize that the introduction of the concept Optimally symmetrized Bloch function Basis (OBB) corresponding to the basis reducing the representation of symmetry operations in spinorial space allows to systematically decompose envelope functions into a minimal set of minimally coupled UREFs (Ultimately Reduced  Envelope Functions) carrying only single group irrep labels which specify their individual transformation properties. It should be pointed out that the coupling in question is through symmetry operations, the Hamiltonian will still couple these functions together, following the single group irrep multiplication table.

From now on we shall leave out the underscore notation on matrices and spinors, assuming that the reader is familiar with the implicit nature of these basic objects of the theory.

\section{Spatial domain reduction technique for scalar functions}
\label{sec_SDR}
In this section we develop a technique which we call spatial domain reduction (SDR), and which allows one to maximally reduce the geometrical domain of solution of eigenproblems linked with partial differential operators for scalar functions. It is the second essential ingredient of the MSRF formalism. The technique is closely related to the general theory developed in ref.~\cite{Boss93}, but identifies from the outset the essential elements of the symmetric domain and the separate role of each special boundary and singular point. Other elements of the underlying theory, based on symmetry adapted basis functions, can be found in~\cite{Fassl92}. Although this SDR method is applicable to any point-group symmetry, we shall again confine ourselves here on the $C_{3v}$  case, keeping a rather intuitive and pedestrian approach, sufficient for grasping easily the method, and using it. A full fleshed mathematical treatment encompassing these developments has also been set up, but will be given elsewhere. In the following section (Sec.~\ref{sec_SDR_spin}) we shall also show how to go beyond the scalar case and reduce the spinorial case, but this requires a simultaneous use of OBB, consistent sets of UREF's, and the SDR. The formulation will apply equally well to a spatial 2D problem or 3D problem, i.e. a quantum wire or a quantum dot (c.f. Eq.~(\ref{eqt_genericH})). 

An important point to stress is that the SDR technique is general enough to simplify the treatment in the same way for {\em all real space methods provided that the discretization scheme respects symmetry}. We have actually used a FE method, but the procedure would be left unaltered for FD methods: one only needs to identify to which domain a given node belongs to separate the blocks in Eqs.~(\ref{eqt_v_psi}) and~(\ref{eqt_HA1}) to follow. A completely similar approach would even hold in the case of tight-binding (TB) methods provided one is able to identify the spatial localization related to each element of the basis. Further along such a line of thought other versions of the SDR, OBB, and MSRF techniques could be developed for Fourier space formulations, using analogous separations of the Fourier part and spinorial part using a symmetrized basis in both space (an example of symmetrized plane wave superpositions covering all sectors of the Fourier space is given in~\cite{OBRE}).

Let us now consider any arbitrary scalar Hamiltonian $H$ displaying the $C_{3v}$ group symmetry and related to a 2D structure like presented in Fig.~\ref{fig_section}. We shall detail in the next three subsections the three different steps of the method which are: 1) decomposition of the spatial domain into minimal disjoint sub-domains, 2) identification of the minimal set of independent sub-domains for every irrep (the reduced domains), 3) computation of the correspondingly reduced Hamiltonian matrices on the reduced domains. The procedure is systematic, and allows to treat non-trivial cases like the minimal domain for the 2D irrep $E$ in $C_{3v}$.

\subsection{Decomposition in disjoint sub-domains, and representation of symmetry transformations by permutation matrices}
\label{subsec_sub_domains}

\begin{figure}[hbt]
\includegraphics[scale=0.6]{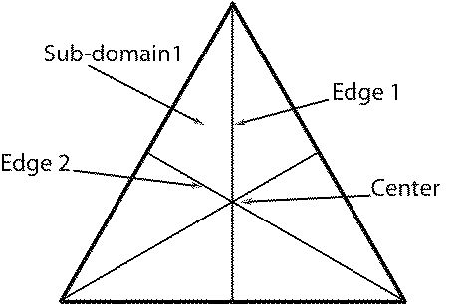}
\caption{Decomposition of the spatial domain in disjoint parts}\label{fig_domain}
\end{figure}

We consider a computational domain ${\cal D}$ with $C_{3v}$ symmetry (Fig.~\ref{fig_domain}), and assume that eigenstates $\psi$ are represented by a vector of dimension $N$, where each vector component $\psi_i$ represent the value of the wavefunction $\psi(\textbf{r}_i)$ on every mesh node $i$ at position $\textbf{r}_i$ in a FE or FD approach. In a TB approach $\psi_i$ would represent the weight of an orbital at each site. 

Since the domain ${\cal D}$ and the system Hamiltonian are $C_{3v}$-symmetric, we can set without restriction a symmetric spatial basis with respect to every operation $g$ of $C_{3v}$. From a real space numerical point of view, it implies the use of a symmetric FE or FD mesh. Let us then decompose the full domain ${\cal D}$ in thirteen disjoint parts (sub-domains) by separating the six interior parts ${\cal S}_i, i=1,\ldots,6$ in-between the three symmetry axes (c.f. Fig.~\ref{fig_domain}), the six borders ${\cal B}_i, i=1,\ldots,6$ between the interior parts, and the remaining central point ${\cal C}$ (this domain decomposition must be performed in such a way that the domains are always the image of one another under every symmetry operations).\\
Let us now formulate the discretized form of Eq.~(\ref{eqt_TransfVect}) under every change in coordinates corresponding to a symmetry operation $g\in C_{3v}$:
\begin{equation}\label{eqt_invMesh}
\textbf{r}'_i = \Re_g\textbf{r}_i = \textbf{r}_{j=\pi_g(i)}
\end{equation}
This condition obviously defines a set of permutations $\pi_g$ of the basis nodes which do form a representation of the symmetry group $C_{3v}$. Let us now assume that within the thirteen sub-domains identified previously the node numbering is always a perfect image of the other similar subdomains under any symmetry operation (this does not restrict generality: a further inner-domain permutation could be added to eventually describe arbitrary non-symmetric node-numbering between the sub-domains). Under this simplifying assumption all the permutation matrices become a block matrix permuting only the subdomains, and leaving unchanged the internal subdomain node numbering. 
We are thus led to consider $13\times 13$ ``domain permutation matrices'' ${\cal P}_g$, with the identity within each block, and defined as ${\cal P}_{ij}(g) = \delta_{i,\pi_g(j)}$. To give a concrete example let us represent the matrix ${\cal P}_{\sigma_{v1}}$ corresponding to the vertical mirror operation $\sigma_{v1}$ (see Fig.~\ref{fig_section}), as well as its transposed action on the set of subdomains:
\begin{equation}\label{eq_msigma1}
\left(
\begin{smallmatrix}
 1 & 0 & 0 & . & . & . &   &   &   & . & . & . & 0 \\
 0 & 1 & 0 &   &   &   &   &   &   &   &   &   & 0 \\
 0 & 0 & 0 &   &   &   &   &   &   &   &   & 0 & 1 \\
 . &   &   & . &   &   &   &   &   &   & 0 & 1 & 0 \\
 . &   &   &   & . &   &   &   &   & 0 & 1 & 0 & . \\
 . &   &   &   &   & . &   &   & 0 & 1 & 0 &   & . \\
   &   &   &   &   &   & 0 & 0 & 1 & 0 &   &   & . \\
   &   &   &   &   &   & 0 & 1 & 0 &   &   &   &   \\
   &   &   &   &   & 0 & 1 & 0 & 0 &   &   &   &   \\
 . &   &   &   & 0 & 1 & 0 &   &   & . &   &   & . \\
 . &   &   & 0 & 1 & 0 &   &   &   &   & . &   & . \\
 . &   & 0 & 1 & 0 &   &   &   &   &   &   & . & . \\
 0 & 0 & 1 & 0 & . & . & . &   &   & . & . & . & 0
\end{smallmatrix}
\right)\;
\left( \begin{smallmatrix}
\cC \\
\cB{1} \\ \cS{1} \\ \cB{2} \\ \cS{2} \\ \cB{3} \\ \cS{3} \\
\cB{4} \\ \cS{4} \\ \cB{5} \\ \cS{5} \\ \cB{6} \\ \cS{6} 
\end{smallmatrix} \right)
=
\left( \begin{smallmatrix}
\cC \\
\cB{1} \\ \cS{6} \\ \cB{6} \\ \cS{5} \\ \cB{5} \\ \cS{4} \\
\cB{4} \\ \cS{3} \\ \cB{3} \\ \cS{2} \\ \cB{2} \\ \cS{1} 
\end{smallmatrix} \right)
\end{equation}
where the big matrix is ${\cal P}^T_{\sigma_{v1}}$. All the other operations and associated matrices can be straightforwardly written down in an analogous way. Considered as simple $13\times 13$ ``domain permutation matrices'', the set of ${\cal P}_g$ permutation matrices obviously forms a {\em reducible} representation of the symmetry group $C_{3v}$. 

The next step is to decompose an arbitrary wavefunction $\psi(\textbf{r}_i)$ into blocks corresponding to the domain decomposition:
\begin{equation}\label{eqt_v_psi}
\psi = 
\left( \begin{smallmatrix}
\psi_{\cC} \\
\psi_{\cB{1}} \\ \psi_{\cS{1}} \\ \psi_{\cB{2}} \\ \psi_{\cS{2}} \\ \psi_{\cB{3}} \\ \psi_{\cS{3}} \\
\psi_{\cB{4}} \\ \psi_{\cS{4}} \\ \psi_{\cB{5}} \\ \psi_{\cS{5}} \\ \psi_{\cB{6}} \\ \psi_{\cS{6}} 
\end{smallmatrix} \right)
\end{equation}
where $\psi_{\cC}$ is the central value, $\psi_{\cB{1}}$ the subvector of values along the border $\cB{1}$ (Fig. \ref{fig_domain}), $\psi_{\cS{1}}$ the subvector representing the values of the wavefunction inside the sub-domain ${\cS{1}}$, etc... In the numerics it is sufficient to group the nodes by domains. Note that in this approach we do not need to refer to the dimensionality of the structure, i.e. {\em the technique will work equally well for a 2D spatial problem (quantum wire), or a 3D problem (a quantum dot)}.\\

It is also important to note that, according to Eq.~(\ref{transfFct}), a passive change in coordinates described by $g\in C_{3v}$ is now equivalent to perform a ${\cal P}_g$ permutation of the $13$ subparts of the wavefunction:
\begin{equation}\label{eqt_transfoP}
\vartheta_{g} \psi = {\cal P}_g \psi
\end{equation}

The decomposition~(\ref{eqt_v_psi}) of the subparts of the wavefunction $\psi$, and the identification of its transformation properties~(\ref{eqt_transfoP}), could seem simplistic, but they are in fact an essential step for the formulation of the theory, allowing in the foregoing section the identification of the independent geometrical parts of the wavefunctions, as well as, much later, a compact solution of the problem in terms of reduced geometrical variables, i.e. on a minimal set of independent sub-domains for each irrep of $C_{3v}$.\\

\subsection{Minimal set of independent sub-domains}
\label{subsec_min_sub_domains}
Let us construct the projection operator $P_\mu^\Gamma$ on the partner function $\mu$ of a given irrep $\Gamma$ of $C_{3v}$, operating on wavefunctions. Group theory provides a closed form expression for $P_\mu^\Gamma$ where we can use the set of ${\cal P}_g$ permutation matrices just defined:
\begin{equation}\label{eqt_proj}
P_\mu^\Gamma = \frac{d_\Gamma}{|\mathcal{G}|} \sum_{g\in \mathcal{G}} \left( D^\Gamma(g) \right)^*_{\mu\mu} \,{\cal P}_g 
\end{equation}
Therefore any arbitrary function on ${\cal D}$, represented by a vector $\psi$, can be written uniquely as a sum of symmetrized components $\psi_\mu^\Gamma$:
\begin{equation}\label{eqt_I_proj}
\psi = \Big\{\sum_{\Gamma, \mu}P^\Gamma_\mu\Big\} \psi= \sum_{\Gamma, \mu} \psi_\mu^\Gamma
\end{equation}
In particular, a function $\psi_\mu^\Gamma$ transforming like the partner function $\mu$ of the irrep $\Gamma$, is necessarily an eigenstate of the projector $P_\mu^\Gamma$ with unit eigenvalue:
\begin{equation}\label{eqt_condition}
P_\mu^\Gamma \psi_\mu^\Gamma =  \psi_\mu^\Gamma
\end{equation}
This condition allows us to identify for every irrep the necessary relations between the values of $\psi_\mu^\Gamma$ on the different subdomains of ${\cal D}$. For example, for $\Gamma = A_1$ or $\Gamma = A_2$, the condition~(\ref{eqt_condition}) leads to
\begin{equation}\label{eqt_Ai}
\psi^{A_1} = 
\left( \begin{smallmatrix}
\psi_{\cC} \\
\psi_{\cB{1}} \\ \psi_{\cS{1}} \\ \psi_{\cB{2}} \\ \psi_{\cS{1}} \\ \psi_{\cB{1}} \\ \psi_{\cS{1}} \\
\psi_{\cB{2}} \\ \psi_{\cS{1}} \\ \psi_{\cB{1}} \\ \psi_{\cS{1}} \\ \psi_{\cB{2}} \\ \psi_{\cS{1}} 
\end{smallmatrix} \right)
\hspace{0.5cm}{\textnormal{or}}\hspace{0.5cm}
\psi^{A_2} = 
\left( \begin{smallmatrix}
0 \\
0 \\  \psi_{\cS{1}} \\ 0 \\ -\psi_{\cS{1}} \\ 0 \\  \psi_{\cS{1}} \\
0 \\ -\psi_{\cS{1}} \\ 0 \\  \psi_{\cS{1}} \\ 0 \\ -\psi_{\cS{1}} 
\end{smallmatrix} \right)
\end{equation}
where the superscript $\Gamma = A_1$ (respectively $A_2$) is implicit on every components $\psi_{\cC},\psi_{\cB{1}},\cdots$ of $\psi^{A_1}$ (respectively $\psi^{A_2}$). The above results~(\ref{eqt_Ai}) for $A_1$ or $A_2$ are very interesting and deserve two comments: first they show that each wavefunction is characterized by its values on a minimal subset of subdomains (which is not the same for all irreps), second for $A_1$ and $A_2$ it should be possible to characterize the wavefunction essentially on one sixth of ${\cal D}$. Note also that the borders are not necessary for $A_2$ since Eq.~(\ref{eqt_Ai}) shows that $\psi^{A_2}$ must necessarily be zero on the symmetry axes between the six main subdomains! For numerical efficiency one is therefore lead to seek a solution for each irrep separately, on the minimal set of subdomains only, and this will be considered in detail in the next subsection, after the treatment of the less trivial case of the 2D irrep $E$.

A novel aspect of the 2D irrep $E$ is that there are a few equivalent matrix representation of $C_{3v}$, and a specific choice must be made beforehand. Since with our axis choice the 3D rotation matrix $\Re$ do already form a reduced 3D representation of the $C_{3v}$ group, it is already in the desired block-form ($A_1\oplus E$). For simplicity we shall take this specific matrix representation for $E$ (explicitly presented in the appendix~\ref{appendix}). One advantage is that this representation is real, another one is that the matrix representing $\sigma_{v1}$ is diagonal, hence the first and second partner functions will be respectively even and odd with respect to $\sigma_{v1}$. The last advantage is that conditions of parity with respect to $\sigma_{v1}$ correspond to local Dirichlet or Neumann boundary conditions (if instead one would have chosen to diagonalize the rotations $C_3^\pm$ one would have got complex wavefunctions with non-local $2\pi/3$-periodic boundary conditions as a function of the azimuthal angle).\\
For each partner function one can now build a different projection operator, called $P^E_1$ and $P^E_2$ according to Eq.~(\ref{eqt_proj}). We can however work with only one partner function, because the independent parameters are related to the irrep and allow to re-built the second partner function from the first. For each partner function one finds:
\begin{equation}\label{eqt_E13_old}
\psi^{E}_1=
\left(
\begin{smallmatrix}
0 \\ \psi_{1,\cB{1}} \\ \psi_{1,\cS{1}} \\ \psi_{1,\cB{2}} \\ \psi_{1,\cS{2}} \\  -\frac{1}{2}\psi_{1,\cB{1}} \\ 
-\psi_{1,\cS{1}}-\psi_{1,\cS{2}} \\ -2\psi_{1,\cB{2}} \\ -\psi_{1,\cS{1}}-\psi_{1,\cS{2}} \\  
-\frac{1}{2}\psi_{1,\cB{1}} \\ \psi_{1,\cS{2}} \\ \psi_{1,\cB{2}} \\ \psi_{1,\cS{1}} 
\end{smallmatrix}
\right)
\hspace{0.3cm}{\textnormal{and}}\hspace{0.3cm}
\psi^{E}_2=
\left(
\begin{smallmatrix}
0 \\ 0 \\ \psi_{2,\cS{1}} \\ \psi_{2,\cB{2}} \\ 
\psi_{2,\cS{2}} \\ \psi_{2,\cB{3}} \\ 
- \psi_{2,\cS{1}} + \psi_{2,\cS{2}} \\ 
0 \\ \psi_{2,\cS{1}} - \psi_{2,\cS{2}} \\  
- \psi_{2,\cB{3}} \\ 
- \psi_{2,\cS{2}} \\ 
- \psi_{2,\cB{2}} \\ - \psi_{2,\cS{1}} 
\end{smallmatrix}
\right)
\end{equation}
We see that the corresponding reduced domain is now larger than a sixth, and its surface/volume exactly corresponds to {\em a third} of the full domain. It is seemingly not possible to rebuilt the full function from a smaller domain as for the non-degenerate $A_i$ irreps. It is however possible to bypass this limitation, and to still reduce to one sixth of the full domain, by noting that, since the first and second partner functions are not independent, one could retain simultaneously the two of them, but only on the first sixth $\cS{1}$. This amounts to replace the $\psi_{1,\cS{2}}$ variable of the first partner function by the $\psi_{2,\cS{1}}$ variable of the second partner function, related to the first internal domain ("first sixth"). It is possible to relate these variables using the $\sigma_{v3}$ symmetry operation. Using the transformation rule~(\ref{eqt_transfFctPart}) we find: 
\begin{eqnarray}
\psi_{1,\cS{2}} &\rightarrow &\left[ D^E(\sigma_{v3}) \right]_{11}\, \psi_{1,\cS{1}} + \left[ D^E(\sigma_{v3}) \right]_{21}\, \psi_{2,\cS{1}} \nonumber \\
& &= -\frac{1}{2} \psi_{1,\cS{1}} + \frac{\sqrt{3}}{2} \psi_{2,\cS{1}} \nonumber \\
\psi_{2,\cB{2}} &\rightarrow & \sqrt{3} \psi_{1,\cB{2}} \nonumber \\
\psi_{2,\cB{3}} &\rightarrow & \frac{\sqrt{3}}{2} \psi_{1,\cB{1}} \nonumber \\
\psi_{2,\cS{2}} &\rightarrow & \frac{1}{2} \psi_{2,\cS{1}} + \frac{\sqrt{3}}{2} \psi_{1,\cS{1}} 
\label{eqt_transfo3to6}
\end{eqnarray}

This leads to the following full expressions for the first partner function as a function of the values of the two partner functions on the first sixth of the domain:
\begin{equation}\label{eqt_E16}
\psi^{E}_1=
\left(
\begin{smallmatrix}
0 \\ \psi_{1,\cB{1}} \\ \psi_{1,\cS{1}} \\ \psi_{1,\cB{2}} \\ 
- \frac{1}{2} \psi_{1,\cS{1}} + \frac{\sqrt{3}}{2} \, \psi_{2,\cS{1}} \\ 
- \frac{1}{2} \psi_{1,\cB{1}} \\ 
- \frac{1}{2} \psi_{1,\cS{1}} - \frac{\sqrt{3}}{2} \, \psi_{2,\cS{1}} \\ -2 \psi_{1,\cB{2}} \\ 
- \frac{1}{2} \psi_{1,\cS{1}} - \frac{\sqrt{3}}{2} \, \psi_{2,\cS{1}} \\  
- \frac{1}{2} \psi_{1,\cB{1}} \\ 
- \frac{1}{2} \psi_{1,\cS{1}} + \frac{\sqrt{3}}{2} \, \psi_{2,\cS{1}} \\ \psi_{1,\cB{2}} \\ \psi_{1,\cS{1}} 
\end{smallmatrix}
\right) , \hspace{0.1cm}
\psi^{E}_2=
\left(
\begin{smallmatrix}
0 \\ 0 \\ \psi_{2,\cS{1}} \\ \sqrt{3} \, \psi_{1,\cB{2}} \\ 
\frac{1}{2} \psi_{2,\cS{1}} + \frac{\sqrt{3}}{2} \, \psi_{1,\cS{1}} \\ 
\frac{\sqrt{3}}{2} \, \psi_{1,\cB{1}} \\ 
- \frac{1}{2} \psi_{2,\cS{1}} + \frac{\sqrt{3}}{2} \, \psi_{1,\cS{1}} \\ 
0 \\ \frac{1}{2} \psi_{2,\cS{1}} - \frac{\sqrt{3}}{2} \, \psi_{1,\cS{1}} \\  
- \frac{\sqrt{3}}{2} \, \psi_{1,\cB{1}} \\ 
- \frac{1}{2} \psi_{2,\cS{1}} - \frac{\sqrt{3}}{2} \, \psi_{1,\cS{1}} \\ 
- \sqrt{3} \, \psi_{1,\cB{2}} \\ - \psi_{2,\cS{1}} 
\end{smallmatrix}
\right)
\end{equation}
For the 2D irrep $E$, $\psi^E_1$ and $\psi^E_2$ are obviously less intuitive than $\psi^{A_1}$ and $\psi^{A_2}$. However the important point is that we have found a systematic procedure to derive them which even allows to find straightforwardly novel analytical results. For instance we can see in Eqs.~(\ref{eqt_E13_old}) and~(\ref{eqt_E16}) that 
\begin{equation}\label{eqt_NTEc}
\psi_{1,\cC}=\psi_{2,\cC}=0
\end{equation}
This is non-trivial and had not be noticed before, even with correct numerical results at hand.\\

\subsection{Reduced domain and non-trivial boundary conditions}
\label{subsec_reduced_domains}
We are now going to use the identification of a minimal number of independent part in the wavefunction to reduce the size of the problem at hand. In addition we show how non-trivial boundary conditions naturally occur at the border of the reduced domain.

For every irreps we may collect the minimal set of independent parameters and define a reduced vector $\psi^\Gamma_{Red}$ on the corresponding reduced domain. For the irreps $A_1$ and $A_2$ they are given by:
\begin{equation}\label{eqt_AiRed}
\psi^{A_1}_{Red}=
\left(
\begin{smallmatrix}
\psi_{\cC} \\ \psi_{\cB{1}} \\ \psi_{\cS{1}} \\ \psi_{\cB{2}} 
\end{smallmatrix}
\right)
\hspace{0.6cm}{\textnormal{and}}\hspace{0.6cm}
\psi^{A_2}_{Red}=
\psi_{\cS{1}}
\end{equation}
and for the irrep $E$ by
\begin{equation}\label{eqt_ERed}
\psi^E_{Red}=
\left(
\begin{smallmatrix}
\psi_{1,\cB{1}} \\ \psi_{1,\cS{1}} \\ \psi_{2,\cS{1}} \\ \psi_{1,\cB{2}} 
\end{smallmatrix}
\right)
\end{equation}
A compact way to summarize Eqs.~(\ref{eqt_AiRed})-(\ref{eqt_ERed}) is to introduce for every irrep $\Gamma$ a set of rectangular reduction matrices $S^\Gamma_\mu$, $\mu=1, \dots d_\Gamma$ such that the reduced vectors can be written:
\begin{equation}\label{eqt_S}
\psi^\Gamma_\mu = S^\Gamma_\mu\psi^\Gamma_{Red}
\end{equation}
The reduction matrices $S^\Gamma_\mu$ will be used in Section~\ref{subsubsec_reduced_A1} to obtain the reduced Hamiltonian corresponding to the independent variables. Note that the $S^\Gamma_\mu$ matrices, obtained here in a rather pedestrian way, are not unitary (for another approach see e.g. ref.~\cite{GALL09} and references therein).

Let us now shortly look at the natural boundary conditions which occur at the edges $\cB{1}$ and $\cB{2}$ of the reduced domain on one sixth $\cS{1}$. For the 1D irrep $A_2$ the boundary conditions are obvious from Eq.~(\ref{eqt_Ai}) and are of the Dirichlet type, since the wavefunction naturally vanishes on the border:
\begin{equation}
\label{eqt_NTBndryA2}
\psi^{A_2}_{\cB{1}} = \psi^{A_2}_{\cB{2}} = 0 
\end{equation}
For the 1D irrep $A_1$ it is not very difficult to show that the corresponding boundary condition is of the Neumann type, and that the normal derivative vanish on the border. To show this we note that the behaviour of the normal derivative on an edge $\cB{j}$ can be connected to the difference between the subdomains on each side: $\psi'^{A_1}_{\cB{j}} \sim  (\psi^{A_1}_{\cB{j-1}} - \psi^{A_1}_{\cB{j}})$ where we assume the definition $\psi^{A_1}_{\cB{0}} = \psi^{A_1}_{\cB{6}}$. Therefore the normal derivatives $\psi'^{A_1}_{\cB{1}}$ and $\psi'^{A_1}_{\cB{2}}$ vanish, i.e.
\begin{equation}
\label{eqt_NTBndryA1}
\psi'^{A_1}_{\cB{1}} = \psi'^{A_1}_{\cB{2}} = 0
\end{equation}
These intuitive results for the 1D irreps $A_1$ and $A_2$ were clearly apparent in the numerical behaviour, have been easy to enforce numerically when computing on the reduced domain. Furthermore they correspond to {\em local boundary conditions} in a real space formulation (FE or FD methods), hence they would not destroy the sparsity of the Hamiltonian and mass matrices. Finally to alleviate the need for node re-numbering the so-called ``penalty method'' can be used to enforce the Dirichlet boundary condition~\cite{COOK}.

Let us now look at the natural boundary conditions for the two partner functions linked with the 2D irrep $E$. We shall see that these natural boundary conditions become not so trivial, and could hardly be guessed by just looking at the numerical result. It may seem easy to directly read valid boundary conditions from Eqs.~(\ref{eqt_E13_old}) or~(\ref{eqt_E16}), i.e. $\psi'^E_{1,\cB{1}} = \psi'^E_{1,\cB{4}} = 0$ and  $\psi^E_{2,\cB{1}} = \psi^E_{2,\cB{4}} = 0$, however these boundary conditions are not on the reduced minimal domain, and using them we would only take advantage of $C_s$ symmetry only, not the full $C_{3v}$ symmetry of the problem.

Therefore let us look more closely at what happens at the border of the reduced domain in the first partner function in Eqs.~(\ref{eqt_E13_old}) or~(\ref{eqt_E16}). Clearly on the border $\cB{3}$ we have
\begin{equation}
\label{eqt_NTBndryE1}
\psi^E_{1,\cB{3}} = - \frac{1}{2} \psi^E_{1,\cB{1}} 
\end{equation}
This beautiful analytical result is not very convenient to use: from a numerical point of view it does correspond to a {\em non-local boundary condition} (coupling far apart points), and a pedestrian implementation would destroy the matrix sparsity pattern linked with nearest neighbor mesh nodes. 

For the second partner function a similar path allows to obtain the corresponding non-local boundary condition: this time the normal derivative of the function with respect to the boundary will be involved. Since along edge $\cB{j}$ the normal derivative of the second partner function $\psi'^E_{2,\cB{i}}$ is related to the difference of the function across the boundary, $\psi'^E_{2,\cB{i}} \sim  (\psi^E_{2,\cB{i-1}}-\psi^E_{2,\cB{i}})$, it is easy to see from the second equation in Eqs.~(\ref{eqt_E13_old}) or~(\ref{eqt_E16}) that the function $\psi^{E}_2$ necessarily obeys the following boundary condition on the border $\cB{3}$:
\begin{equation}
\label{eqt_NTBndryE2}
\psi'^E_{2,\cB{3}} = - \frac{1}{2} \psi'^E_{2,\cB{1}} 
\end{equation}
The non-trivial character of this boundary condition is further evidenced by the fact that on the border $\cB{3}$ the value of the function $\psi^{E}_2$ is not vanishing, by contrast with the value on the border $\cB{1}$.

The problems linked with boundaries have been studied in detail by ref.~\cite{Boss93} with the help of a general approach, valid for any symmetry group, but also applied specifically to $C_{3v}$ as an example. Eqs.~(\ref{eqt_NTBndryA1}) and~(\ref{eqt_NTBndryA2}) were straightforwardly found, but not directly Eqs.~(\ref{eqt_NTEc}),(\ref{eqt_NTBndryE1})-(\ref{eqt_NTBndryE2}), because their approach considers only coupling the two partner functions on the boundaries of the so-called ``fundamental symmetry cell'' $\cS{1}$, as could be read directly from Eq.~(\ref{eqt_E16}).

To summarize, in the present section we have found the boundary conditions~(\ref{eqt_NTBndryA2}-\ref{eqt_NTBndryE2}) which do allow to solve numerically in an unambiguous way the eigenproblems related to each irrep on a minimal domain whose size vary from one sixth to one third. However for the 2D $E$ representation they are both not very trivial and non-local in space. Later, in subsection~\ref{subsec_reduced_Hs}, we shall see that the use of SDR-reduced Hamiltonians will avoid the explicit use of boundary conditions whilst keeping the natural sparsity patterns, i.e. it will be possible to solve all issues at once.

\subsection{The structure of the full Hamiltonian}
\label{subsec_full_H}
Having developed the SDR technique for every type of wavefunction, we shall now consider the form of the full Hamiltonian (i.e the Hamiltonian on the full domain). This will enable us to reduce the full eigenvalue problem to distinct reduced eigenvalue problems for each type of irreducible representation. We recall that we deal for the moment with a scalar Hamiltonian. In the case of $C_{3v}$ such a Hamiltonian can be seen as an invariant $13\times 13$ block operator, operating on the decomposition~(\ref{eqt_v_psi}) of the subparts of the wavefunction $\psi$. Then using group theory it is easy to construct the most general $13\times 13$ block-form matrix which would satisfy the following two conditions : 1) respect of the sub-domain connectivity induced by differential operators, and 2) invariance with respect to the symmetry operations of the group (i.e. the full Hamiltonian operator belong to the irrep $A_1$). The discretized form of the Hamiltonian will be represented by a similar block-matrix.

The first condition, concerning the most general Hamiltonian on the full domain respecting the connectivity between sub-domains, can be easily enforced: the Hamiltonian must have a block-form corresponding to the the most general $13\times 13$ matrix $H_{c}$ respecting the connectivity, which is written down {\em by crossing out off-diagonal elements that would couple non-contiguous domains in real space}. Here, since we use only first order FE elements, we can also take into account the further simplification that the center $\cC$ couples with every edge $\cB{i}$ but not with the internal domains $\cS{i}$ (but this restriction can be easily lifted). 

The second condition, concerning the invariance, can be satisfied by enforcing the fact that $H$ must be invariant under projection on the irrep $A_1$ (the projector on $A_1$ for {\em operators}, a ``superprojector'', is built analogously to~(\ref{eqt_proj}), where the linear operator ${\cal P}_g$ is replaced by a linear ``superoperator'', whose action on an operator ${\cal O}$ is ${\cal P}_g {\cal O} {\cal P}_g^{-1}$). 

One can then find that the corresponding scalar Hamiltonian matrix $H$ is necessarily of the form
\begin{widetext}
\begin{equation}\label{eqt_HA1}
H\equiv H_c^{A_1} = 
\left(
\begin{smallmatrix}
 H_{\cC} & H_{\cC,\cB{1}} & H_{\cC,\cS{1}} & H_{\cC,\cB{2}} & H_{\cC,\cS{1}} & H_{\cC,\cB{1}} & H_{\cC,\cS{1}} & H_{\cC,\cB{2}} & H_{\cC,\cS{1}} & H_{\cC,\cB{1}} & H_{\cC,\cS{1}} & H_{\cC,\cB{2}} & H_{\cC,\cS{1}} \\ 
 H^{\dag}_{\cC,\cB{1}} & H_{\cB{1}} & H_{\cB{1},\cS{1}} & H_{\cB{1},\cB{2}} & 0 & 0 & 0 & 0 & 0 & 0 & 0 & H_{\cB{1},\cB{2}} & H_{\cB{1},\cS{1}} \\
 H^{\dag}_{\cC,\cS{1}} & H^{\dag}_{\cB{1},\cS{1}} & H_{\cS{1}} & H_{\cS{1},\cB{2}} & 0 & 0 & 0 & 0 & 0 & 0 & 0 & 0 & 0 \\
 H^{\dag}_{\cC,\cB{2}} & H^{\dag}_{\cB{1},\cB{2}} & H^{\dag}_{\cS{1},\cB{2}} & H_{\cB{2}} & H^{\dag}_{\cS{1},\cB{2}} & H^{\dag}_{\cB{1},\cB{2}} & 0 & 0 & 0 & 0 & 0 & 0 & 0 \\
 H^{\dag}_{\cC,\cS{1}} & 0 & 0 & H_{\cS{1},\cB{2}} & H_{\cS{1}} & H^{\dag}_{\cB{1},\cS{1}} & 0 & 0 & 0 & 0 & 0 & 0 & 0 \\
 H^{\dag}_{\cC,\cB{1}} & 0 & 0 & H_{\cB{1},\cB{2}} & H_{\cB{1},\cS{1}} & H_{\cB{1}} & H_{\cB{1},\cS{1}} & H_{\cB{1},\cB{2}} & 0 & 0 & 0 & 0 & 0 \\
 H^{\dag}_{\cC,\cS{1}} & 0 & 0 & 0 & 0 & H^{\dag}_{\cB{1},\cS{1}} & H_{\cS{1}} & H_{\cS{1},\cB{2}} & 0 & 0 & 0 & 0 & 0 \\
 H^{\dag}_{\cC,\cB{2}} & 0 & 0 & 0 & 0 & H^{\dag}_{\cB{1},\cB{2}} & H^{\dag}_{\cS{1},\cB{2}} & H_{\cB{2}} & H^{\dag}_{\cS{1},\cB{2}} & H^{\dag}_{\cB{1},\cB{2}} & 0 & 0 & 0 \\
 H^{\dag}_{\cC,\cS{1}} & 0 & 0 & 0 & 0 & 0 & 0 & H_{\cS{1},\cB{2}} & H_{\cS{1}} & H^{\dag}_{\cB{1},\cS{1}} & 0 & 0 & 0 \\
 H^{\dag}_{\cC,\cB{1}} & 0 & 0 & 0 & 0 & 0 & 0 & H_{\cB{1},\cB{2}} & H_{\cB{1},\cS{1}} & H_{\cB{1}} & H_{\cB{1},\cS{1}} & H_{\cB{1},\cB{2}} & 0 \\
 H^{\dag}_{\cC,\cS{1}} & 0 & 0 & 0 & 0 & 0 & 0 & 0 & 0 & H^{\dag}_{\cB{1},\cS{1}} & H_{\cS{1}} & H_{\cS{1},\cB{2}} & 0 \\
 H^{\dag}_{\cC,\cB{2}} & H^{\dag}_{\cB{1},\cB{2}} & 0 & 0 & 0 & 0 & 0 & 0 & 0 & H^{\dag}_{\cB{1},\cB{2}} & H^{\dag}_{\cS{1},\cB{2}} & H_{\cB{2}} & H^{\dag}_{\cS{1},\cB{2}} \\
 H^{\dag}_{\cC,\cS{1}} & H^{\dag}_{\cB{1},\cS{1}} & 0 & 0 & 0 & 0 & 0 & 0 & 0 & 0 & 0 & H_{\cS{1},\cB{2}} & H_{\cS{1}}
\end{smallmatrix}
\right)
\end{equation}
\end{widetext}
In this equation the square diagonal blocks $H_{ii}$ are simply noted $H_i$ and represent the restriction of the Hamiltonian operator within every subdomain. The off-diagonal rectangular blocks $H_{i j}=H^{\dag}_{j i}$ couple the $i$ and $j$ sub-domains. One can see in Eq.~(\ref{eqt_HA1}) that there are many redundant blocks: indeed this is a consequence of symmetry and it was one of our primary goals to identify them. FD or FE schemes will automatically generate such a block structure for the Hamiltonian, provided the mesh does respect $C_{3v}$ symmetry. Each block can be either computed numerically only once (to save computer time, but it requires some care), or can be obtained by sampling a posteriori the blocks in the full discretized Hamiltonian. To this end it is sufficient to tag every node with its domain. 

It should be pointed out that the reduction technique presented in the next section allows to go even further in the reduction by building the smallest possible reduced Hamiltonians, where every Hamiltonian block appears a minimum number of times. The proposed procedure will amount to analytically pre-diagonalize by block our full Hamiltonian using group theory. Then the final numerical diagonalization of each block on the diagonal will directly produce the independent relevant parts of each type of eigenfunction without redundancy.
\subsection{The reduced Hamiltonians and explicit results for the $C_{3v}$ group}
\label{subsec_reduced_Hs}
For every irrep, it is useful to obtain a corresponding reduced Hamiltonians related to the minimum number of independent part on the sub-domains. To this end the rectangular reduction matrices $S^\Gamma_\mu$, $\mu=1, \dots d_\Gamma$ defined earlier in Eq.~(\ref{eqt_S}) allow one to write:
\begin{equation}\label{eqt_HRed}
H_{Red}^\Gamma = \frac{1}{d_{Red}} \left(S^\Gamma_\mu\right)^{\dag} H \, S^\Gamma_\mu
\end{equation}
where $d_{Red}$ represent a ``normalization'' factor, chosen conveniently so that the diagonal blocks $H_{\cS{1}}$ of the Hamiltonian, corresponding to the interior parts $\cS{1}-\cS{6}$, would be unaffected by reduction. By definition the number of blocks in such a reduced Hamiltonian is minimal and its solutions correspond to a given irrep $\Gamma$ of the symmetry group of the full Hamiltonian. It should be pointed out that it is enough to solve only once within each irrep $\Gamma$ to obtain all eigenvalues and reconstruct all eigenspaces by using symmetry transformations. This is also reflected in the fact that $H_{Red}^\Gamma$ is independent on $\mu$.

In a FE formulation of the eigenvalue problem one should also take into account that the original differential equation in real space is mapped to a generalized eigenvalue problem 
\begin{equation}\label{eqt_geneig}
H\,\psi^\Gamma_{\mu,n} = E^\Gamma_n \, M \, \psi^\Gamma_{\mu,n}
\end{equation}
where $H$ and $M$ are respectively the so-called ``stiffness'' and ``mass matrices'', and $E^\Gamma_n, n=1,...$ are different eigenvalues all labelled by the $\Gamma$ irrep. The stiffness matrix $H$ correspond to the Hamiltonian expressed in the FE basis and has also the block-form presented in~(\ref{eqt_HA1}). Similarly the mass matrix $M$ which represent the non-orthogonality of FE basis is also invariant (i.e. is associated with the $A_1$ irrep) and has the same block-structure as $H$. The scalar product in the FE approach simply reduces to $\left< \psi \right| \left. \!\phi \right> = \psi^+ M \, \phi$, where $\psi$ and $\phi$ are the vectors corresponding to the coefficients of the decomposition on the FE basis. Hence in the FE approach one also needs to define a reduced mass matrix similar to~(\ref{eqt_HRed}):
\begin{equation}\label{eqt_MRed}
M_{Red}^\Gamma = \frac{1}{d_{Red}} \left(S^\Gamma_\mu\right)^{\dag} M \, S^\Gamma_\mu
\end{equation}
so that the set of reduced problems then reads
\begin{equation}\label{eqt_PbRed}
H^\Gamma_{Red} \, \psi_{Red}^\Gamma = E^\Gamma M^\Gamma_{Red} \, \psi_{Red}^\Gamma
\end{equation}
where the factor $d_{Red}$ cancels. It may however be interpreted: since both matrices $M$ and $M_{Red}^\Gamma$ can be considered as measures involved in the definition of scalar products of vectors of type $\psi$ and $\psi_{Red}$ appearing in~(\ref{eqt_geneig}) and~(\ref{eqt_PbRed}) respectively, the value of $d_{Red}$ corresponds to the ratio of the measures of the full and the reduced domain (but would be unity for unitary $S^\Gamma_\mu$ matrices, see e.g. ref.~\cite{GALL09} and references therein). In our $C_{3v}$ case our procedure led naturally to the value $d_{Red}=6$, which indeed is linked with a domain reduction by a factor $6$. Eq.~(\ref{eqt_HRed}) shows that there are as many separate eigenproblems as the number of irreps in the group $\mathcal{G}$. In a FD scheme the mass matrix is reduces simply to the identity. 

We are now going to review the specific forms of the reduced Hamiltonians for our $C_{3v}$ specific example. Similar forms will also hold for mass matrices.

\subsubsection{Reduced Hamiltonian for $A_1$}
\label{subsubsec_reduced_A1}
\begin{figure}[hbt]
\includegraphics[scale = 0.6]{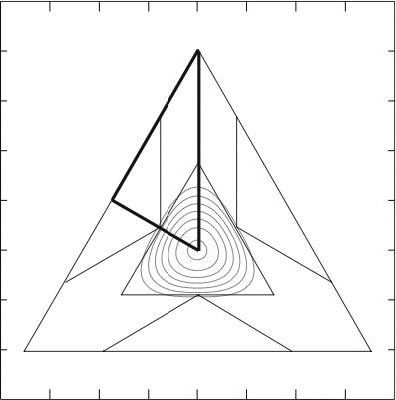}
\caption{Eigenstate with $A_1$ symmetry}\label{fig_A1}
\end{figure}

For $A_1$ the reduced Hamiltonian reads
\begin{equation}
\label{eqt_HA1Red}
H^{A_1}_{Red}=
\left(
\begin{smallmatrix}
 \frac{1}{6} H_{\cC} & \frac{1}{2} H_{\cC,\cB{1}} & 0 & \frac{1}{2} H_{\cC,\cB{2}} \\
 {} & \frac{1}{2}H_{\cB{1}} & H_{\cB{1},\cS{1}} & H_{\cB{1},\cB{2}} \\
 {} & {} & H_{\cS{1}} & H_{\cS{1},\cB{2}} \\
 c.c. & {} & {} & \frac{1}{2} H_{\cB{2}}
\end{smallmatrix}
\right)
\end{equation}
The factor $1/2$ appearing in front of the $H_{\cB{1}}$ and $H_{\cB{2}}$ in the diagonal can be interpreted simply in a FE scheme : it is equivalent to fill the matrix with only the contribution of the FEs on the interior side between the symmetry planes $\sigma_{v1}$ and $\sigma_{v2}$. It can be proved that such a procedure implicitely enforces a Neumann condition (with a zero normal derivative on the interface), evidencing clearly that indeed the boundary conditions corresponding to $A_1$ symmetry are implicitely incorporated! The same intuitive argument can explain the factor $1/6$ for the central block $H_{\cC}$. In Fig.~\ref{fig_A1} we show the ground electronic eigenstate with $A_1$ symmetry in $C_{3v}$ quantum wire. The highlighted sixth correspond to the reduced spatial domain used in the numerical solution.

\subsubsection{Reduced Hamiltonian for $A_2$}
\label{subsubsec_reduced_A2}

\begin{figure}[hbt]
\includegraphics[scale = 0.6]{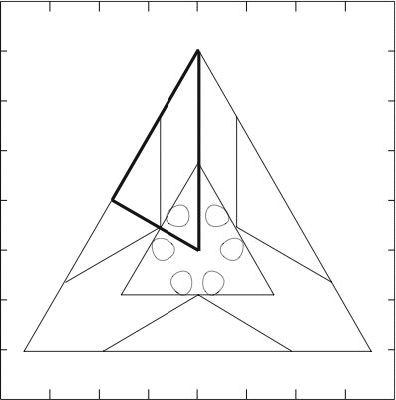}
\caption{Eigenstate with $A_2$ symmetry}\label{fig_A2}
\end{figure}

For the $A_2$ irrep, the reduced Hamiltonian will obviously have only one block (see Eq.~(\ref{eqt_AiRed})). 
\begin{equation}\label{eqt_HA2Red}
H^{A_1}_{Red} = H_{\cS{1}} 
\end{equation}
The eigenstates then naturally vanish on every border (Dirichlet condition) and again one only needs to solve on the first internal sub-domain. Fig.~\ref{fig_A2} displays such an eigenfunction, found as a highly excited state in $C_{3v}$ quantum wire.
For convenience in our numerics we still use the same grid as for $A_1$ and treat the border nodes, but ``cross-out'' the border points and enforce the Dirichlet boundary condition with the ``penalty method''. This avoids a cumbersome node-renumbering task between the two irreps $A_1$ and $A_2$, at a negligible numerical cost.

\subsubsection{Reduced Hamiltonian for $E$}
\label{subsubsec_reduced_E}
Finally, let us consider the more complicated degenerate irrep $E$ and apply the same procedure. We obtain the following reduced Hamiltonian
\begin{equation}
\label{eqt_HERed}
H^{E}_{Red}=
\frac{1}{2}\left(
\begin{smallmatrix}
 \frac{1}{2} H_{\cB{1}} & H_{\cB{1},\cS{1}} & 0 & H_{\cB{1},\cB{2}} \\
 {} & H_{\cS{1}} & 0 & H_{\cS{1},\cB{2}} \\
 {} & {} & H_{\cS{1}} & \sqrt{3} \, H_{\cS{1},\cB{2}} \\
 C.C. & {} & {} & 2 \, H_{\cB{2}}
\end{smallmatrix}
\right)
\end{equation}
corresponding to the ``mixed'' reduced partner wavefunction~(\ref{eqt_ERed}). The $\sqrt{3} \, H_{\cS{1},\cB{2}}$ block represent the non-trivial coupling between the two partner functions (coupling of $\psi_{1,\cB{2}}$ with $\psi_{2,\cS{1}}$). As explained before, in the same way we could have started with the second partner function and obtain a similar reduced Hamiltonian with respect to a different but similar set of reduced variables (see the dependence of $S^\Gamma_\mu$ on $\mu$ in Eq.~(\ref{eqt_S})), but we do not need to do so because the new Hamiltonian would carry exactly the same information, and starting from the reduced variables in $\psi^E_{Red}$ one is already able to reconstruct the full domain vectors $\psi^E_1$ and $\psi^E_2$ given by Eqs.~(\ref{eqt_E16}) or Eq.~(\ref{eqt_E13_old}) and its corresponding form for $\psi^E_2$.
\begin{figure}[hbt]
\begin{tabular}{cc}
\includegraphics[scale = 0.6]{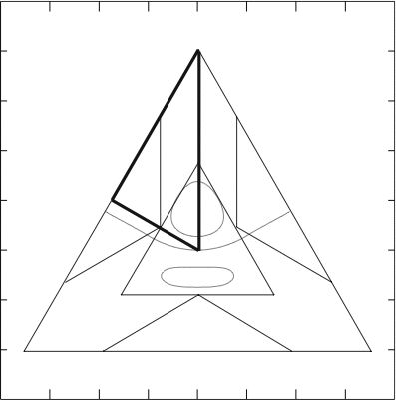} & 
\includegraphics[scale = 0.6]{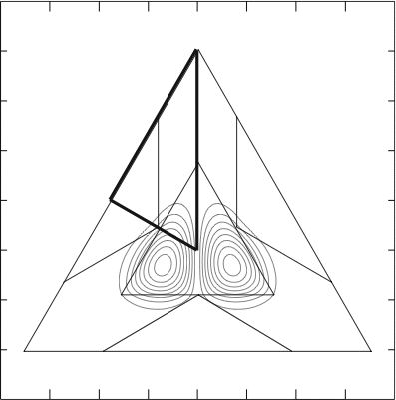} \\
\textbf{(a)} & \textbf{(b)} \\
\end{tabular}
\caption{Eigenstate with $E$ symmetry (\textbf{(a)} partner function 1 (even), 
\textbf{(b)} partner function 2 (odd))}
\label{fig_E}
\end{figure}
As an illustration we show in Fig.~\ref{fig_E} the corresponding degenerate even/odd partner eigenfunctions linked with the first excited electronic state in a $C_{3v}$ quantum wire. 

From the numerical point of view the advantage of the ``mixed'' reduced partner wavefunction~(\ref{eqt_ERed}) on one sixth of the domain is that one can use as elementary variables at every node the two values corresponding  the first and second partner functions. This allows to keep the same sparsity pattern linked with the connectivity on one sixth, but with blocks of two variables. In this way node-renumbering can be avoided also for the $E$ representation.

Among the important points of this section, let us mention that now there is no need to incorporate explicitely any of the non-trivial boundary conditions in the set of reduced problems~(\ref{eqt_PbRed}) with $H^{\Gamma}_{Red}$ (and eventually $M^{\Gamma}_{Red}$) given by Eqs.~(\ref{eqt_HA1Red}) or~(\ref{eqt_HA2Red}) or~(\ref{eqt_HERed}), they will be automatically satisfied, whilst the hermiticity of the matrix problem will be preserved.

\section{The case of spinorial sets of functions}
\label{sec_SDR_spin}
In the last sections, we presented two essential methods allowing to choose an optimal bloch function basis for a spin dependent problem~(Sec.~\ref{sec_OBB}) and to perform a spatial domain reduction for a spinless problem~(Sec.~\ref{sec_SDR}) .
A combination of these two approaches allows to easily reduce spin-dependent problems, as we show in the present section. For convenience we shall restrict again our discussion to the $C_{3v}$ case, although the procedure is general and quite simple: one starts with the spinorial eigenstates~(\ref{eqt_2E32})-(\ref{eqt_E12}), labelled by the double group irreps $\widetilde{\Gamma}$, and, using the transformation defined by Eq.~(\ref{eqt_ultEnvFctComp}), one can construct the corresponding vector of UREF's for each irrep $\widetilde{\Gamma}={}^2\!E_{3/2},{}^1\!E_{3/2},E_{1/2}$. Since every UREF can be considered as a scalar function labelled by a single group composite label (irrep + partner labels), one is able to treat them with the SDR technique by dividing each one into $13$ pieces corresponding to the $13$ subdomains. The reduction must then be carried out simultaneously in the sets of UREF's linked with every irrep $\widetilde{\Gamma}$, by selecting the minimum number of independent parts. 

This composite procedure leads to the following three reduced sets of UREF's:
\begin{equation}
\label{eqt_psi_red1}
\psi_{Red}^{{}^2\!E_{3/2}}=
\left(
\begin{array}{c}
\phi_{Red}^{A_2} \vspace{.15cm}\\
\phi_{Red}^{E} \vspace{.15cm}\\
\phi_{Red}^{A_1} 
\end{array}
\right)
\hspace{1cm}
\psi_{Red}^{{}^1\!E_{3/2}}=
\left(
\begin{array}{c}
\phi_{Red}^{A_1} \vspace{.15cm}\\
\phi_{Red}^{E} \vspace{.15cm}\\
\phi_{Red}^{A_2} 
\end{array}
\right)
\end{equation}
\begin{equation}
\label{eqt_psi_red2}
\psi_{Red}^{E_{1/2}} =
\left(
\begin{array}{c}
\phi_{Red}^E \vspace{.15cm}\\
\phi_{Red}^{A_1} \vspace{.15cm}\\
\Phi_{Red}^E \vspace{.15cm}\\
\phi_{Red}^{A_2} \vspace{.15cm}\\
\varphi_{Red}^E 
\end{array}
\right)
\end{equation}
To simplify the notations we have omitted in Eq.~(\ref{eqt_psi_red1}) the implicit double group label ${}^2\!E_{3/2}$ which should also be beared by every component function $\phi_{Red}^{A_2},\phi_{Red}^{E}$ and $\phi_{Red}^{A_1}$ of the reduced spinor $\psi_{Red}^{{}^2\!E_{3/2}}$ (and similarly for $\psi_{Red}^{{}^1\!E_{3/2}}$ and $\psi_{Red}^{E_{1/2}}$). Despite the fact that in Eq.~(\ref{eqt_psi_red2}) the degenerate irrep $E$ appears three times in the UREF decomposition (c.f. Eq.~(\ref{eqt_E12})), we outline that the corresponding UREFs $\phi_{Red}^E, \Phi_{Red}^E$, and $\varphi_{Red}^E$ are {\em distinct} functions and correspond to independent variables. Using the reduced spinors~(\ref{eqt_psi_red1}) and~(\ref{eqt_psi_red2}) one can construct in each case the corresponding $S^{\widetilde{\Gamma}}_i$ matrices (analogously to Eq.~(\ref{eqt_S})) which allow to reduce as in~(\ref{eqt_PbRed}) the full Luttinger Hamiltonian on a sixth of the structure for each double group irrep $\widetilde{\Gamma}$.

The last step to achieve the full reduction of the Hamiltonian on one sixth of the domain is to construct the most general $52\times 52$ block-form valence-band Luttinger Hamiltonian operating on every subdomain separately. We shall take advantage of the fact that, as discussed in Sec.~\ref{sec_OBB}, the $p,q,r,s$ operators can be decomposed with respect to the single group irreps in a similar way as for the spinorial eigenstates. In particular, for $C_{3v}$, $p, q$ are $A_1$-type operators whilst the couple $(r, s)$ forms an ITO transforming like the $E$ irrep.\\ 
Hence since $p$ and $q$ are $A_1$, their most general form is identical to the form of the scalar Hamiltonian appearing in Eq.~(\ref{eqt_HA1}), whilst the block form of $r$ and $s$ must be specifically computed using the restriction that they are partner operators in $E$. To this end one must simply restart from the $13 \times 13$ connectivity block matrix $H_c$ and use the projectors on $E$: $H_c^{E,i} = \frac{2}{|\mathcal{G}|} \sum_{g\in \mathcal{G}} D^{E}_{ii}{}^*(g) \, {\cal P}_g H_c {\cal P}_g^{-1}$, with $i=1,2$ for $r,s$ respectively. In this way the full $52\times 52$ block Hamiltonian can be constructed. 

The reduced Hamiltonian and mass matrices $H_{Red}^{\widetilde{\Gamma}}$ and $M_{Red}^{\widetilde{\Gamma}}$ for each irrep $\widetilde{\Gamma}={}^2\!E_{3/2},{}^1\!E_{3/2},E_{1/2}$ can be obtained using the new reduction matrices $S^{\widetilde{\Gamma}}_i$ resulting from Eqs.~(\ref{eqt_psi_red1}) and~(\ref{eqt_psi_red2}). The form of these reduced Hamiltonian is discussed in the foregoing subsections.

\subsection{Reduced Hamiltonian for the non-degenerate irreps ${}^i\!E_{3/2}$}
\label{subsec_reduced_H_iE32}
For the non-degenerate irreps ${}^i\!E_{3/2}$, $i=1,2$, the reduced Hamiltonians read
\begin{equation}\label{eqt_Hred_2E32}
H^{{}^2\!E_{3/2}}_{Red} =
\left(
\begin{array}{ccc}
H_{Red}^{A_2}(p+q) & C_{A_2-E}(r,s) & 0 \\ 
{} & 2 \, H_{Red}^E(p-q) & C_{E-A_1}(r,s)\\
h.c. & {} & H_{Red}^{A_1}(p+q)
\end{array}
\right)
\end{equation}
\begin{equation}\label{eqt_Hred_1E32}
H^{{}^1\!E_{3/2}}_{Red} =
\left(
\begin{array}{ccc}
H_{Red}^{A_1}(p+q) & C_{A_1-E}(r,s) & 0 \\ 
{} & 2 \, H_{Red}^E(p-q) & C_{E-A_2}(r,s)\\
h.c. & {} & H_{Red}^{A_2}(p+q)
\end{array}
\right)
\end{equation}
On the diagonal we find that $H_{Red}^{A_1}, H_{Red}^{A_2}, H_{Red}^E$ have a form exactly similar to the reduced Hamiltonian for the spinless conduction band problem~(\ref{eqt_HA1Red},\ref{eqt_HA2Red}) and~(\ref{eqt_HERed}), and depend only on the $p+q$ and $p-q$ differential operators.

The non-diagonal coupling terms $C_{\Gamma_1-\Gamma_2}$, $\Gamma_i=A_1, A_2$ or $E$ have the following forms
\begin{eqnarray}
\lefteqn{C_{E-A_1}(r,s) =} \nonumber \\
&= &\left(
\begin{smallmatrix}
\frac{1}{2}\,r_{\cB{1},\cC} & \frac{1}{2}\,r_{\cB{1}} & r_{\cB{1},\cS{1}} & r_{\cB{1},\cB{2}} \\ 
0 & r_{\cS{1},\cB{1}} & r_{\cS{1}} & r_{\cS{1},\cB{2}} \\ 
0 & s_{\cS{1},\cB{1}} & s_{\cS{1}} & s_{\cS{1},\cB{2}} \\ 
2\,r_{\cB{2},\cC} & r_{\cB{2},\cB{1}} + \sqrt{3}\,s_{\cB{2},\cB{1}} & r_{\cB{2},\cS{1}} + \sqrt{3}\,s_{\cB{2},\cS{1}} & 2\,r_{\cB{2}} 
\end{smallmatrix}
\right)\nonumber \\
\label{CEA1}
\end{eqnarray}
\begin{equation}
C_{E-A_2}(r,s) = \left(
\begin{smallmatrix}
-s_{\cB{1},\cS{1}}\\
-s_{\cS{1}}\\
r_{\cS{1}}\\
\sqrt{3}\,r_{\cB{2},\cS{1}} - s_{\cB{2},\cS{1}}
\end{smallmatrix}
\right) 
\label{CEA2}
\end{equation}
and depend only on the $r$ and $s$ operators. These terms are a consequence of band coupling between different envelope functions which is a feature of the Luttinger Hamiltonian~(\ref{eqt_lutt}). This band coupling is  {\em fully compatible} with the wavefunction decoupling during symmetry operations achieved by the OBB (c.f. Eq.~(\ref{eqt_TrsfUltEnvFct})). If one would neglect band coupling, i.e. band-mixing, the reduced Hamiltonian would be a block-diagonal set of decoupled scalar Hamiltonians. 

Since $r$ and $s$ are not self-adjoint operators, we have $C_{A_i-E} \neq C^+_{E-A_i}$, but the natural correspondence is $C_{A_i-E} = C^T_{E-A_i}(r_{ij}\rightarrow r_{ji},s_{ij}\rightarrow s_{ji}),$ where $r_{ij}$ and $s_{ij}$ are the blocs appearing in the $r,s$ operators (c.f. Eqs.~(\ref{CEA1}-\ref{CEA2})).

It is apparent in the structure of the reduced Hamiltonians~(\ref{eqt_Hred_2E32}) and~(\ref{eqt_Hred_1E32}) that they are linked to each other by time-reversal invariance. Nevertheless in the case of quantum wires one should pay attention to the fact that the parallel wavevector $\textbf{k}$ is reversed by time reversal, so that for the same $\textbf{k}$-value the Hamiltonians~(\ref{eqt_Hred_2E32}) and~(\ref{eqt_Hred_1E32}) give rise to different eigenstates. In quantum dots the eigenstates of~(\ref{eqt_Hred_2E32}) and~(\ref{eqt_Hred_1E32}) form Kramers  degenerate pairs and it is enough to solve for one of them only.

\subsection{Reduced Hamiltonian for the degenerate irrep $E_{1/2}$}
\label{subsec_reduced_H_iE12}
Let us now discuss the most complex case, the 2D $E_{1/2}$ irrep, with the same reduction technique. We obtain the reduced Hamiltonian:
\begin{widetext}
\begin{equation}\label{eqt_Hred_E12}
H^{E_{1/2}}_{Red} =
\frac{1}{2}\left(
\begin{smallmatrix}
2H^E_{Red}(p+q) & \frac{1}{\sqrt{2}}C_{E-A_1}& -\frac{1}{\sqrt{2}}C_{E-E}& \frac{1}{\sqrt{2}}C_{E-A_2}&0\\
{} & H^{A_1}_{Red}(p-q) & 0 & 0 & \frac{1}{\sqrt{2}}C_{A_1-E}\\
{} & {} & 2H^{E}_{Red}(p-q) & 0 & \frac{1}{\sqrt{2}}C_{E-E}\\
{} & {} & {} & H^{A_2}_{Red}(p-q) & \frac{1}{\sqrt{2}}C_{A_2-E}\\
c.c. & {} &{} & {} & 2H^{E}_{Red}(p+q) 
\end{smallmatrix}
\right)
\end{equation}
It is interesting to note that the coupling terms $C_{E-A_i}$ and $C_{A_i-E}, i=1,2$, are the same operators which appeared in the previously reduced Hamiltonians~(\ref{eqt_Hred_2E32}) and~(\ref{eqt_Hred_1E32}). However there is only one additional block, $C_{E-E}$, given by
\begin{equation}\label{eqt_block_EE}
C_{E-E}(r,s)=
\left(
\begin{smallmatrix}
\frac{1}{2}r_{\cB{1}} & r_{\cB{1},\cS{1}} & -s_{\cB{1},\cS{1}} & r_{\cB{1},\cB{2}}-\sqrt{3}s_{\cB{1},\cB{2}} \\ 
r_{\cS{1},\cB{1}} & r_{\cS{1}} & -s_{\cS{1}} & r_{\cS{1},\cB{2}}-\sqrt{3}s_{\cS{1},\cB{2}} \\ 
-s_{\cS{1},\cB{1}} & -s_{\cS{1}} & -r_{\cS{1}} & -s_{\cS{1},\cB{2}}-\sqrt{3}r_{\cS{1},\cB{2}} \\ 
r_{\cB{2},\cB{1}}+\sqrt{3}s_{\cB{2},\cB{1}} & r_{\cB{2},\cS{1}}-\sqrt{3}s_{\cB{2},\cS{1}} & -s_{\cB{2},\cS{1}}-\sqrt{3}r_{\cB{2},\cS{1}} & -4r_{\cB{2}} 
\end{smallmatrix}
\right)
\end{equation}
\end{widetext}
\subsection{MSRF post-symmetrization technique}
In the previous subsections we have simultaneously applied the OBB and the SDR techniques, what we call the MSRF technique, to decompose totally the envelope functions into ``Ultimately Reduced Envelope Functions'' (UREF) on a minimal domain, for a problem involving spinors and subject to double group symmetry. As result of the explicit separation of the spinorial and spatial part, in the double group Hamiltonians, all UREF's were shown to depend only on the single group irreps, for $C_{3v}$ the $A_1$, $A_2$ and $E$ irreps. We have also obtained reduced Hamiltonians and showed that they have in this formalism fully symmetrized matrix elements. The reduced Hamiltonians are block-diagonal and their structure is as follows: diagonal blocks correspond to scalar Hamiltonians $H^\Gamma (p\pm q)$ and off-diagonal coupling blocks $C_{\Gamma_1,\Gamma_2}(r,s)$ which only depend on the operators $(r,s)$ and where the irreps $\Gamma_1$ and $\Gamma_2$ are single group representations which correspond to a coupling between UREF's with symmetry $\Gamma_1$ and $\Gamma_2$.\\

One may however wonder about the complexity of the MSRF technique, which could be cumbersome numerically because of the need to recode existing $k\cdot p$ codes. Actually it is not absolutely necessary to modify an existing $k\cdot p$ code to benefit from most of the advantages of MSRF. We shall therefore conclude this section by outlining how the MSRF symmetrization technique can be used as a ``data post-processing technique''. 

The steps of MSRF post-symmetrization can be described are as follows: 
\begin{enumerate}
\item Identification of the main symmetry elements: a) the symmetry group by taking the common symmetry elements between the $k\cdot p$ Hamiltonian and the heterostructure, and b) the subdomains for the SDR decomposition and buildup ${\cal P}_g$ matrices.
\item Construction of the OBB, this requires the following steps: a) identification of the set of $\underline{\underline{V}}(\tilde{g})$ matrices linked with the original $k\cdot p$ envelope function formulation and check of Eq.~(\ref{eqt_invHL}), b) analytical block-diagonalization of the representation $\underline{\underline{V}}(\tilde{g})$, and c) definition of the OBB via Eqs.~(\ref{eq_newBasis}, \ref{eqt_transfoVmat}, \ref{eqt_blocksVB}).
\item Numerical computation of all eigenstates of interest with existing $k\cdot p$ code and interpolation of the envelope wavefunctions on a symmetrized grid corresponding to the SDR subdomain decomposition
\item Symmetry classification of numerical eigenstates. This requires the following steps: a) for all eigenstates perform numerically the change in basis towards the OBB, b) symmetry classification of the interpolated eigenstates by computing the expectation value of the projectors on every irrep $\widetilde{\Gamma}$ (identify also accidental degeneracies), c) identify partner functions belonging to the same eigenspace with the help of symmetry operations (and average the corresponding eigenvalues if there would be a slight numerical lifting of the degeneracy), d) change in basis within every eigenspace to further fix the form of the representation matrices linked with partner functions, e) eventual check the transformation laws~(\ref{eqt_newReconnect}). 
\item Construction of the UREF's using Eq.~(\ref{eqt_ultEnvFctComp}), and eventual check of UREF's transformation laws~(\ref{eqt_TrsfUltEnvFct}).
\end{enumerate}
Note that the order of some of these steps can be interchanged. All analytical steps can be easily computed and automated on a personal computer. In~\cite{GALL09,DUPGALDUP} we have put into practice the automatic recognition of mode symmetries, as well as the techniques of postsymmetrization and postconstruction of UREFs (called URCFs in the case of~\cite{GALL09}). 

We see here that MRSF symmetrization can be viewed as a ``data post-processing technique'', but which still does allow to benefit from all subsequent advantages. 

In the next section, we shall demonstrate in practice the potential of MSRF by computing matrix elements linked with optical transitions. We shall show that one can obtain in our $C_{3v}$-example novel analytical expressions for optical polarization anisotropy which were far from obvious beforehand, and which go beyond the standard use of selection rules for eigenstates. Typically such specific advantage of MSRF is due to the formulation in terms of UREF's, and the possibility to take advantage of selection rules at the intermediate level of UREF's. Another big numerical advantage of MSRF symmetrization will be that in any type of subsequent matrix element computations the reduced domain can also be used, speeding tremendously computation. This can be especially important in heavy cases, for example when computing multidimensional Coulomb integrals for exciton complexes in quantum dots, or any complex object like polaron, etc\ldots Like any symmetrized theory, MSRF enables the use of minimal coupling scenarios on a wide scale, even for possible fine-structure splittings due to symmetry breaking effects of all kinds.\\

\section{Selection rules and matrix elements with the MSRF formalism}
\label{sec_SR}
In this section, we apply the MSRF formalism to computation of matrix elements of operators. Wigner-Eckart Theorem (WET) allows to obtain selection rules for operators, with certain symmetry, sandwiched between symmetry-classified states. With our MSR formalism, one can of course recover the general selection rules predicted by WET, but in addition we can further obtain systematically simpler expressions for the amplitude of transitions through the help of UREF's.

The WET gives selection rules for an operator $A^{\Gamma_c}$ transforming like an irreducible representation $\Gamma_c$ between two states bearing the representation $\wGam_l$ on the left and $\wGam_r$ on the right: $\left< \psi^{\wGam_l} \right| A^{\Gamma_c}\left|\psi^{\wGam_r} \right> =0$ if $\wGam_l$ does not appear in the reduction of $\Gamma_c\otimes\wGam_r$ (or, in a symmetric way, if $A_1$ does not appear in $\wGam_l^*\otimes\Gamma_c\otimes\wGam_r$). If $A^{\Gamma_c}$ is a member of an ITO $A^{\Gamma_c}_{\mu_c}$, the generalized Wigner-Eckart Theorem (gWET) gives more information about the amplitudes by factorizing $\left< \psi^{\wGam_l}_{\mu_l} \right| A^{\Gamma_c}_{\mu_c}\left|\psi^{\wGam_r}_{\mu_r} \right>$ into the product of ``reduced matrix elements'', which only depend on irreps $\wGam_l,\Gamma_c,\wGam_r$ and on physics-independent Clebsch-Gordan (CG) coefficients which involve also partner indices $\mu_l,\mu_c,\mu_r$~\cite{COR}. With our formalism, it is easy to further separate the spinorial and spatial part of $\left< \psi^{\wGam_l}_{\mu_l} \right| A^{\Gamma_c}_{\mu_c}\left|\psi^{\wGam_r}_{\mu_r} \right>$ and  simplify the result using only single group selection rules linked with UREF's. This procedure amounts to use further gWET selection rules at an intermediate level in the theory.

\subsection{Selection rules at the intermediate level of UREF's}
\label{subsec_SR_intermediate}

Equation~(\ref{eqt_blocksPsiV}) is formally equivalent to write $\left| \psi^{\wGam_i}_{\mu_i}\right> = \sum_{\wGam_b,\mu_b} \psi^{\wGam_i,\wGam_b}_{\mu_i,\mu_b}(\textbf{r}) \, \left| \wGam_b,\mu_b\right>$ where the ket belongs to the OBB ~(\ref{eq_newBasis}). The envelope functions $\psi^{\wGam_i,\wGam_b}_{\mu_i,\mu_b}(\textbf{r})$ of section~\ref{sec_OBB} can be systematically decomposed into UREF's $\phi^{\wGam_i,\Gamma_a}_{\wGam_b,\mu_a}(\textbf{r})$, which bear a single group irrep index $\Gamma_a$ and a corresponding partner function index $\mu_a$ (c.f. Eq.~(\ref{eqt_fixedVarDev})).\\

Let us now return to the main problem of finding the matrix elements of ITO operators $A^{\Gamma_c}_{\mu_c}$ with, for simplicity, the restriction that $A^{\Gamma_c}_{\mu_c}$ operates only on spinorial coordinates. The matrix element can be written as
\begin{eqnarray}
\lefteqn{\left< \psi^{\wGam_l}_{\mu_l} \right| A^{\Gamma_c}_{\mu_c} \left| \psi^{\wGam_r}_{\mu_r} \right> = } \nonumber \\
& & \sum_{\wGam_{b'},\mu_{b'},\wGam_b,\mu_b} \, a(\wGam_l,\mu_l,\wGam_{b'},\mu_{b'};\wGam_r,\mu_r,\wGam_{b},\mu_{b}) \nonumber \\
& &\qquad\qquad\qquad\,\times \left< \wGam_{b'},\mu_{b'} \right| A^{\Gamma_c}_{\mu_c} \left| \wGam_{b},\mu_{b} \right>
\label{eq_SR}
\end{eqnarray}
where the gWET can be used to evaluate $\left< \wGam_{b'},\mu_{b'} \right| A^{\Gamma_c}_{\mu_c} \left| \wGam_{b},\mu_{b} \right>$ (this is nothing else than the matrix elements of a $\Gamma_c$-ITO expressed in the OBB). The coefficients $a(\wGam_l,\mu_l,\wGam_{b'},\mu_{b'};\wGam_r,\mu_r,\wGam_{b},\mu_{b})$ can be evaluated in turn as
\begin{eqnarray}
\lefteqn{ a(\wGam_l,\mu_l,\wGam_{b'},\mu_{b'};\wGam_r,\mu_r,\wGam_{b},\mu_{b}) = } \nonumber \\
& & \sum_{\Gamma_a,\mu_a} \, C^{\wGam_l,\wGam_{b'}^*;\Gamma_a}_{\mu_l,\mu_{b'};\mu_a} \, \left[ C^{\wGam_r,\wGam_{b}^*;\Gamma_{a}}_{\mu_r,\mu_b;\mu_{a}} \right]^*  \,\, \left|\left| \phi^{\wGam_l,\Gamma_a}_{\wGam_{b'}} \left|\right. \phi^{\wGam_r,\Gamma_a}_{\wGam_b} \right|\right| \nonumber \\
\label{eq_SR2}
\end{eqnarray}
where we have used the gWET (at the intermediate level of UREF's) to evaluate $\left< \phi^{\wGam_l,\Gamma_{a'}}_{\wGam_{b'},\mu_{a'}} \left|\right. \phi^{\wGam_r,\Gamma_a}_{\wGam_b,\mu_a} \right>$. As a result the so-called ``reduced matrix elements'' appear, they are denoted $\left|\left| \phi^{\wGam_l,\Gamma_a}_{\wGam_{b'}} \left|\right. \phi^{\wGam_r,\Gamma_a}_{\wGam_b} \right|\right|$ and defined by
\begin{eqnarray}
\left|\left| \phi^{\wGam_l,\Gamma_a}_{\wGam_{b'}} \left|\right. \phi^{\wGam_r,\Gamma_a}_{\wGam_b} \right|\right| &= & \left< \phi^{\wGam_l,\Gamma_a}_{\wGam_{b'},\mu_a} \left|\right. \phi^{\wGam_r,\Gamma_a}_{\wGam_b,\mu_a} \right> \nonumber \\
&= & \int d^d\textbf{r} \left( \phi^{\wGam_l,\Gamma_a}_{\wGam_{b'},\mu_a}(\textbf{r}) \right)^* \phi^{\wGam_r,\Gamma_a}_{\wGam_b,\mu_a}(\textbf{r}) \nonumber \\
\label{eq_scalprod}
\end{eqnarray}
Note that, according to the gWET, the reduced matrix elements are independent of the partner function index $\mu_a$ appearing in the right hand side.

To conclude let us emphasize that such use of gWET and corresponding selection rules ``at an intermediate level'' is a salient feature enabled by the use of OBB and the resulting appearance of UREF's. Note that it allows in addition to minimize strictly the number of integrals to evaluate, furthermore the use of the SDR technique is enabled so as to compute all such integrals on the minimum domain.\\

Besides these numerical advantages, we shall show in the example treated in the foregoing section that novel strong {\em analytical} features, physically observable, can also be most straighforwardly calculated with the use of the gWET ``at an intermediate level''.\\

\subsection{Polarization anisotropy for ground state momentum matrix elements in $C_{3v}$ symmetry}
\label{subsec_SR_momentum}

Our example will be an operator important for optical transitions in semiconductors: the momentum operator  $\vec{P} = P_x\, \hat{e}_x + P_y\, \hat{e}_y + P_z\, \hat{e}_z$ between a ground conduction band state and valence-band states in $C_{3v}$ quantum wires and quantum dots ($\hat{e}_d$ ($d=x,y,z$) are unit vectors along the main polarization directions~\cite{BAST}, c.f.~Fig.~\ref{fig_section}). A small complication is that one must include here the conduction band spin, by considering the products of the envelope function single group irreps $A_1,A_2,E$ with the irrep $E_{1/2}$ for spin. The representation $\underline{\underline{V}}^B(\tilde{g})$ corresponding to the OBB for valence band is given in Eq.~(\ref{eq_W32}). For conduction band we choose $\underline{\underline{V}}^B(\tilde{g}) = D^{E}(g) \chi^{{}^2\!E_{3/2}}(g)$, corresponding to the central $2\times 2$ block of $\underline{\underline{V}}^B(\tilde{g})$. Since $P_x$ is  $A_1$ and $(P_y,P_z)$ form an ITO linked with $E$, the application of the gWET at the intermediate level on $\left< \wGam_{b'},\mu_{b'} \right| P_d \left| \wGam_{b},\mu_{b} \right>, d=x,y,z$ leads to the following $2\times 4$ matrix representation for $P_x\, ,\, P_y\, ,\, P_z$:
\begin{eqnarray}
P_x &= & P_0\left(
\begin{smallmatrix}
0 & \sqrt{2/3} & 0 & 0 \\
0 &  0 & \sqrt{2/3} & 0 
\end{smallmatrix}
\right)
\nonumber\\
P_y &= & P_0\left(
\begin{smallmatrix}
0 & -i/\sqrt{6} & 0 & 1/\sqrt{2} \\
1/\sqrt{2} & 0 & i/\sqrt{6} & 0 
\end{smallmatrix}
\right)
\nonumber\\
P_z &= & P_0\left(
\begin{smallmatrix}
-1/\sqrt{2} & 0 & i/\sqrt{6} & 0 \\
0 & i/\sqrt{6} & 0 & 1/\sqrt{2} 
\end{smallmatrix}
\right)
\label{eqt_kane}
\end{eqnarray}
where $P_0 = \left<S \left| P_x \right| X\right> = \left<S \left| P_y \right|  Y\right> = \left<S  \left| P_z \right|Z\right>$ is the Kane matrix element. Another equivalent way to obtain them is to apply the $\underline{\underline{U}}(c_2)$ transformation on the standard Kane matrices~\cite{BAST}.

Let us now consider the momentum  operator matrix elements related to optical transitions from the valence band to ground conduction band state, assuming that the latter has $A_1$ symmetry. This is generally the case, for common QWRs or QDs. Neglecting electron spin, we can use single group representations. Including electron spin, this ground state becomes twice degenerate and transforms like the product irrep $A_1\otimes E_{1/2} = E_{1/2}$, the conduction band spinor can then be written simply as
\begin{equation}\label{eq_A1_E12}
\underline{\psi}_{c,1}^{E_{1/2}(A_1)}\!\!\left(\mathbf{r}\right) = \left( \!\!\begin{array}{c} \psi_c^{A_1}\left(\mathbf{r}\right) \\ 0 \end{array} \!\!\right)
\,\text{,}\,\,
\underline{\psi}_{c,2}^{E_{1/2} (A_1)}\!\!\left(\mathbf{r}\right) = \left( \!\!\begin{array}{c} 0 \\ \psi_c^{A_1}\left(\mathbf{r}\right) \end{array} \!\!\right)
\end{equation}
where the index $c$ recalls the nature of the state and may scan all conduction band states of symmetry $A_1$.

Using the conduction band spinors~(\ref{eq_A1_E12}) we shall evaluate the squared momentum matrix elements between such $E_{1/2}$ ($A_1$) ground conduction band state and the $v$-th valence band state with irrep $\wGam_v$ (either ${}^1\!E_{3/2}, {}^2\!E_{3/2}$ or $E_{1/2}$). The squared momentum matrix elements are defined by 
\begin{equation}\label{eq_SQmomME}
M_{d}(E_{1/2}(A_1)-\wGam_v) = \sum_{\mu_c,\mu_v} \left|  \left< \psi_{c,\mu_c}^{E_{1/2}(A_1)}\right|P_d\left| \psi_{v,\mu_v}^{\wGam_v} \right> \right|^2
\end{equation}
where $\mu_c$ and $\mu_v$ stand respectively for the partner functions of the conduction band state $E_{1/2}$ and the valence-band state $\wGam_v$. The scalar product can be decomposed as  
\begin{eqnarray}
\lefteqn{\left< \psi_{c,\mu_c}^{E_{1/2}(A_1)}\right|P_d\left| \psi_{v,\mu_v}^{\wGam_v} \right> = } \nonumber \\
& &\sum_{\mu_{b'},\wGam_b,\mu_b} \left< E_{1/2},\mu_{b'} \right| P_d \left| \wGam_{b},\mu_{b} \right>   \nonumber \\
& &\quad \, \times\quad a(c:E_{1/2}(A_1),\mu_c,E_{1/2},\mu_{b'};v:\wGam_v,\mu_v,\wGam_b,\mu_{b}) \nonumber \\
\label{eq_momME}
\end{eqnarray}
where $\left< E_{1/2},\mu_{b'} \right| P_d \left| \wGam_{b},\mu_{b} \right>$ and the coefficient $a(c:E_{1/2}(A_1),\mu_c,E_{1/2},\mu_{b'};v:\wGam_v,\mu_v,\wGam_b,\mu_{b})$ are given by Eqs.~(\ref{eqt_kane}) and~(\ref{eq_scalprod}) respectively.

The explicit results are as follows:
\begin{eqnarray}
M_x(E_{1/2}(A_1)-{}^i\!E_{3/2}) &= &0 \; , \;\; i=1,2   \nonumber \\
M_x(E_{1/2}(A_1)-E_{1/2}) &= &\frac{2}{3} \left|\left| \psi_c^{A_1} \left|\right.\Phi_v^{A_1} \right|\right|^2 \nonumber \\
M_y(E_{1/2}(A_1)-{}^i\!E_{3/2}) &=& \frac{1}{2} \left|\left| \psi_c^{A_1} \left|\right.\psi_v^{A_1} \right|\right|^2 \nonumber \\
&=& M_z(E_{1/2}(A_1)-{}^i\!E_{3/2}) \nonumber \\
M_y(E_{1/2}(A_1)-E_{1/2}) &=& \frac{1}{6} \left|\left| \psi_c^{A_1} \left|\right.\Phi_v^{A_1} \right|\right|^2 \nonumber \\
&=& M_z(E_{1/2}(A_1)-E_{1/2}) \nonumber \\
\label{eq_el_mtx}
\end{eqnarray}
The first two results can be easily understood by recalling the fact that $P_x$ is an $A_1$-ITO: according to WET only $\wGam-\wGam$ transitions are allowed since $A_1\otimes\wGam \equiv \wGam$.
Then, coming back to a single group labelling for the conduction band, we find that the $A_1-{}^i\!E_{3/2}, i=1,2$ transitions are forbidden in the $x$-direction.

However eqs.~(\ref{eq_el_mtx}) contain a more striking result: from the second and fourth line we see that for the transition $A_1-E_{1/2}$ there is a constant analytical ratio of oscillator strength (squared momentum matrix element) between the $x$- and $y/z$-directions. This entirely novel analytical prediction is in principle accessible to experimental verification, since the ground valence-band state (in the realistic QWR's investigated in~\cite{DAL1}) is precisely of $E_{1/2}$ symmetry. It has the same value of polarization anisotropy as in quantum wells~\cite{BAST}, but it should be pointed out that this result is still a completely new result, much stronger than saying that the ground transition would be ``quantum well light-hole-like''. Indeed {\em this result holds exactly in presence of band-mixing as well}, like at zone-center of the ground $E_{1/2}$ QWR subband, {\em and also for any transition with same initial and final symmetries, whether it would be in $C_{3v}$ QWRs or $C_{3v}$ QDs, i.e. independently of dimensionality}. In~\cite{Karlsson06} a fairly detailed analysis of optical polarization anisotropy in $C_{3v}$  QDs has been carried out, on the basis of the quantum well heavy/light-hole-like analogy, but the much stronger results obtained here are complementary and bring more light. The complete analysis, including electron excited states as well as other kinds of hole states, will be reported elsewhere~\cite{DAL1}.

To summarize, we have illustrated in this section that, with the new formalism, the matrix elements of operators do take a very simple analytical form, and that only overlaps of scalar UREF functions with the same symmetry are involved. Besides this new advantage, we do not expect additional selection rules at the global level. All the intermediate level calculations become minimal from a computational point of view, and much more transparent. It is the reason why we were able to find novel analytical ratio of oscillator strength between $x$- and $y/z$-polarization directions for certain optical transitions, which hold even in the presence of valence-band mixing.

\section{Application of the MSRF formalism to a variety of symetries}
\label{sec_autres_ex}
Up to now we have discussed only the example of nanostructures with $C_{3v}$ symmetry. In this section we shall shortly discuss the application of MSRF formalism to other specific symmetries:
$C_{6v}$, $D_{3h}$, $C_n$ and $C_s$. $C_{6v}$ is a higher symmetry group presenting a 6 fold axis, some real QD structures like the wurtzite-based GaN/AlN QDs~\cite{SIM,TRO} do display hexagonal symmetry like in Fig.~\ref{fig_C6v}. $D_{3h}$ symmetry can be found at zone-center in $C_{3v}$ QWR's~\cite{DAL1}. $C_n$ are pure rotational sub-groups (without roto-inversions), they typically correspond to the reduced symmetry occurring in previous structures under a magnetic field. Our last example will be the $C_s$ case, since it corresponds to the first type of structure - and the simplest - that we studied in the first section of this paper~\ref{lowCSsymm} (see also~\cite{MAD00,MAD05_2}).\\
In the following, we shall only discuss the symmetry of envelope functions obtained by taking into account the optimal choice of Bloch function basis and vectorial basis in real space.
\subsection{$C_{6v}$ symmetry}
\label{subsec_C6v}

One can explicitly construct $C_{6v}$ from $C_{3v}$ by adding a single $\pi$-rotation $C_2$: $C_{6v}=\left\{ g, gC_2, \quad \forall g\,\in C_{3v}\right\}$. The single group $C_{6v}$ has four 1D irreps $A_i,B_i, i=1,2$, and two 2D irreps $E_i$ , $i=1,2$. Both the 1D irreps and the two 2D irreps are half even and half odd with respect to the new operation $C_2$. For the double group, one has three 2D irreps $E_{i/2}, i=1,3,5$.

Let us now consider the $p,q,r,s$ operators appearing in the valence band Luttinger Hamiltonian (c.f. subsection~\ref{sub_form_env}), and which are second order polynomials in $\textbf{k}$. We note $P,Q,R,S$ the corresponding $3\times 3$ "matrices" of coefficients in such way that $p = \textbf{k}^t P\textbf{k}$ , ... , $s = \textbf{k}^t S\textbf{k}$. For $C_{3v}$ we recalled in subsection~\ref{sub_form_env} that $(r,s)$ form an ITO transforming with the $E$ irrep, indeed it can be shown that the corresponding $R$ and $S$ matrices can be written as
\begin{equation}
R^{C_{3v}}=
\left(\begin{smallmatrix}
0 & a & 0\\
a & b & 0\\
0 & 0 & -b
\end{smallmatrix}\right)
\qquad
S^{C_{3v}}=
\left(\begin{smallmatrix}
0 & 0  & a\\
0 & 0  & -b\\
a & -b & 0
\end{smallmatrix}\right)
\end{equation}
where $a$ and $b$ are some spatially dependent constant depending on the Luttinger parameters linked with the underlying material.

For $C_{6v}$ further decompositions occur, and we can write $r=r_1+r_2$ and $s=s_1+s_2$ where $(r_i,s_i)$ are ITO's transforming like $E_i$ irrep of $C_{6v}$. We obtain 
\begin{equation}\label{eqt_c6v1}
R_1^{C_{6v}}=
\left(\begin{smallmatrix}
0 & \tilde{a} & 0\\
\tilde{a} & 0 & 0\\
0 & 0 & 0
\end{smallmatrix}\right)
\qquad
S_1^{C_{6v}}=
\left(\begin{smallmatrix}
0 & 0 & \tilde{a} \\
0 & 0 & 0\\
\tilde{a} & 0 & 0
\end{smallmatrix}\right)
\end{equation}
\begin{equation} \label{eqt_c6v2}
R_2^{C_{6v}}=
\left(\begin{smallmatrix}
0 & 0 & 0\\
0 & \tilde{b} & 0\\
0 & 0 & -\tilde{b}
\end{smallmatrix}\right)
\qquad
S_2^{C_{6v}}=
\left(\begin{smallmatrix}
0 & 0 & 0\\
0 & 0 & -\tilde{b}\\
0 & -\tilde{b} & 0
\end{smallmatrix}\right)
\end{equation}
where $\tilde{a}=Re (a)$ and $\tilde{b}=i Im (b)$ are respectively associated to the $E_1$ and the $E_2$ ITO's.
In Eqs.~(\ref{eqt_c6v1}) and~(\ref{eqt_c6v2}) the effects of increasing the symmetry appear clearly: 1) the parameters $a$ and $b$ are simplified, 2) every parameter is related to a different irrep $E_i$ in this new decomposition. \\
Another way to explicitly obtain the $C_{6v}$ Luttinger Hamiltonian starting from $C_{3v}$ would be to symmetrize the Hamiltonian with respect to the $C_2$ operation, as follows
\begin{equation}\label{eqt_sym_C6v}
H^{C_{6v}}(\textbf{r},\textbf{k}) =
\frac{1}{2}\left(H^{C_{3v}}(\textbf{r},\textbf{k})+ \vartheta_{C_2} 
H^{C_{3v}}(\textbf{r},\textbf{k})\vartheta^{-1}_{C_2}\right)
\end{equation}
Of course we have to assume in addition that the spatially dependent Luttinger parameters in Eq.~(\ref{eqt_sym_C6v}) are now invariant with respect to the $C_{6v}$ symmetry operations.

Switching to the OBB for $C_{6v}$, we find that the OBB reduces to $E_{1/2}\oplus E_{3/2}$, and that the Luttinger Hamiltonian expressed in the $1/2,-1/2,3/2,-3/2,$ basis (special order) reads
\begin{equation}\label{eqt_HL_PQRS_D3h}
H_L = -\frac{\hbar^2}{m_0}
\left(
\begin{smallmatrix}
(P-Q) I_2 & A \\
A^+ & (P+Q)I_2
\end{smallmatrix}
\right)
\quad 
\textrm{;  } A=
\left(
\begin{smallmatrix}
-s^+ & r \\
r^+ & s
\end{smallmatrix}
\right)
\end{equation}
The decomposition of envelope functions into UREF's, expressed naturally with respect to the $+3/2,+1/2,-1/2,-3/2$ basis (standard order), leads to the following single group irreps (the argument $\mathbf{r}$ of functions is omitted in the following):
\begin{equation*}
\underline{\psi}_1^{E_{1/2}}= \tfrac{1}{\sqrt{2}}\left(
\begin{smallmatrix}
-\phi^{E_1}_2-\phi^{E_2}_2 \\
\phi^{A_1}+\Phi_1^{E_1} \\
\phi^{A_2}-\Phi_2^{E_1} \\
\varphi^{E_1}_1+\varphi^{E_2}_1 
\end{smallmatrix}
\right)
,\,
\underline{\psi}_2^{E_{1/2}}= \tfrac{1}{\sqrt{2}}\left(
\begin{smallmatrix}
\phi^{E_1}_1-\phi^{E_2}_1 \\
-\phi^{A_2}-\Phi_2^{E_1} \\
\phi^{A_1}-\Phi_1^{E_1} \\
\varphi^{E_1}_2-\varphi^{E_2}_2 
\end{smallmatrix}
\right)
\end{equation*}
\begin{equation*}
\underline{\psi}_1^{E_{3/2}}= \tfrac{1}{\sqrt{2}}\left(
\begin{smallmatrix}
-\phi^{A_1}-\phi^{B_1} \\
-\phi^{E_1}_2-\phi_2^{E_2}  \\
\phi^{E_1}_1-\phi_1^{E_2} \\
\phi^{A_2}+\phi^{B_2} 
\end{smallmatrix}
\right)
,\,
\underline{\psi}_2^{E_{3/2}}= \tfrac{1}{\sqrt{2}}\left(
\begin{smallmatrix}
-\phi^{A_2}+\phi^{B_2} \\
\phi^{E_1}_1+\phi_1^{E_2} \\
\phi^{E_1}_2-\phi_2^{E_2} \\
-\phi^{A_1}+\phi^{B_1} 
\end{smallmatrix}
\right)
\end{equation*}
\begin{equation}\label{eq_fcts_C6v}
\underline{\psi}_1^{E_{5/2}}= \tfrac{1}{\sqrt{2}}\left(
\begin{smallmatrix}
\phi^{E_1}_2+\phi^{E_2}_2 \\
\phi^{B_1}+\Phi_1^{E_2} \\
\phi^{B_2}-\Phi_2^{E_2} \\
-\varphi^{E_1}_1-\varphi^{E_2}_1 
\end{smallmatrix}
\right)
,\,
\underline{\psi}_2^{E_{5/2}}= \tfrac{1}{\sqrt{2}}\left(
\begin{smallmatrix}
\phi^{E_1}_1-\phi^{E_2}_1 \\
-\phi^{B_2}-\Phi_2^{E_2} \\
\phi^{B_1}-\Phi_1^{E_2} \\
\varphi^{E_1}_2-\varphi^{E_2}_2 
\end{smallmatrix}
\right)
\end{equation}
The different components in Eqs.~(\ref{eq_fcts_C6v}) indicate, according to the subduction table $C_{6v}\to C_{3v}$~\cite{ALT}, that $E_{1/2}$ and $E_{5/2}$ are related to $E_{1/2}(C_{3v})$ while $E_{3/2}$ is related to ${}^1\!E_{3/2} \oplus {}^2\!E_{3/2}$. As before reduced Hamiltonians can also be calculated.
\subsection{$D_{3h}$ symmetry}

The symmetry $D_{3h}$ is not common in semiconductors nanostructures. However it can be shown in $C_{3v}$-symmetry Quantum Wires (QWR) that it is an approximate symmetry at zone center~\cite{DAL1}. It is obtained from $C_{3v}$ by adding a symmetry operation $\sigma_h$, a planar reflection with respect to an ``horizontal'' symmetry plane, i.e. perpendicular to the rotation axis of $C_3^\pm$ operations. Indeed $D_{3h} = C_{3v} \otimes C_s = \left\{g,g\sigma_h,\, \forall g\in C_{3v}\right\}$, and the single group irreps of $D_{3h}$ are simply the double of those of $C_{3v}$, and correspond to even and odd irreps with respect to the new horizontal symmetry plane $\sigma_h$~\cite{ALT}. For the double group irreps, the situation is slghtly more complex: $D_{3h}$ involves then three 2D irreps $E_{i/2}\, , \, i=1,3,5$. The descent of symmetry tables~\cite{ALT} for $D_{3h} \to C_{3v}$ give the correspondance $E_{3/2} \to {}^1\!E_{3/2}\oplus {}^2\!E_{3/2}$, $E_{1/2} \to E_{1/2}$ and $E_{5/2} \to E_{1/2}$. 

In the case of the Luttinger Hamiltonian, the eigenstate decompositions obtained with MSRF are as follows:
\begin{equation}\label{eqt_D3h_E32}
\underline{\psi}^{E_{3/2}}_1(\textbf{r})=
\left(
\begin{smallmatrix}
\phi^{A'_1}(\textbf{r}) \\
-\phi^{E'}_2(\textbf{r}) \\
-\phi^{E'}_1(\textbf{r}) \\
\phi^{A'_2}(\textbf{r}) 
\end{smallmatrix}
\right)
\, , \quad
\underline{\psi}^{E_{3/2}}_2(\textbf{r})=
\left(
\begin{smallmatrix}
-\phi^{A'_2}(\textbf{r}) \\
\phi^{E'}_1(\textbf{r}) \\
-\phi^{E'}_2(\textbf{r}) \\
\phi^{A'_1}(\textbf{r}) 
\end{smallmatrix}
\right)
\end{equation}
\begin{equation}\label{eqt_D3h_E12}
\underline{\psi}^{E_{1/2}}_1(\textbf{r})=
\left(
\begin{smallmatrix}
-\phi^{E'}_2(\textbf{r}) \\
\Phi^{E'}_1(\textbf{r}) \\
-\Phi^{E'}_2(\textbf{r}) \\
\varphi^{E'}_1(\textbf{r}) 
\end{smallmatrix}
\right)
\, , \quad
\underline{\psi}^{E_{1/2}}_2(\textbf{r})=
\left(
\begin{smallmatrix}
\phi^{E'}_1(\textbf{r}) \\
-\Phi^{E'}_2(\textbf{r}) \\
-\Phi^{E'}_1(\textbf{r}) \\
\varphi^{E'}_2(\textbf{r}) 
\end{smallmatrix}
\right)
\end{equation}
\begin{equation}\label{eqt_D3h_E52}
\underline{\psi}^{E_{5/2}}_1(\textbf{r})=
\left(
\begin{smallmatrix}
\phi^{E'}_2(\textbf{r}) \\
\phi^{A'_1}(\textbf{r}) \\
\phi^{A'_2}(\textbf{r}) \\
-\varphi^{E'}_1(\textbf{r}) 
\end{smallmatrix}
\right)
\, , \quad
\underline{\psi}^{E_{5/2}}_2(\textbf{r})=-
\left(
\begin{smallmatrix}
\phi^{E'}_1(\textbf{r}) \\
-\phi^{A'_2}(\textbf{r}) \\
\phi^{A'_1}(\textbf{r}) \\
\varphi^{E'}_2(\textbf{r}) 
\end{smallmatrix}
\right)
\end{equation}
in the $\left|E_{3/2},1\right>,\left|E_{5/2},1\right>,\left|E_{5/2},2\right>,\left|E_{3/2},2\right>$ basis.

We shall naturally find additional selections rules in $D_{3h}$, they are related to the different "$D_{3h}$ labels" corresponding to otherwise similar $E_{1/2}$ eigenstates when $C_{3v}$-symmetry is only taken into account. Comparing the partner functions of $E_{1/2}(D_{3h})$ and $E_{5/2}(D_{3h})$ presented in Eqs.~(\ref{eqt_D3h_E12}) and~(\ref{eqt_D3h_E52}) to the $E_{1/2}(C_{3v})$ partner functions given in Eq.~(\ref{eqt_E12}), we note that the superpositions of $A_i$ and $E$ functions in the $+1/2,-1/2$ components, separate into two different irreps in $D_{3h}$ (only even scalar function respect to $\sigma_h$ are involved).\\
When the deviation from $D_{3h}$ symmetry is very small, every $E_{1/2}(C_{3v})$ gets only a small $A_i$ or $E$ part in agreement with the partner function decomposition linked to the corresponding irrep $E_{i/2}(D_{3h})\, , \, i=1,5$. In such a case (small symmetry breaking), pure $D_{3h}$-selection rules indicate which transitions have only very small matrix elements due to the approximate symmetry (see e.g.~\cite{DAL1}).
\subsection{The $C_n$ groups : subgroups of the rotations group}
\label{subsec_Cn}
\begin{figure}[hbt]
\begin{tabular}{cc}
\includegraphics[scale = 0.3]{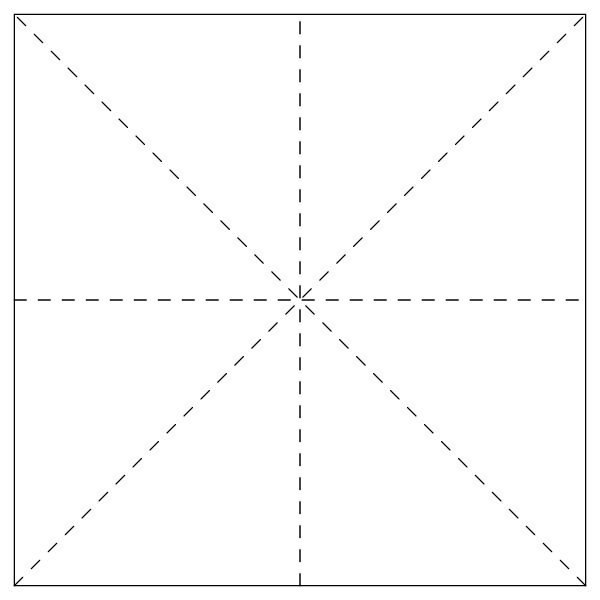} &
\includegraphics[scale = 0.3]{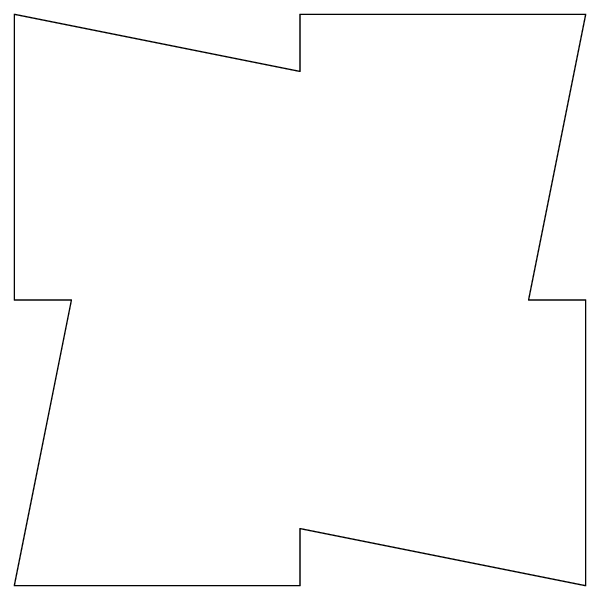} \\
\textbf{(a)} & \textbf{(b)} \\
\end{tabular}
\caption{General shape of quantum structure with symmetry,
\textbf{(a)} $C_{4v}$ ; \textbf{(b)} $C_4$}\label{fig_carre}
\end{figure}

The subgroups $C_n\, , \, n\in \mathbb{N}$ of the pure rotation group $SO(3)$ result from a descent of symmetry from $C_{nv}$. As an illustration we compare in Fig.~\ref{fig_carre} (a) and (b) the shape of a $C_{4v}$ and a $C_{4}$ quantum structure. They are also of particular interest in the important case of an applied magnetic field on a $C_{nv}$ structure (see e.g. the case of $InAs$ quantum dots~\cite{VUK05}). 

Since $C_n$ groups are cyclic and abelian groups, all their single and double group irreps are one-dimensionnal. Therefore in the $C_n$ case, the reduction of representations is equivalent to a diagonalization of the representations. This makes rather trivial the approach in~\cite{VUK05,VUK06}, where the authors study $C_{4v}$ and $C_{6v}$ QDs with and without magnetic field using only the properties due to $C_n$ symmetry. To compare with the present work no additionnal separation in spatial and spinorial representation is performed, there is only a block-diagonalization of the full Hamiltonian into 1D irreps of $C_n$ before numerical computation. We shall show that even in such a simple $C_n$ case, the use of a fully symmetrized basis together with spatial reduction sheds some new light and allows some further analytical and numerical simplifications. 
 
Let us study the $C_4$ symmetry, which is schematically represented in Fig.~\ref{fig_carre} by comparison with $C_{4v}$. The four single group irreps of the group $C_4$ are $A,B,{}^i\!E$, whilst its four double group irreps are ${}^i\!E_{1/2},{}^i\!E_{3/2}\, , \,i=1,2$. The ${}^i\!E, {}^i\!E_{1/2},{}^i\!E_{3/2}$ irreps with $i=1,2$ form three sets of mutually conjugated one-dimensional irreps. Therefore in the OBB the $\underline{\underline{V}}^B(g)$ representations matrices are necessarily diagonal and would correspond to ${}^1\!E_{3/2}\oplus{}^1\!E_{1/2}\oplus{}^2\!E_{1/2}\oplus{}^2\!E_{3/2}$ in the case of our Luttinger Hamiltonian. By contrast, one can see that, despite the fact that the $3\times 3$ representation matrices constructed with the rotation matrices~(\ref{eqt_mtx_R}) are block-diagonal,  these blocks are not naturally labelled by irreps of $C_4$. The $1\times 1$ block is trivially linked with the irrep $A$, and the reduction of the $2\times 2$ block leads to ${}^2\!E\oplus {}^1\!E$. The required switch from the cartesian vectorial basis to the ``Optimal Vectorial Basis'' (OVB) is completely analogous to the OBB, and is given by 
\begin{eqnarray}\label{eqt_3D_OVB}
\hat{e}^A&\equiv & \hat{e}_x \nonumber \\
\hat{e}^{{}^1\!E}&\equiv & \hat{e}_{\sigma-} =\frac{1}{\sqrt{2}}\left( \hat{e}_y - i\hat{e}_z\right)\nonumber \\
\hat{e}^{{}^2\!E}&\equiv & \hat{e}_{\sigma+}=\frac{1}{\sqrt{2}}\left( \hat{e}_y + i\hat{e}_z\right)
\end{eqnarray}
The last two vectors correspond, in optics, to complex unit vectors for circular polarization in the $y-z$ plane. The OVB has been used in ref.~\cite{GALL09} to treat a vectorial problem linked with the eigenmodes of photonic crystal microcavities.

Using such fully symmetrized OVB, simultaneously with the OBB basis, one can also obtain for $C_4$ the ultimate decomposition of spinorial eigenstates into UREF's;
\begin{eqnarray}\label{eqt_C4_32}
\underline{\psi}^{{}^1\!E_{1/2}}=
\left(
\begin{smallmatrix}
\phi^{B} \\
\phi^{A} \\
\phi^{{}^1\!E} \\
\phi^{{}^2\!E} 
\end{smallmatrix}
\right)
&\, , \quad&
\underline{\psi}^{{}^2\!E_{1/2}}=
\left(
\begin{smallmatrix}
\phi^{{}^2\!E} \\
\phi^{{}^1\!E} \\
\phi^{A} \\
\phi^{B} \\
\end{smallmatrix}
\right)\nonumber
\\
\underline{\psi}^{{}^1\!E_{3/2}}=
\left(
\begin{smallmatrix}
\phi^{A} \\
\phi^{B} \\
\phi^{{}^2\!E} \\
\phi^{{}^1\!E} 
\end{smallmatrix}
\right)
&\, , \quad&
\underline{\psi}^{{}^2\!E_{3/2}}=
\left(
\begin{smallmatrix}
\phi^{{}^1\!E} \\
\phi^{{}^2\!E} \\
\phi^{B} \\
\phi^{A} \\
\end{smallmatrix}
\right)
\end{eqnarray}

As we now would like to discuss selection rules for optical transitions between the conduction and valence band, one should remark that the use of the ``optimal vectorial basis'' calls for new corresponding Kane matrices. To this end a preliminary remark is that for conduction band states with spin ($j=1/2$), the $\underline{\underline{V}}^B(g)$ representation reduces to diagonal matrices according to ${}^1\!E_{1/2}\oplus{}^2\!E_{1/2}$. We also note that, in a similar way as for $C_{3v}$, the spinors decomposition correspond to the central part of the corresponding valence band  decomposition ($j=3/2$).

The new Kane matrices are then
\begin{eqnarray}\label{eqt_kane_C4}
P^A_x &= & P_0\left(
\begin{smallmatrix}
0 & \sqrt{2/3} & 0 & 0 \\
0 & 0 & \sqrt{2/3} & 0
\end{smallmatrix}
\right)
\nonumber\\
P^{{}^1\!E}_{\sigma-} &= & P_0\left(
\begin{smallmatrix}
0 & 0 & 0 & 1 \\
0 &  -1/\sqrt{3} & 0 & 0 
\end{smallmatrix}
\right)
\nonumber\\
P^{{}^2\!E}_{\sigma+}&=&P_0\left(
\begin{smallmatrix}
0 & 0 & 1/\sqrt{3} & 0 \\
1 & 0 & 0 & 0 
\end{smallmatrix}
\right)
\end{eqnarray}

As far as selection rules are concerned, it can be checked term by term with Eqs.~(\ref{eqt_C4_32}) and~(\ref{eqt_kane_C4}) that the allowed transitions between a conduction and a valence band state follow indeed the $C_4$ irrep multiplication table~\cite{ALT}, i.e. only diagonal $\Gamma-\Gamma$ transitions are permitted in $x$ direction (where $\Gamma$ is any irrep), and ${}^i\!E_n-{}^{(3-i)}\!E_{(2-n)}$ (with $i=1,2$ and $n =1/2,3/2$) are only allowed in $\sigma_-$ polarisation, whilst only ${}^i\!E_n-{}^{(3-i)}\!E_n$ are only allowed in $\sigma_+$ polarization.
\subsection{Return to the $C_s$ group} 
Although we developped the new formalism to study the more difficult HSH, it is interesting to return to $C_s$ low symmetry heterostructures, like T or V-shaped QWRs seen in~\ref{lowCSsymm} and treated by the Luttinger Hamiltonian, to see how the novel formalism reduces to the old solution technique: the choice of an "optimal quantization axis" perpendicular to the symmetry plane $\sigma$~\cite{MAD00} (see also~\cite{MAD05_2}). This will also definitely put all cases on the same firm ground.

The ``optimal quantization axis choice'' was only a way to choose the OBB which diagonalized the $\underline{\underline{V}}(\sigma)$ related to the planar reflection with respect to the $\sigma$ symmetry plane. With this we obtained even/odd envelope functions~\cite{MAD00} which we can now associate to UREF's labelled by the one-dimensional single group irreps $A',A''$. However we see that the complete reduction of the $4 \times 4$ matrix representation $\underline{\underline{V}}(g)$ of $C_s$ may read as ${}^1\!E_{1/2}\oplus{}^2\!E_{1/2}\oplus{}^1\!E_{1/2}\oplus{}^2\!E_{1/2}$. A few remarks are here in order: 1) since all double group irreps of $C_s$ are one-dimensional it amounts, this particular case, to a diagonalization of the representation (like in $C_4$), 2) the unitary transformation is accomplished by a rotation matrix parametrized by the three Euler angles $(\alpha, \beta, \gamma)$ corresponding to a "3D rotation" $\Re (\alpha, \beta, \gamma)$ of the so-called 
quantization axis direction, 3) $j=3/2$ labels can be kept for the new basis and refer to rotated Bloch functions, 4) we see that for the first time some irreps appear twice in the decomposition (${}^1\!E_{1/2}$ and ${}^2\!E_{1/2}$). Considering the cartesian vectorial basis presented in Fig.~\ref{fig_section}, and a vertical plane normal to the $\hat e_z$ direction, the vectorial basis $\hat e_x,\hat e_y,\hat e_z$ is already optimal and reduces to $A'\oplus A'\oplus A''$. To summarize, the Luttinger Hamiltonian eigenstates can be decomposed in the basis above as
\begin{equation}\label{eqt_Cs_32}
\underline{\psi}^{{}^1\!E_{1/2}} = 
\left(
\begin{smallmatrix}
\phi^{A'} \\
\phi^{A''} \\
\varphi^{A'} \\
\varphi^{A''}
\end{smallmatrix}
\right)
\, , \quad
\underline{\psi}^{{}^2\!E_{1/2}} = 
\left(
\begin{smallmatrix}
\varphi^{A''} \\
\varphi^{A'} \\
\phi^{A''} \\
\phi^{A'}
\end{smallmatrix}
\right)
\end{equation}
where, for every ${}^i\!E_{1/2}$ irrep, $\phi^\Gamma$ and $\varphi^\Gamma$ are different functions with the same symmetry. The functions in $\underline{\psi}^{{}^1\!E_{1/2}}$ and $\underline{\psi}^{{}^2\!E_{1/2}}$ are identical only if $\underline{\psi}^{{}^1\!E_{1/2}}$ and $\underline{\psi}^{{}^2\!E_{1/2}}$ are related by time reversal symmetry. As already seen for $C_4$, for the conduction band with spin, the $j=1/2, m=\pm 1/2$ envelope functions would just be analogous to the central part of the $j=3/2, m=\pm 1/2$ valence band envelope functions. The momentum operators $P_x$ and $P_y$ in the symmetry plane are even with respect to $\sigma$ ($A'$) and the operator $P_z$ perpendicular to the axis is odd ($A''$). The latter $A''$ operator only couples mutually conjugated bands ($M_x^{\Gamma,\Gamma} =M_y^{\Gamma,\Gamma} =0$, selection rule) whilst the former $A'$ operators only allow diagonal $\Gamma-\Gamma$ transitions.
\section{Wide potential applicability and originality of the MSRF formalism}
\label{orig_MSRF}
The foundation of our new MSRF formalism can be summarized, for most general spin dependent problems, as first choosing a really optimal basis (fully symmetrized basis) for the orbital and spinorial space and treating every spinorial component as a sum of symmetrized UREFs. Second, on this set of UREFs the spatial domain reduction (SDR) technique can be applied.
With the fully symmetrized basis, the coupling between different envelope functions becomes minimized and every spinorial function can be decomposed with respect to symmetry using only single group irreps. With the SDR applied to every UREFs one can then achieve domain reduction.
In this context it should therefore be clear for the reader that in band-structure problems any number of bands could easily be treated by the MSRF at the price of more complexity, strain~\cite{BIRPIKUS} could also be treated (treating also by MSRF the separate strain equations in a self-consistent way), as well as the Burt-Foreman interface terms~\cite{BURT92,BURT99, FOR93}, or the presence of an external field like a magnetic field (if one takes into account the lower global symmetry in this case). The choice of fully symmetrized basis presented in Sec.~\ref{sec_OBB} only depends on the symmetry group considered (some other examples have been given in Sec.~\ref{sec_autres_ex}). 

The SDR technique used in Sec.~\ref{sec_SDR} can be easily generalized, but one must realize that it may need slight tuning depending on the problem at hand. For example for higher order FEs (like second order FEs which are frequently used), the general SDR procedure would be exactly the same, provided one is careful in considering anew the connectivity between sub-domains (a few new blocks must be included). 

It is important to note that the MSRF formalism is not restricted to our example of a 2D $k.p$ spinorial problem but can be easily generalized for any other more general vectorial, spinorial or tensorial-like problems, also with different dimensionality (1D-3D), and a numerical solution with other techniques, whether they would be in the spatial domain or the Fourier domain, or in any other sensible basis. The approach is even not limited to linear problems provided one accounts for proper products of representations. The essence is to symmetrize separately the bases in the spinorial-like space and the real ``coordinate-like'' space. Real space resolution methods, are usually the most powerful for the study of spatially isolated heterostructures (i.e. occurrence of band matrices or sparse matrices), whilst Fourier space decomposition of the envelope functions, is most powerful for periodic structures (i.e. superlattices). Indeed for a non-periodic problem a formulation in a plane wave basis, which give rise to full matrices during the numerical solution, would lead to a significant waste of computer time and memory space compared to real space methods like FD of FE methods. 

In real space formulations the MSRF method sketched above has the further potential to {\em keep the sparse matrix structure} of the numerical problem (when formulated with a FD or FE approach), therefore it can be married with convenient and maximally efficient methods for sparse matrices.

The MSRF formalism is also plainly applicable to the study of strongly coupled periodic structures where symmetrized plane waves decomposition in the way of Vukmirovic et al.~\cite{VUK05} are efficient. For a $k\!\cdot\! p$ Hamiltonian the optimal Bloch function basis would of course be the same, the only part that has to be modified is the spatial domain reduction approach, where one would then introduce the concept of a "reduced plane waves basis". We have already achieved steps in this direction in~\cite{OBRE}, where a procedure to work with a reduced domain in Fourier space has been introduced (see Fig. 6c of~\cite{OBRE} to see the reduced Fourier domain for every irreps of the  $C_{3v}$ group). Once this is done the way to proceed for strongly coupled periodic structures is along the same lines, and would again lead to optimal algorithms and detailed symmetry analyses.

It is also worth to point out that once a calculation is done the completely symmetrized envelope functions provided by the MSRF formalism may also be very useful to further build in a maximally efficient way more complex symmetrized objects. A typical example in semiconductor heterostructures in building excitons, or excitonic complexes, which gives rise to multidimensional Coulomb integrals involving products where not only additional single group selection rules can be used, but where also the domain reduction would allow drastic speedups of the computation of integrals. Of course in this domain too MRSF also has the potential to provide further natural physical insight by the resulting minimal coupling scenario at the individual envelope function level. 

To which extent does the MSRF links with previous work using heterostructure symmetry to simplify the calculation of electronic levels? Little work has in fact been done up to now on this topic. Let us compare our method to the work of Vukmirovic et al.~\cite{VUK05,VUK06} which appears to be closest of ours, in particular in~\cite{VUK05}, a method using symmetrized plane waves decomposition is suggested for a $C_4$ pyramidal QD. Our first comment is that such a Fourier decomposition is most useful only for {\em strongly coupled periodic structures}, like found for example in~\cite{VUK06} for hexagonal QD superlattices. In the other limit of spatially isolated heterostructures the approach is {\em not} numerically optimal because it leads to full matrices instead of band matrices like in real space approaches like FD or FE schemes. Our second comment is that symmetry is far from fully exploited in their approach. Let us discuss this last point in more details. First the authors define an N-dimensional representation of the rotation operators on the full coupled Fourier-spin space (no separate symmetrization), and they show that a "symmetry-adapted basis" can be found (diagonalization of the representation). As expected this allows to block-diagonalize the Hamiltonian, but really it should be pointed out that the block-diagonalization achieved is nothing else than the main decoupling that one should obtain by separating the eigenproblem into its main different solutions, indexed by double-group irreps. Second, the authors show that their method allow to separate the Hamiltonian into $n$ blocks of equivalent sizes which does reduce the necessary computation time, but this is obvious for a $C_n$ group. The limitations of the approach~\cite{VUK05,VUK06} are therefore evident. First, it does not separate orbital and spinorial contributions (only double group irreps appear in the reduction), hence the symmetry properties of the structure are taken into account only at the last step before numerical resolution, no single group classification of envelope functions, no selection rules at the envelope function level can be obtained, nor any domain reduction can be achieved. Second, this approach is well-adapted only for the case of rotation sub-groups $C_n$ where every irrep of the single and double group are 1D. As we have seen earlier already for small dihedral groups like $C_{2v}$, degenerate eigenvalues could appear. Indeed in~\cite{VUK06} the author study a $C_{6v}$ problem (every double group irreps is 2D) but only exploit the $C_6$ symmetry (with only 1D irreps) by neglecting the vertical symmetry planes, since in their framework they cannot describe optimally the further degeneracies of the eigenstates. Of course in the presence of a magnetic field this would be the correct approach, since additional lifting of degeneracies would occur.

\section{Conclusion}
\label{sec_concl}
We have presented a new formalism called MSRF (Maximal Symmetrization and Reduction of Fields) widely applicable to the study of high symmetry heterostructures of various dimensionality (e.g.QWR's or QD's), and well adapted to treat both scalar and spinorial problems. The use of this formalism has been illustrated with particular examples, mainly a $C_{3v}$ QWR.

For scalar functions (e.g. conduction band in the frame of a single band spinless Hamiltonian) it reduces to the spatial domain reduction (SDR) technique. For every irreducible representation (irrep), the independent parameters provide systematically the reduced minimal domain for significant description of every quantum state, and allows to throw away the computation of all the redundant parts by identifying the reduced Hamiltonian for the independent parameters. The method was developed in a pedestrian way for $C_{3v}$, further mathematical developments of the SDR technique will be reported elsewhere.

For spinorial sets of functions (e.g. valence band with a four bands Luttinger Hamiltonian), it is necessary to separate the spinorial and orbital parts and to compute a spatial domain reduction of the spinorial problem. For this purpose we introduced the concept of optimal fully symmetrized basis for the spinorial field (every basis member transforms like a partner function of an irrep of the group). The first advantage is the simplification in spin space, since the Optimal Bloch function Basis (OBB) completely reduces the matrix representation of the transformation laws in spin space into block-diagonal form, which minimizes the complex coupling between spinorial components by symmetry operations. Moreover we showed that every spinorial component can be decomposed further into ultimately reduced envelope functions (UREFs), labelled by single group irreps, which finally allows the application of the SDR technique.

The advantages of such a formalism are manifold. In particular the Hamiltonians take usually a much more simple form. Operators and spinorial component can simply be labelled with single group irreps, and with the help of the SDR technique one can solve a smaller specific problem for every irrep on a reduced spatial domain. This procedure leads to simpler analytical expression for operator matrix elements. Just as with any symmetrized formalism, any symmetry breaking mechanism can be understood more easily, both in a qualitative and quantitative way. Hidden approximate higher symmetries can also be quickly detected by the analysis of selection rules. From the numerical point of view the formalism leads to highly advantageous algorithms, and applies in a flexible way to a wide variety of methods (real or Fourier space), so that the most practical one for the case at hand can be chosen without restriction. The reduction factor in computer time for diagonalization can be estimated to be about $20$ for $C_{3v}$. Last, but not least the MSRF technique can be applied as a post-symmetrization technique on {\em existing numerical results} (no absolute need to recode), giving access to the full power of symmetry analysis at the envelope function level, and greatly increased performance in subsequent numerical computations. In~\cite{GALL09,DUPGALDUP} we have implemented the automatic recognition of mode symmetries, as well as the postsymmetrization technique.

In a forthcoming paper~\cite{DAL1} extensive analytical and numerical results on the electronic and optical properties of a real $C_{3v}$ QWR will be presented and will further demonstrate the power of the MSR approach. 

As already stressed, straightforward generalizations of the method may be developed for arbitrary tensorial fields obeying a set partial differential equations (even non-linear!). Therefore in a forthcoming paper~\cite{GALL09} we shall apply it to Maxwell's equations in a case corresponding to photonic bandgap microcavities~\cite{Painter03}. Other possible application could be connected heterostructure problems (definition of strain fields, phonon fields, etc...). 

A further high potential of MSRF also lies in possible subsequent calculations. Often, in a second stage, one is also interested to build more complex objects, like excitons, polaritons, polarons, etc... To build such objects the symmetrized field components provided by our technique really represent {\em optimal} building blocks, for which selection rules and well-defined transformation properties are readily available by construction, hence MSRF should be of great use to an even larger community. Last, but not least, it is worth pointing out that the spirit of the MSR approach can also be applied to other widely used models for heterostructures like tight-binding, pseudo-potential, etc..., with the same potential analytical and computational advantages.

\textbf{Acknowledgments}
We would like to thank Dr. Fabienne Michelini, Prof. Guy Fishman, and Prof. Ramdas L. Ram-Mohan for useful discussions, as well as Dr. Dmitri Boiko for a thorough critical reading of the manuscript. We acknowledge financial support from Swiss NF project No.200020-109523.

\appendix
\section{$C_{3v}$ point group tables} 
\label{appendix}
In the first part of this appendix we recall the most important $C_{3v}$ point-group tables which are extensively used in the main body of the paper (additional informations and tables can be found in~\cite{ALT}, note that the multiplication table $(c)$ is transposed with  respect to ref.~\cite{ALT} since we use the passive point of view).\\ 

In the second part, since infinitely many equivalent matrix representations can be used for the 2D degenerate irreps of $C_{3v}$, we explicitly specify which 2D matrix representations are used. 
\begin{table}[hbtp]
\begin{center}
\begin{tabular}{c|rrr}
$C_{3v}$ & $E$ & $C_3^\pm$ & $\sigma_{vi}$\\
\hline
$A_1$ & $1$ & $1$ & $1$\\ 
$A_2$ & $1$ & $1$ & $-1$\\ 
$E$ & $2$ & $-1$ & $0$\\ 
\hline
$E_{1/2}$ & $2$ & $1$ & $0$\\ 
${}^1\!E_{3/2}$ & $1$ & $-1$ & $i$\\ 
${}^2\!E_{3/2}$ & $1$ & $-1$ & $-i$\\ 
\end{tabular} 
\hspace{0.5cm}\textbf{(a)}\vspace{.5cm}\\
\begin{tabular}{c|cccccc}
$C_{3v}$&$A_1$ & $A_2$ & $E$ & $E_{1/2}$& ${}^1\! E_{3/2}$& ${}^2\! E_{3/2}$ \\
\hline
$A_1$&$A_1$ & $A_2$ & $E$ & $E_{1/2}$ & ${}^1\! E_{3/2}$& ${}^2\! E_{3/2}$\\
$A_2$&& $A_1$ & $E$ & $E_{1/2}$ & ${}^2\! E_{3/2}$& ${}^1\! E_{3/2}$\\
$E$&& & $A_1\oplus \{A_2\}\oplus E$ & $E_{1/2}\oplus{}^1\! E_{3/2}\oplus{}^2\! E_{3/2}$& $E_{1/2}$ & $E_{1/2}$\\
$E_{1/2}$&&&& $\{A_1\}\oplus A_2\oplus E$& $E$ & $E$\\
${}^1\! E_{3/2}$&&&&& $A_2$ & $A_1$\\
${}^2\! E_{3/2}$&&&&&&$A_2$\\
\end{tabular}
\hspace{0.5cm}\textbf{(b)}\vspace{.5cm}
\\
\begin{tabular}{ccc}
\begin{tabular}{c|cccccc}
$C_{3v}$& $E$ & $C_3^+$ & $C_3^-$ & $\sigma_{v1}$ & $\sigma_{v2}$& $\sigma_{v3}$\\
\hline
$E$& $E$ & $C_3^+$ & $C_3^-$ & $\sigma_{v1}$ & $\sigma_{v2}$& $\sigma_{v3}$\\
$C_3^+$& $C_3^+$ & $C_3^-$ & $E$& $\sigma_{v2}$ & $\sigma_{v3}$& $\sigma_{v1}$\\
$C_3^-$& $C_3^-$ & $E$ & $C_3^+$ & $\sigma_{v3}$ & $\sigma_{v1}$& $\sigma_{v2}$\\
$\sigma_{v1}$& $\sigma_{v1}$ & $\sigma_{v3}$& $\sigma_{v2}$& $E$&$C_3^-$ & $C_3^+$\\
$\sigma_{v2}$& $\sigma_{v2}$ & $\sigma_{v1}$& $\sigma_{v3}$& $C_3^+$ & $E$& $C_3^-$\\
$\sigma_{v3}$& $\sigma_{v3}$ & $\sigma_{v2}$& $\sigma_{v1}$& $C_3^-$ & $C_3^+$ & $E$\\
\end{tabular} 
&\qquad&
\begin{tabular}{c|rrrrrr}
$C_{3v}$& $E$ & $C_3^+$ & $C_3^-$ & $\sigma_{v1}$ & $\sigma_{v2}$& $\sigma_{v3}$\\
\hline
$E$& $1$&$1$&$1$&$1$&$1$&$1$\\
$C_3^+$&$1$&$-1$&$1$&$-1$&$-1$&$-1$\\
$C_3^-$& $1$&$1$&$-1$&$-1$&$-1$&$-1$\\
$\sigma_{v1}$&$1$&$-1$&$-1$&$-1$&$1$&$1$\\
$\sigma_{v2}$& $1$&$-1$&$-1$&$1$&$-1$&$1$\\
$\sigma_{v3}$&$1$&$-1$&$-1$&$1$&$1$&$-1$\\
\end{tabular} \\
\textbf{(c)}&\quad&\textbf{(d)}\\
\end{tabular}
\caption{Character \textbf{(a)}, direct product of irreps \textbf{(b)}, multiplication \textbf{(c)} and factor \textbf{(d)} tables for $C_{3v}$ group}\label{table_char_C3v}
\end{center}
\end{table} 
For the $E$ irrep we have chosen the following 2D matrix representation:
\begin{equation}
\begin{array}{ll}
D^E(E) = \left(
\begin{array}{ll}
 1 & 0 \\
 0 & 1
\end{array}
\right)
&
D^E(C_3^+) = \left(
\begin{array}{ll}
 -\frac{1}{2} & \frac{\sqrt{3}}{2} \\
 -\frac{\sqrt{3}}{2} & -\frac{1}{2}
\end{array}
\right)
\\
D^E(C_3^-) = \left(
\begin{array}{ll}
 -\frac{1}{2} & -\frac{\sqrt{3}}{2} \\
 \frac{\sqrt{3}}{2} & -\frac{1}{2}
\end{array}
\right)
&
D^E(\sigma_{v1}) = \left(
\begin{array}{ll}
 1 & 0 \\
 0 & -1
\end{array}
\right)
\\
D^E(\sigma_{v2}) = \left(
\begin{array}{ll}
 -\frac{1}{2} & -\frac{\sqrt{3}}{2} \\
 -\frac{\sqrt{3}}{2} & \frac{1}{2}
\end{array}
\right)
&
D^E(\sigma_{v3}) = \left(
\begin{array}{ll}
 -\frac{1}{2} & \frac{\sqrt{3}}{2} \\
 \frac{\sqrt{3}}{2} & \frac{1}{2}
\end{array}
\right)
\end{array}
\end{equation}
For the other 2D $E_{1/2}$ irrep, the chosen 2D matrix representation can be constructed with the help of the choice $D^{E_{1/2}}(\tilde{g})=\chi^{{}^2\!E_{3/2}}(\tilde{g})\,D^E(\tilde{g})$.

\end{document}